\documentclass[twocolumn]{aastex63}
\usepackage{epsfig}

\newcommand{\Nmm} 	{$N_\mathrm{mm}$}

\newcommand{\mum}   	{$\mu$m}
\newcommand{\kms}   {km~s$^{-1}$}

\newcommand{\cmt}   {cm$^{-3}$}
\newcommand{\jpb}   {$\rm Jy~beam^{-1}$}    
\newcommand{\lo}    {$L_{\sun}$}
\newcommand{\mo}    {$M_{\sun}$}
\newcommand{\htcop}    {H$^{13}$CO$^+$}

\newcommand{\et}    {et al.}
\newcommand{\eg}    {e.\,g.,}
\newcommand{\ie}    {i.\,e.,}

\newcommand{\supa}  {$^\mathrm{a}$}
\newcommand{\supb}  {$^\mathrm{b}$}
\newcommand{\supc}  {$^\mathrm{c}$}
\newcommand{\supd}  {$^\mathrm{d}$}
\newcommand{\supe}  {$^\mathrm{e}$}

\definecolor{RED}{rgb}{1.0,0.0,0.0}

\received{June 9, 2020}
\revised{January 11, 2021}
\accepted{\today}
\submitjournal{ApJ}

\shorttitle{Fragmentation and magnetic fields}
\shortauthors{Palau et al.}
\begin{document}

\title{Does the magnetic field suppress fragmentation in massive dense cores?}

\correspondingauthor{Aina Palau}
\email{a.palau@irya.unam.mx}
\author[0000-0002-9569-9234]{Aina Palau}
\affiliation{Instituto de Radioastronom\'ia y Astrof\'isica, Universidad Nacional Aut\'onoma de M\'exico, P.O. Box 3-72, 58090, Morelia, Michoac\'an, Mexico}

\author{Qizhou Zhang}
\affiliation{Center for Astrophysics | Harvard \& Smithsonian , 60 Garden Street, Cambridge, MA 02138, USA}

\author{Josep M. Girart}
\affiliation{Institut de Ci\`encies de l'Espai (CSIC-IEEC), Campus UAB, Carrer de Can Magrans S/N, E-08193 Cerdanyola del Vall\`es, Catalonia, Spain}
\affiliation{Institut d’Estudis Espacials de Catalunya, E-08030 Barcelona, Catalonia, Spain}

\author{Junhao Liu}
\affiliation{Center for Astrophysics | Harvard \& Smithsonian , 60 Garden Street, Cambridge, MA 02138, USA}
\affiliation{School of Astronomy and Space Science, Nanjing University, 163 Xianlin Avenue, Nanjing 210023, People's Republic of China}
\affiliation{Key Laboratory of Modern Astronomy and Astrophysics (Nanjing University), Ministry of Education, Nanjing 210023, People's Republic
of China}

\author{Ramprasad Rao}
\affiliation{Center for Astrophysics | Harvard \& Smithsonian , 60 Garden Street, Cambridge, MA 02138, USA}

\author{Patrick M. Koch}
\affiliation{Institute of Astronomy and Astrophysics, Academia Sinica, No.1, Sec. 4, Roosevelt Rd, Taipei 10617, Taiwan}

\author{Robert Estalella}
\affiliation{Departament de F\'isica Qu\'antica i Astrof\'isica, Institut de Ci\`encies del Cosmos, Universitat de Barcelona, (IEEC-UB) Mart\'i i Franqu\`es, 1, E-08028 Barcelona, Spain}

\author{Huei-Ru Vivien Chen}
\affiliation{Institute of Astronomy and Department of Physics, National Tsing Hua University, Hsinchu 30013, Taiwan}

\author{Hauyu Baobab Liu}
\affiliation{Institute of Astronomy and Astrophysics, Academia Sinica, No.1, Sec. 4, Roosevelt Rd, Taipei 10617, Taiwan}

\author{Keping Qiu}
\affiliation{School of Astronomy and Space Science, Nanjing University, 163 Xianlin Avenue, Nanjing 210023, People's Republic of China}
\affiliation{Key Laboratory of Modern Astronomy and Astrophysics (Nanjing University), Ministry of Education, Nanjing 210023, People's Republic
of China}

\author{Zhi-Yun Li}
\affiliation{Astronomy Department, University of Virginia, Charlottesville, VA 22904, USA}

\author{Luis A. Zapata}
\affiliation{Instituto de Radioastronom\'ia y Astrof\'isica, Universidad Nacional Aut\'onoma de M\'exico, P.O. Box 3-72, 58090, Morelia, Michoac\'an, Mexico}

\author{Sylvain Bontemps}
\affiliation{Universit\'e de Bordeaux, Laboratoire d'Astrophysique de Bordeaux, CNRS/INSU, UMR 5804, BP 89, 33271 Floirac Cedex, France}

\author{Paul T. P. Ho}
\affiliation{Institute of Astronomy and Astrophysics, Academia Sinica, No.1, Sec. 4, Roosevelt Rd, Taipei 10617, Taiwan}

\author{Henrik Beuther}
\affiliation{Max Planck Institute for Astronomy, K\"onigstuhl 17, 69117 Heidelberg, Germany}

\author{Tao-Chung Ching}
\affiliation{Institute of Astronomy and Astrophysics, Academia Sinica, No.1, Sec. 4, Roosevelt Rd, Taipei 10617, Taiwan}

\author{Hiroko Shinnaga}
\affiliation{Department of Physics and Astronomy, Graduate School of Science and Engineering, Kagoshima University, Amanogawa Galaxy Astronomy Research Center, 1-21-35 Korimoto, Kagoshima 890-0065, Japan}

\author{Aida Ahmadi}
\affiliation{Leiden Observatory, Leiden University, PO Box 9513, 2300 RA Leiden, The Netherlands}

%\collaboration{1}{(AAS Journals Data Scientists collaboration)}



\begin{abstract}
Theoretical and numerical works indicate that a strong magnetic field should suppress fragmentation in dense cores. However, this has never been tested observationally in a relatively large sample of fragmenting massive dense cores. Here we use the polarization data obtained in the Submillimeter Array Legacy Survey of Zhang et al. (2014) to build a sample of 18 massive dense cores where both fragmentation and magnetic field properties are studied in a uniform way. We measured the fragmentation level, \Nmm, within the field of view common to all regions, of $\sim0.15$~pc, with a mass sensitivity of $\sim0.5$~\mo, and a spatial resolution of $\sim1000$~AU. In order to obtain the magnetic field strength using the Davis-Chandrasekhar-Fermi method, we estimated the dispersion of the polarization position angles, the velocity dispersion of the H$^{13}$CO$^+$\,(4--3) gas, and the density of each core, all averaged within 0.15~pc. A strong correlation is found between \Nmm\ and the average density of the parental core, although with significant scatter. When large-scale systematic motions are separated from the velocity dispersion and only the small-scale (turbulent) contribution is taken into account, a tentative correlation is found between \Nmm\ and the mass-to-flux ratio, as suggested by numerical and theoretical works.
\end{abstract}

%https://journals.aas.org/keywords-2013/#Astronomical_instrumentation,_methods_and_techniques
%techniques: polarimetric


\section{Introduction} \label{si}

\begin{deluxetable*}{lRRCDCCCCC}
%\tablenum{1}
\tablecaption{Properties of the observations used to assess the fragmentation level in the sample of massive dense cores at 0.87 and 1.3~mm\label{tsample}}
\tablewidth{0pt}
\tablehead{
\colhead{} &
\colhead{D} & 
\colhead{$L_\mathrm{bol}$\supb} & 
\colhead{$M_\mathrm{core}$\supb} & 
\multicolumn2c{Rms\supc} & 
\colhead{$M_\mathrm{min}$\supc} &
\colhead{$uv-$range} &
\colhead{Spat. res.\supd} & 
\colhead{LAS\supd} & 
\colhead{}  \\
\colhead{Source\supa} & 
\colhead{(kpc)} & 
\colhead{(\lo)} & 
\colhead{(\mo)}&
\multicolumn2c{(mJy)} &
\colhead{(\mo)} & 
\colhead{(k$\lambda$)} & 
\colhead{(AU)} &
\colhead{(AU)} &
\colhead{Refs.\supe}
}
\decimalcolnumbers
\startdata
1-W3IRS5          		&1.95	&140000		&510		&2.4 		&0.31	&22-210	&1600  	&8050	&1\\
2-W3H2O 			&1.95	&36000		&540		&2.6* 	&1.22	&55-585	&682		&3230	&2,3\\
3-G192				&1.52	&2700		&40		&3.3 		&0.27	&40-260	&1070	&3460	&4,5\\
4-NGC\,6334V			&1.30	&40000		&370		&7.0 		&0.42	&25-210	&1140	&4730	&4,6\\
5-NGC\,6334A(IV)		&1.30	&1000		&600		&6.6 		&0.40	&25-210	&1200	&4730	&4\\
6-NGC\,6334I			&1.30	&48000		&300		&12 		&0.72	&25-210	&1130	&4730	&4\\
7-NGC\,6334In			&1.30	&1300		&730		&17 		&1.03	&25-210	&1190	&4730	&4\\
8-G34.4.0				&1.57	&2300		&150		&7.0		&0.62	&40-260	&1110	&3570	&4\\ 
9-G34.4.1				&1.57	&1100		&310		&2.0 		&0.17	&40-260	&1160	&3570	&4\\ 
10-G35.2N			&2.19	&15000		&1060	&1.3 		&0.22	&40-260	&1620	&4980	&4\\ 
11-IRAS\,20126+4104 	&1.64	&8900		&60		&0.7* 	&0.46	&48-551	&629     	&3120	&7,8\\
12-CygX-N3(DR17) 		&1.40	&200			&400		&1.0* 	&0.26	&20-200	&1400	&6370	&9,10\\
13-W75N(CygX-N30) 	&1.40	&20000		&270		&2.6* 	&0.72	&20-160	&1650	&6370	&11\\
14-DR21OH(CygX-N44)	&1.40	&10000		&490		&4.7 		&0.33	&40-260	&1050	&3150	&4,12\\
15-CygX-N48(DR21OHS) &1.40	&4400		&610		&1.5* 	&0.39	&20-200	&1400	&6370	&9,13\\
16-CygX-N53			&1.40	&300			&240		&1.9* 	&0.49	&20-200	&1400	&6370	&9,13\\
17-CygX-N63(DR22) 	&1.40	&470			&70		&3.0* 	&0.77	&20-200	&1400	&6370	&9,10\\
18-NGC\,7538S		&2.65	&12000		&1120	&0.4* 	&0.43	&68-765	&848		&3520	&2,14\\
%19-CygX-N38(48N,DR21OHW)	&1.40&				&		&		&4200	&		&		&		&Qizhou	\\
%CygX-48N is not included because too poor angular resolution
%20-NGC\,2264C	&0.40	&590				&2.0		&0.08	&1160	&		&$\sim9$	&8		&Cunningham+16 at 1mm\\
%20-NGC\,2264C	&0.40	&590				&0.35	&		&140		&		&9		&Watanabe+17, ALMA band 7 beam 0.35\\
%NGC2264C: distance from Dzib+16?; Lbol estimated from Cunningham+16, who use 2000 Lsun for CMM5 at a distance of 738 pc\\
%NGC2264C is not included to keep a uniform sample with similar distances. Without this source, the distance range from 1.4 to 2.6, which is very good!
%CygX-N53 only presents 3 pol vectors data in Ching+17:
\enddata
\tablecomments{
\\
$^\mathrm{a}$ Complete names commonly used for each massive dense core. In the following tables and figures a short version of the name will be used.\\
$^\mathrm{b}$ $L_\mathrm{bol}$ is calculated with the flux densities used to build the spectral energy distribution for the model described in Section~\ref{sadensity}.  $M_\mathrm{core}$ is calculated by integrating our modeled density structure for each core (Section~\ref{sadensity}) up to the observed radius with the JCMT.\\
$^\mathrm{c}$ Rms at 870~\micron\ (from SMA observations) for all sources, except for those marked with an asterisk, for which the rms corresponds to the image at $\sim1.3$~mm (Plateau de Bure and/or NOEMA observations). $M_\mathrm{min}$, the mass sensitivity, is taken at 6 times the rms noise of each image (identification threshold), assuming a dust temperature of 20~K, and a dust (+gas) mass opacity coefficient at 0.87(1.3)~mm of 0.0175(0.00899) cm$^2$\, g$^{-1}$ (column 6 of Table~1 of Ossenkopf \& Henning 1994, corresponding to agglomerated dust grains with thin ice mantles at densities of $10^6$~cm$^{-3}$).\\
$^\mathrm{d}$ Spatial resolution taken from the synthesized beam of each image and the distance to the source. LAS stands for Largest Angular Scale, estimated using the smallest $uv$-distance given in column (7), and following equation A5 of Palau et al. (2010). This corresponds to the maximum spatial scale the interferometer was able to recover.\\
$^\mathrm{e}$ References for the continuum emission used to assess the fragmentation level $N_\mathrm{mm}$ and for the polarization data used to study the polarization angle dispersion: (1) H.-R. V. Chen et al., in preparation; (2) Beuther et al. (2018) and Ahmadi et al. (2018): this region is part of the CORE Large Project carried out with NOEMA; (3) Chen et al. (2012a); (4) Zhang et al. (2014); (5) Liu et al. (2013); (6) Ju\'arez et al. (2017); (7) Cesaroni et al. (2014); (8) H. Shinnaga et al. (in preparation); (9) Bontemps et al. (2010); (10) SMA archive; (11) F. O. Alves et al. (in preparation); (12) Girart et al. (2013); (13) Ching et al. (2017); (14) Beuther et al. (2012).
}
\end{deluxetable*}

How stellar clusters form and what determines their number of objects and stellar densities is a long-standing question, intimately related to the fragmentation properties of molecular clouds. It is thought that a number of properties of molecular clouds could influence and determine how clouds fragment. First, their density and temperature structures determine the balance between thermal support and gravity required for pure thermal Jeans fragmentation (\eg\ Myhill \& Kaula 1992; Burkert et al. 1997; Girichidis et al. 2011). There are a number of additional properties however which could play a crucial role as well. The most important ones are the properties of turbulence (solenoidal/compressive and Mach number; \eg\ V\'azquez-Semadeni et al. 1996; Padoan \& Nordlund 2002; Schmeja \& Klessen 2004; Federrath et al. 2008; Girichidis et al. 2011; Keto et al. 2020), stellar feedback (\eg\ Myers et al. 2013, Cunningham et al. 2018), initial angular momentum (\eg\ Boss \& Bodenheimer 1979; Boss 1999; Hennebelle et al. 2004; Machida et al. 2005; Forgan \& Rice 2012; Chen et al. 2012b, 2019), and magnetic fields.

A number of theoretical and numerical studies suggest that magnetic fields could be a key ingredient in the fragmentation process of molecular clouds, because it is a form of support against gravitational contraction (\eg\ Boss 2004; V\'azquez-Semadeni et 2005, 2011; Ziegler 2005; Price \& Bate 2007; 
%Hennebelle et al. 2008: fragmentation depends more strongly on the amplitude A of the perturbations seeded initially than on the mass-to-flux over critical mass-to-flux ratio...
Commer\c con et al. 2011; Peters et al. 2011; Bailey \& Basu 2012; Myers et al. 2013; Boss \& Keiser 2013, 2014; Girichidis et al. 2018; Hennebelle \& Inutsuka 2019). 
%Banerjee \& Pudritz 2006: not exactly focused on fragmentation but on outflows: Outflows and Jets from Collapsing Magnetized Cloud Cores
Therefore, it is expected that those cores with stronger magnetic fields should present a smaller degree of fragmentation, along with fragment masses larger than the pure thermal Jeans mass, compared to cores with weaker magnetic fields. This should hold at least for cores with similar densities and turbulence.

Massive dense cores are excellent targets to study the formation of stellar clusters. These are dense cores embedded within molecular clumps, with large masses ($\gtrsim50$~\mo) and typical sizes of 0.1--0.5~pc, which do not necessarily collapse into one star but can fragment into compact condensations and form a small cluster of stars\footnote{Strictly speaking, the entity which will contain the entire cluster should be the molecular clump (with sizes $\sim1$~pc), while probably the massive dense core will contain only the central or most embedded part of the stellar cluster (Zhang et al. 2009, 2015;  Csengeri et al. 2011).} (Williams et al. 2000; Bontemps et al. 2010;  Motte et al. 2007). This makes massive dense cores excellent candidates to study forming clusters, which are usually associated with intermediate/high-mass stars. 
The fragmentation properties in samples of about $\sim20$ massive dense cores have been studied by a number of authors (\eg\ Bontemps et al. 2010; Palau et al. 2014, 2015; Beuther et al. 2018; Fontani et al. 2018; Sanhueza et al. 2019; Svoboda et al. 2019). In these works, relations between the fragmentation level and density structure, turbulence and initial angular momentum were searched for, but none of these works studied if there is the expected relation between the fragmentation level and the magnetic field strength from an observational point of view.

Observational studies of fragmentation vs magnetic fields are very scarce.  Most of the studies approaching this key question are based on a comparison of observations of dust continuum emission to the outputs of magneto-hydrodynamical simulations. For example, for the low-mass case, Maury et al. (2010) find that magneto-hydrodynamical models agree much better with their observations. And for the intermediate/high-mass case, Peretto \et\ (2007) find difficulties matching the observed masses and number of fragments with the results of hydrodynamical simulations, suggesting that an extra support such as protostellar feedback or magnetic fields is at play. This is supported by the more recent works of Palau et al. (2013) and Fontani et al. (2016, 2018), where the number, mass and spatial distribution of the fragments of particular regions are consistent with simulations of fragmenting cores with different mass-to-flux ratios (Commer\c con et al. 2011). However, the extreme fragmentation in the DR21OH core cannot be fully reproduced in these simulations because its measured mass-to-flux ratio is 20 times smaller than the one used in the simulations for the highly fragmenting cores (Girart et al. 2013).

Regarding studies reporting a direct measure of the magnetic field strength compared to fragmentation levels, Santos et al. (2016) present polarimetric data at optical and near-infrared wavelengths towards an infrared dark cloud with different fragmentation levels in two hubs (Busquet et al. 2016), and find no significant differences between the magnetic field at each hub at clump scales, while submillimeter polarization observations at core scales for the same two hubs reveal hints of a stronger magnetic field in the non-fragmenting case  (A\~nez-L\'opez et al. 2020b). On the other hand, in the mini-starburst star-forming region W43, very recent Atacama Large Millimeter/submillimeter Array observations show similar magnetic field strengths for cores with different fragmentation levels (Cort\'es et al. 2019).  Also in the G34.43+00.24 region the three cores studied by Tang et al. (2019) present different fragmentation levels, but they seem to result from an interplay between gravity, turbulence and magnetic field, with no clear evidence for a unique role of the magnetic field.
%, for a sample of 6 cores of this region, that there seems to be a trend of higher fragmentation in the cores with polarization intensities, but they could not infer magnetic fields strengths for the cores with higher fragmentation level 
In a high resolution polarimetric imaging study of a massive infrared dark cloud, Liu et al. (2020) find that magnetic fields play a role at the early stages of cluster formation, similar to the conclusion of Pillai et al. (2015). On the other hand, more sensitive recent works show that the magnetic field seems to be dragged by flows of material inflowing towards the hubs of hub-filament systems (\eg\ Beuther et al. 2020, Pillai et al. 2020, Wang et al. 2020).
However, all the aforementioned observational works do not perform a uniform study in a relatively large sample of regions, but focus only on a single or a handful of regions at most. %and a uniform study of the fragmentation and magnetic field properties in a relatively large sample of massive dense cores is imperative. 
Galametz et al. (2018) study the submillimeter polarized emission in a sample of 12 low-mass Class 0 protostars, and find that the morphology of the magnetic field could be related to the rotational energy and the formation of single or multiple systems, with the magnetic field being aligned along the outflow direction for single sources. Actually, in a subsequent study of a sample of 20 low-mass protostellar cores, Galametz  et al. (2020) found a positive correlation between the angular momentum in the envelope and the misalignment between the outflow axis and the magnetic field, indicating that the magnetic field could be regulating some of the processes of low-mass star formation. However the Galametz works focus on the low-mass regime.
Thus, a uniform study of the fragmentation and magnetic field properties in a relatively large sample of massive dense cores is lacking and therefore imperative.

\renewcommand{\thefigure}{1a}
\begin{figure*}[ht]
\begin{center}
\begin{tabular}[b]{c}
    \epsfig{file=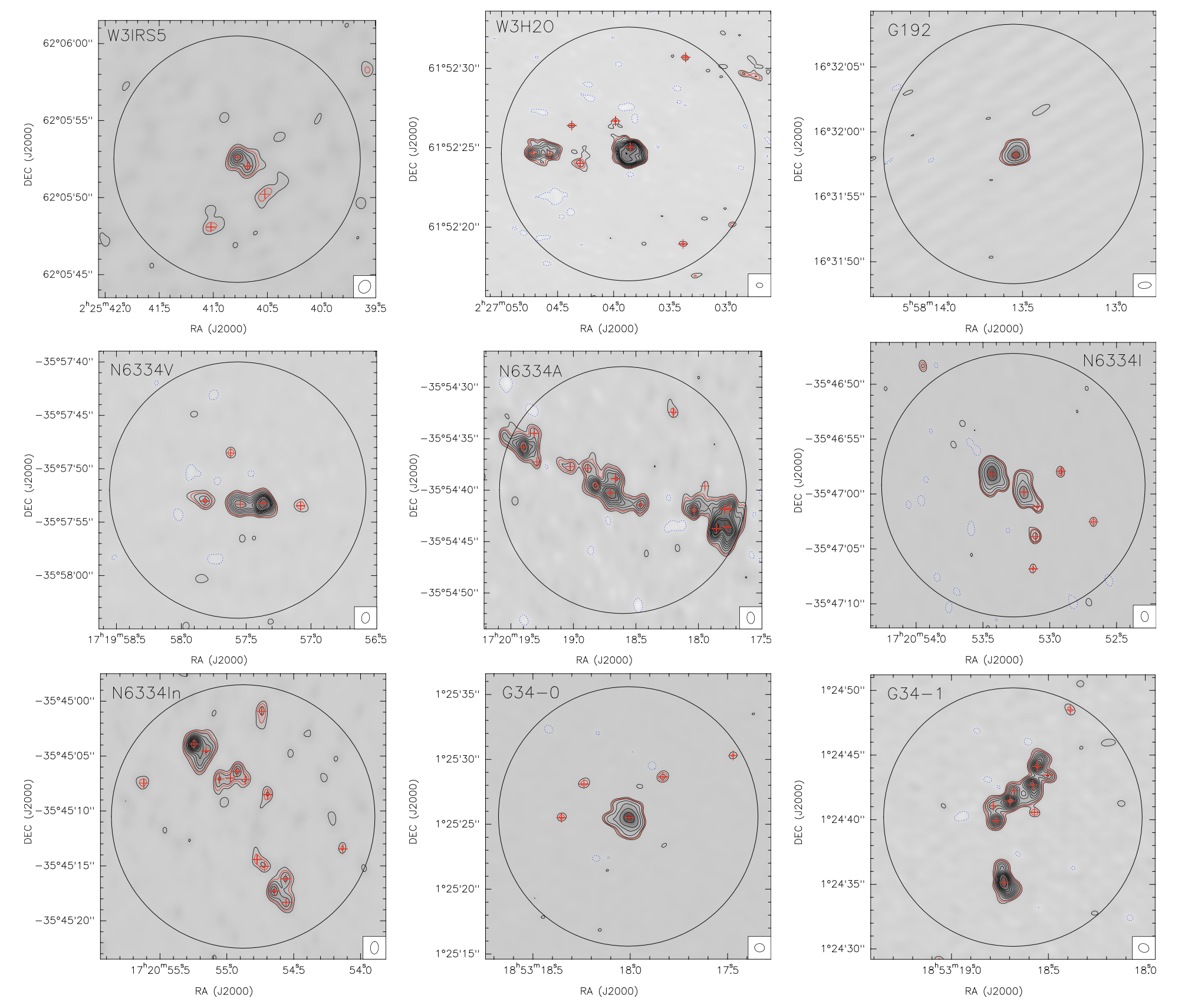, width=15cm, angle=0}\\
\end{tabular}
\caption{0.87 or 1.3~mm continuum high angular resolution maps. 
Contours for all regions are $-4$, 4, 8, 12, 16, 20, 24, 28, 36, 40, 50, 60, 70, 80, 100 and 120 times the rms noise, listed in Table~\ref{tsample}, except for G192, N6334I,  and G34-0, for which contours are $-4$, 4, 8, 16, 32, 64, and 128 times the rms noise. Synthesized beams are plotted in the bottom-right corner of each panel, and the black circle corresponds to a field of view of 0.15~pc diameter (the field of view common to all the regions, given their primary beams). 
%Crosses correspond to Spitzer/IRAC sources at 3.6~\mum, and plus signs indicate the 2MASS sources. 
In all panels the red contour corresponds to the identification level of $6\sigma$, and the plus signs correspond to the identified fragments.
}
\label{fcont1}
\end{center}
\end{figure*}

\renewcommand{\thefigure}{1b}
\begin{figure*}[ht]
\begin{center}
\begin{tabular}[b]{c}
    \epsfig{file=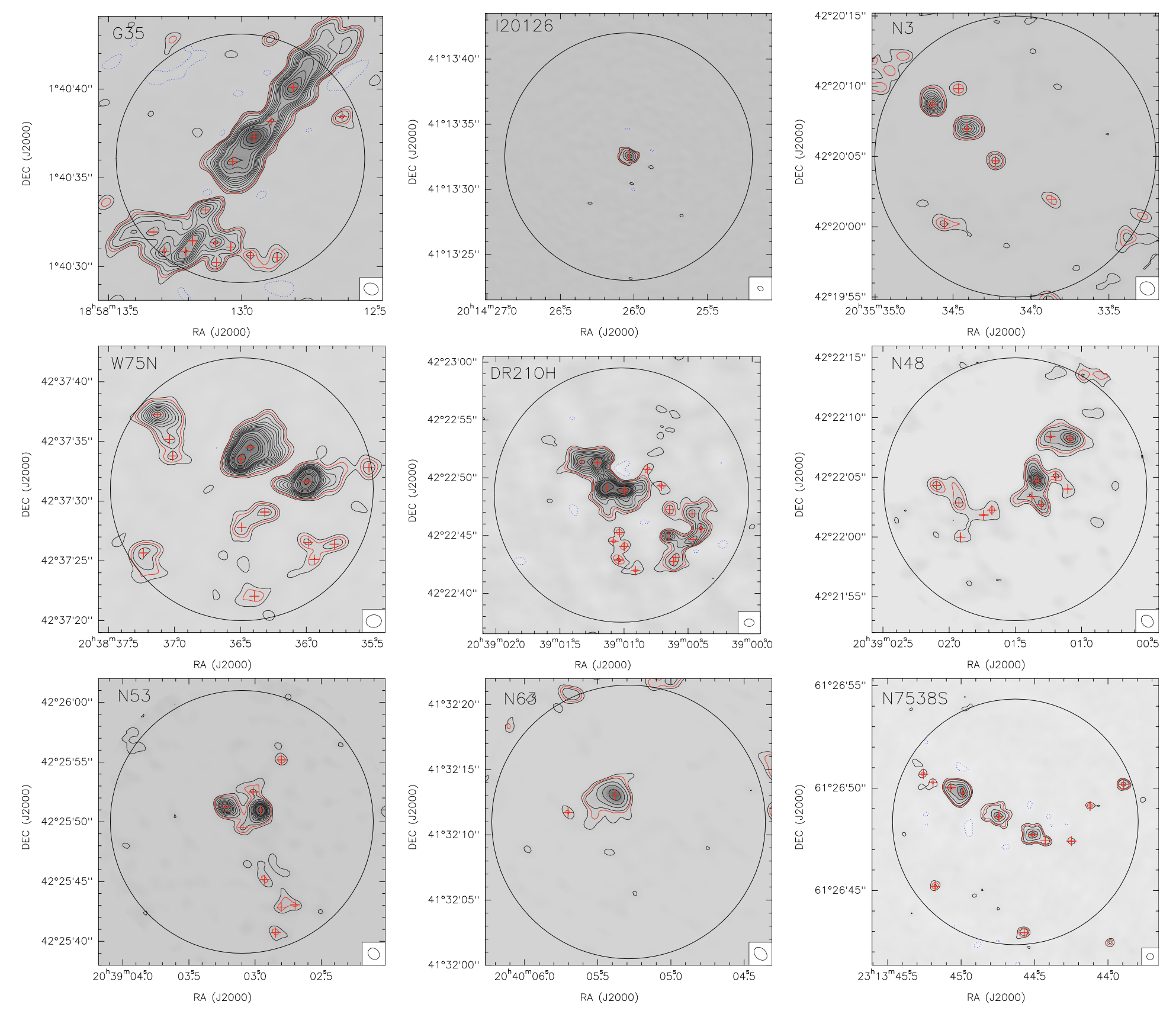, width=15cm, angle=0}\\
\end{tabular}
\caption{0.87 or 1.3~mm continuum high angular resolution maps. 
Contours for all regions are $-4$, 4, 8, 12, 16, 20, 24, 28, 36, 40, 50, 60, 70, 80, 100 and 120 times the rms noise, listed in Table~\ref{tsample}, except for I20126, N63, and N7538S for which contours are $-4$, 4, 8, 16, 32, (48), 64, and 128 times the rms noise, and for G35, for which contours are 
$-4$, 4, 8, 12, 20, 28, 36, 44, 52, 60, 80, 100, 120, 150 and 200 times the rms noise. Synthesized beams are plotted in the bottom-right corner of each panel, and the black circle corresponds to a field of view of 0.15~pc diameter (the field of view common to all the regions, given their primary beams).
%Crosses correspond to Spitzer/IRAC sources at 3.6~\mum, and plus signs indicate the 2MASS sources. 
In all panels the red contour corresponds to the identification level of $6\sigma$, and the plus signs correspond to the identified fragments.
}
\label{fcont2}
\end{center}
\end{figure*}

Here we used the submillimeter polarization data of the Submillimeter Array (SMA) Legacy Survey of Zhang et al. (2014), together with regions from the literature with similar observational properties to build a sample of 18 massive dense cores. In Section 2, we describe the sample and observations, in Section 3, we present the continuum, polarization and \htcop\ data, in Section 4 we analyze the polarization data, determine density profiles for all the sample, measure line widths, and perform the Angular Dispersion Function analysis to finally infer magnetic field strengths. In Section 5 a discussion of the results is presented and in Section 6 our main conclusions are given.

\begin{deluxetable*}{lccCCCCCCC}
%\tablenum{1}
\tablecaption{Fragmentation properties of the massive dense cores\label{tfrag}}
\tablewidth{0pt}
\tablehead{
\colhead{} & 
\colhead{} & 
\colhead{Frag.} & 
\colhead{$\langle R_\mathrm{fragm}\rangle$\supc} & 
\colhead{$\sigma(R_\mathrm{fragm})$\supc} &
\colhead{$\langle M_\mathrm{fragm}\rangle$\supc} & 
\colhead{$\sigma(M_\mathrm{fragm})$\supc} & 
\colhead{$\langle n_\mathrm{frag}\rangle$\supc} & 
\colhead{$M_\mathrm{Jeans}$\supd} & 
\colhead{$M_\mathrm{crit}$\supd} \\
\colhead{Source} & 
\colhead{\Nmm\supa} & 
\colhead{type\supb} & 
\colhead{(AU)} &
\colhead{(AU)} &
\colhead{(\mo)} & 
\colhead{(\mo)} & 
\colhead{($10^7\,$cm$^{-3}$)} & 
\colhead{(\mo)} & 
\colhead{(\mo)}
}
\colnumbers
\startdata
1-W3IRS5          		&4	&cl		&1490	&70		&0.1		&0.1		&0.2		&1.3-19	&0.8  \\
2-W3H2O 			&8	&cl		&730		&430		&2.8		&3.9		&15		&0.8-6.8	&0.6  \\
3-G192				&1	&no		&1770	&$-$		&2.7		&-		&1.8		&1.9-5.7	&2.7	\\
4-N6334V				&5	&cl		&1310	&460		&1.1		&1.3		&1.1		&0.9-4.1	&1.6	\\
5-N6334A				&16	&al		&1270	&370		&1.3		&1.1		&2.0		&0.9-2.6	&1.0	\\
6-N6334I				&7	&cl		&1070	&620		&3.8		&6.8		&3.4		&0.7-4.8	&3.3 \\
7-N6334In				&15	&al		&1000	&200		&1.0		&1.2		&3.1		&0.7-2.8	&4.4	 \\
8-G34-0				&5	&cl		&1090	&840		&2.9		&5.8		&3.2		&0.9-3.3	&1.4	\\ 
9-G34-1				&10	&al		&1170	&440		&0.7		&1.1		&1.0		&1.0-2.0	&3.4	 \\ 
10-G35				&15	&al		&1820	&690		&2.8		&4.3		&0.9		&1.0-3.6	&3.6	\\ 
11-I20126			 	&1	&no		&$-$		&$-$		&-		&-		&-		&1.4-7.2	&- 	\\
12-N3 				&6	&al		&1220	&270		&0.7		&0.7		&1.1		&1.2-1.4	&0.4	\\
13-W75N 				&14	&cl    	&1980	&660		&2.1		&2.5		&0.7		&0.9-5.5	&3.5	\\
14-DR21OH			&18	&cl		&1250	&440		&1.2		&1.3		&1.7		&0.6-1.9	&1.7	\\
15-N48 				&12	&cl		&1320	&410		&0.7		&0.6		&0.4		&0.8-1.6	&0.7	\\
16-N53				&9	&al		&$-$		&$-$		&-		&-		&-		&0.9-1.1	&-	 \\
17-N63 				&2	&no		&1950	&1380	&7.8		&10		&2.4		&1.4-2.2	&1.1	\\
18-N7538S			&12	&al		&1100	&440		&1.6		&1.7		&3.3		&0.7-2.8	&1.3	\\
\enddata
\tablecomments{
\\
$^\mathrm{a}$ $N_\mathrm{mm}$ is the fragmentation level, estimated counting the number of millimeter sources above a 6$\sigma$ threshold, covering at least half a beam at 4$\sigma$, and closing at least one contour, within the common field of view for all the regions of 0.15~pc of diameter. \\
%For N63, \Nmm\ was inferred from new Plateau de Bure observations in AB configuration (S. Bontemps, priv. communication).
$^\mathrm{b}$ Fragmentation type according to Tang et al. (2019). `cl' corresponds to `clustered fragmentation'; `al' corresponds to `aligned fragmentation'; and `no' corresponds to `no fragmentation' (see Section~\ref{sr}).\\
$^\mathrm{c}$ $\langle R_\mathrm{fragm}\rangle$ and $\langle M_\mathrm{fragm}\rangle$ correspond to the average radius and mass (at the 3$\sigma$ level), respectively, of all fragments in a given massive dense core. $\sigma(R_\mathrm{fragm})$ and $\sigma(M_\mathrm{fragm})$ correspond to the standard deviation of the radius and mass of the fragments in each massive dense core.
The mass of the fragments was calculated using the flux density within the 3$\sigma$ contour, and considering the temperature corresponding to the temperature power-law derived in Section~\ref{sadensity}, using as distance the projected distance measured from the fragment to the peak of the single-dish submillimeter source. The opacity law used is the same as the opacity given in the notes of Table~\ref{tsample}.
$\langle n_\mathrm{frag}\rangle$ is the average density of the fragments in each region. For each fragment, its density was calculated as $n_\mathrm{frag}=M_\mathrm{frag}/\frac{4\pi}{3}\,R_\mathrm{frag}^3)$.\\
$^\mathrm{d}$ $M_\mathrm{Jeans}$ corresponds to the Jeans mass calculated following equation~\ref{eqMjeans} and using the values of $n_\mathrm{0.15pc}$ listed in Table~\ref{tfit}. The range of values corresponds to the range of temperatures assumed, 20~K (lower limit) or $T_\mathrm{0.15pc}$ from Table~\ref{tfit} (upper limit).
$M_\mathrm{crit}$ is the magnetic critical mass (the Jeans mass analog in the magnetic support case) calculated following equation~\ref{eqMcrit}, using $\langle R_\mathrm{fragm}\rangle$ and the magnetic field strength $B_\mathrm{stdev}$ (Table~\ref{tB}) scaled in density to the average density of all the fragments, of $3\times10^7$~\cmt, assuming that $B_\mathrm{frag}=B_\mathrm{stdev}\big[\frac{3\times10^7~\mathrm{cm}^{-3}}{n_\mathrm{0.15pc}}\big]^{0.4}$ (Li et al. 2015).
}
\end{deluxetable*}

\renewcommand{\thefigure}{2a}
\begin{figure*}[ht]
\begin{center}
\begin{tabular}[b]{c}
    \epsfig{file=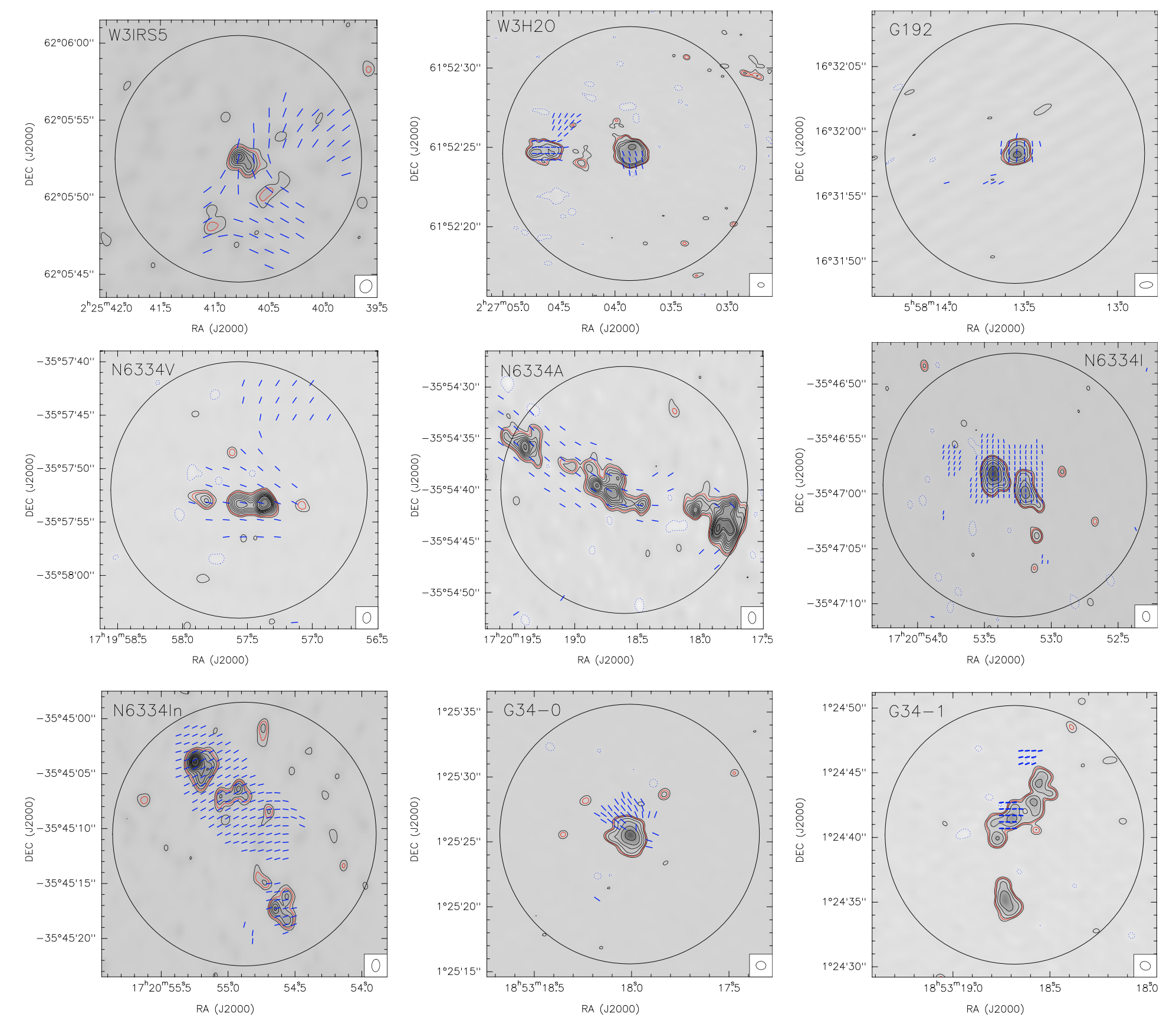, width=15cm, angle=0}\\
\end{tabular}
\caption{0.87 or 1.3~mm continuum high angular resolution maps with the magnetic field segments overplotted in blue. 
Contours for all regions are $-4$, 4, 8, 12, 16, 20, 24, 28, 36, 40, 50, 60, 70, 80, 100 and 120 times the rms noise, listed in Table~\ref{tsample}, except for G192, N6334I,  G34-0, and G34-1, for which contours are $-4$, 4, 8, 16, 32, 64, and 128 times the rms noise. 
For W3H2O contours are $-4$, 4, 8, 16, 24, 32, 64, 120 times the rms noise.
Synthesized beams are plotted in the bottom-right corner of each panel, and the black circle corresponds to the common field of view of 0.15~pc diameter. 
}
\label{fpol1}
\end{center}
\end{figure*}

\renewcommand{\thefigure}{2b}
\begin{figure*}[ht]
\begin{center}
\begin{tabular}[b]{c}
        \epsfig{file=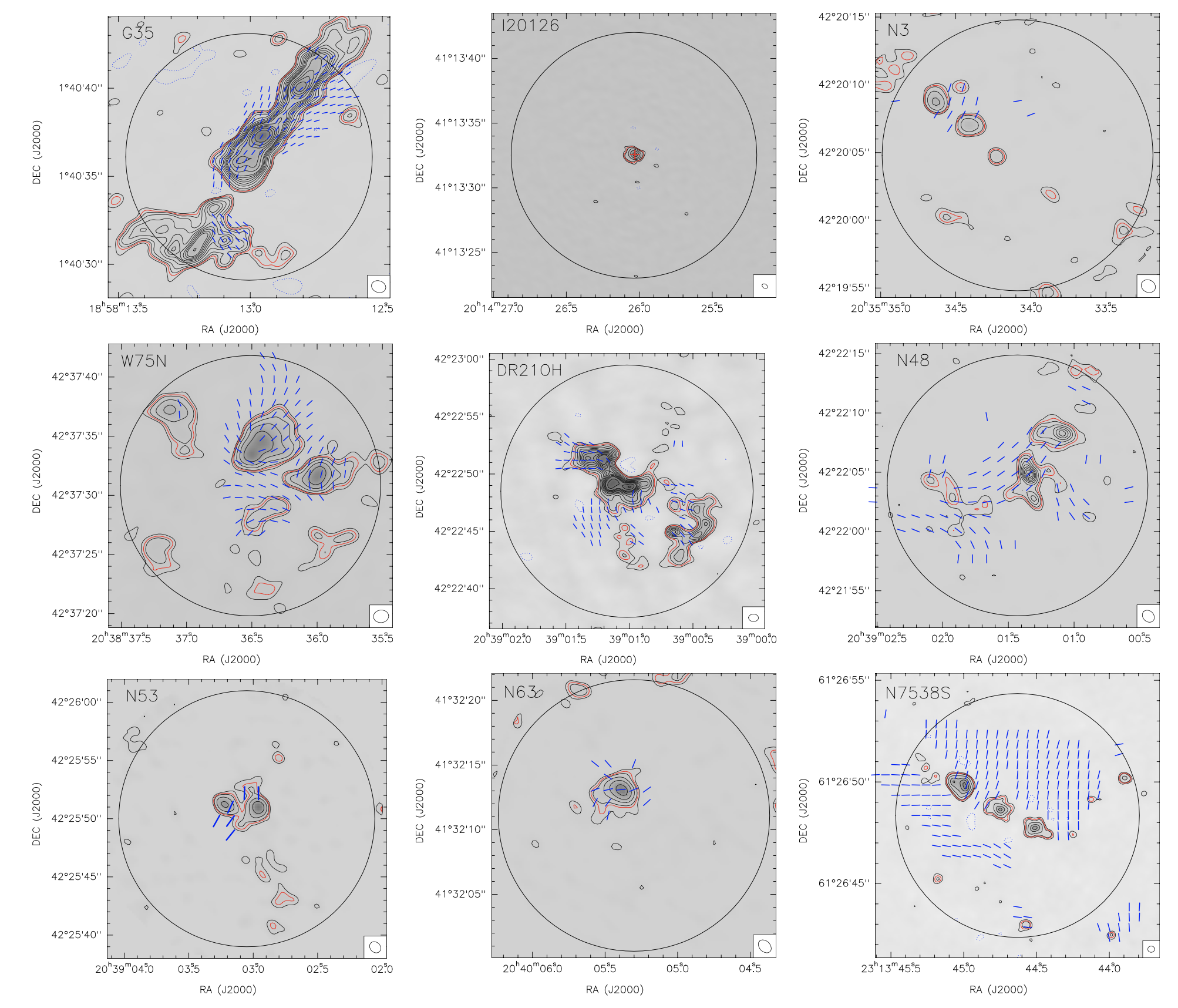, width=15cm, angle=0}\\
\end{tabular}
\caption{0.87 or 1.3~mm continuum high angular resolution maps with the magnetic field segments overplotted in blue. 
Contours for all regions are $-4$, 4, 8, 12, 16, 20, 24, 28, 36, 40, 50, 60, 70, 80, 100 and 120 times the rms noise, listed in Table~\ref{tsample}, except for I20126, W75N, N53, N63 and N7538S for which contours are $-4$, 4, 8, 16, 32, 64, and 128 times the rms noise, and for G35, for which contours are $-4$, 4, 8, 12, 20, 28, 36, 44, 52, 60, 80, 100, 120, 150 and 200 times the rms noise. Synthesized beams are plotted in the bottom-right corner of each panel, and the black circle corresponds to the common field of view of 0.15~pc diameter.
}
\label{fpol2}
\end{center}
\end{figure*}

%%%%%%%%%%%%%%%%%%%%%%%%%%%%%%%%%%%%%%%%%%%%%%%%%%%%%%%%%%%%%%%%%%%%%%%%%%%%%%%%
\section{The sample \label{sobs}}

In Table~\ref{tsample} we present the sample of 18 massive dense cores\footnote{The radii of our core/clumps estimated from the FWHM in the JCMT images is about 0.2--0.4 pc. Thus, they are structures in the transition between cores ($\sim0.1$~pc) and clumps ($\sim1$~pc). Since in this work we focus on the inner parts of these structures of ~0.075 pc of radius, the term `core' has been chosen to emphasize that the study is at these scales. 
%We note that most of them are not isolated but part of larger structures and their borders are ill-defined.
} studied in this work. Among the 18 regions, the 0.87~mm polarization data of 11 were presented in the Submillimeter Array Legacy Survey of Zhang et al. (2014; see also Ching et al. 2017). The 0.87~mm polarization data of the remaining regions were taken from the literature or the SMA archive (see last column of Table~\ref{tsample}). We thus refer to these works for the details of the polarization observations. In general, the typical 1$\sigma$ rms noise in the Stokes $Q$ and $U$ images is of $\sim2$~~m\jpb. It is worth noting that we took special care to build a sample as uniform as possible. Thus, to avoid biases with distance, we restricted our sample to regions in the range 1.4--2.6 kpc, \ie\ there is less than a factor of 2 in distance for the sources of our sample. Two of the regions were observed down to similar rms noises as the other regions, but no signal of polarized emission was detected: N53 from Ching et al. (2017), and I20126 from H. Shinnaga et al. (in preparation).

%The primary beam of the SMA is $36.4''$.
%0.869*1.22*206265/6000 = 36.44633794999999

%We required a minimum of polarization detected so that it makes sense to calculate PA dispersions. This excludes I20126 and NGC2264C and CygX-N53.
%NOTE: the rms noise obtained w extended config for I22198 in stokes Q and U was 3.5 mJy/beam, comparable to the noise in Zhang+14, but we did not include the region because it is too close in distance compared to the others.

Regarding the continuum images used to assess the fragmentation level, Table~\ref{tsample} provides the properties of the images along with the references. For most of the regions we used the 0.87~mm continuum emission observed with the SMA using the extended configuration only from Zhang et al. (2014). Only those marked with an asterisk in Table~\ref{tsample} were observed with the Plateau de Bure(PdBI)/NOrthern Extended Millimeter Array (NOEMA) at 1.3~mm. 
In order to build a uniform sample, we specifically checked to ensure that the $uv$-coverage of both the SMA and PdBI/NOEMA are comparable (see column (6) of Table~\ref{tsample}). This implied in some cases re-imaging using the visibilities only from the extended configuration.
This will ensure not only a similar spatial resolution of $\sim1000$~AU for all the regions, but also that the largest angular scale filtered out by the interferometers is similar for all regions (see columns (7), (8), and (8) of Table~\ref{tsample}). Finally, special care was also taken regarding the sensitivity, which was required to be around $\sim0.5$~\mo\ (at $6\sigma$) or better. For this purpose, we self-calibrated some of the regions, with the final rms noises listed also in the table. We note that 6 of the regions included in the present sample (W3IRS5, I20126, DR21OH, N48, N53, and N63) overlap with the sample of Palau et al. (2014). 

%\vspace{2cm}

%%%%%%%%%%%%%%%%%%%%%%%%%%%%%%%%%%%%%%%%%%%%%%%%%%%%%%%%%%%%%%%%%%%%%%%%%%%%%%%%
\section{Results}\label{sr}

Figs.~\ref{fcont1} and \ref{fcont2} present the resulting continuum images (with extended configuration only) used to assess the fragmentation level. The fragmentation level is estimated by counting the number of submillimeter sources above a 6$\sigma$ threshold and within a region of 0.15~pc of diameter, which corresponds to the smallest field of view in our sample (given by the primary beam of the PdBI/NOEMA observations and the distance for each region). Table~\ref{tfrag} lists the fragmentation level \Nmm\ estimated for each region. As can be seen from the table, the measured fragmentation level ranges from almost no fragmentation (1--2 fragments for G192, I20126 and N63) to highly-fragmenting regions (with up to 18 fragments, such as DR21OH). 
A total number of 160 fragments were identified within the 18 massive dense cores. For each core, the mass and size (at the 3$\sigma$ level) for each fragment was calculated (see details in Table~\ref{tfrag}), along with the average mass and size in each core, and their standard deviations.
In addition, each massive dense core has been classified according a `fragmentation type', following Tang et al. (2019): `clustered fragmentation' corresponds to cores with fragments distributed more or less homogeneously within the core (8 regions); `aligned fragmentation' corresponds to cores with fragments predominantly aligned along a particular direction (7 regions); `no fragmentation' corresponds to cores with only 1 or 2 fragments (3 regions).

In order to ensure that the fragmentation level is not affected by biases with distance, sensitivity, or evolutionary stage, in Fig.~\ref{fbias} of Appendix~\ref{apbias} we present \Nmm\ vs spatial resolution, mass and column density sensitivity, and the evolutionary indicator $L_\mathrm{bol}/M_\mathrm{core}$ (Molinari et al. 2016; with $L_\mathrm{bol}$ being the bolometric luminosity of the massive dense core and $M_\mathrm{core}$ being its mass estimated from our modeling, see Section~\ref{sadensity} and Table~\ref{tsample}). The figure shows that there are no trends and thus the sample is well suited to compare the fragmentation properties between the different regions.
%Mention Sadaghiani+19 about fragmentation in N6334: they also find more fragmentation towards In than towards I.

\begin{deluxetable*}{lccc cccC Cc}
%\tablenum{1}
\tablecaption{Fitted parameters of the density and temperature structure of the massive dense cores, and inferred properties\label{tfit}}
\tablewidth{0pt}
\tablehead{
\colhead{} & 
\colhead{} & 
\colhead{$T_0$\supa} & 
\colhead{$\rho_0$\supa} & 
\colhead{} & 
\colhead{} & 
\colhead{} & 
\colhead{$M_\mathrm{0.15pc}$\supb} & 
\colhead{$n_\mathrm{0.15pc}$\supb} & 
\colhead{$T_\mathrm{0.15pc}$\supb} \\
%%%%%%%%%%%%%%%%%%%%%%%%%
\colhead{ID-Source} & 
\colhead{$\beta$\supa} & 
\colhead{(K)} & 
\colhead{(g cm$^{-3}$)} & 
\colhead{$p$\supa} & 
\colhead{$\chi_r$\supa} & 
\colhead{$q$\supb} & 
\colhead{(\mo)} & 
\colhead{(10$^5$\,cm$^{-3}$)} & 
\colhead{(K)} 
}
\colnumbers
\startdata
1-W3IRS5	    		&$1.04\pm0.12$	&$260\pm30$	&$(2.4\pm0.3)\times10^{-17}$	&$1.46\pm0.04$	&0.602	&0.40	&22\pm3	&1.8\pm0.2	&118		\\
2-W3H2O			&$1.30\pm0.16$	&$152\pm16$	&$(1.5\pm0.2)\times10^{-16}$	&$1.90\pm0.05$	&0.532	&0.38	&59\pm8	&4.8\pm0.6	&82		\\
3-G192			&$1.36\pm0.23$	&$66\pm6$	&$(4.1\pm0.6)\times10^{-17}$	&$2.13\pm0.08$	&0.338	&0.37	&11\pm2	&0.9\pm0.1	&42		\\
4-N6334V			&$2.19\pm0.25$	&$96\pm11$	&$(1.3\pm0.3)\times10^{-16}$	&$1.89\pm0.08$	&0.334	&0.32	&51\pm12	&4.2\pm1.0	&56		\\
5-N6334A			&$2.18\pm0.15$	&$78\pm8$	&$(4.6\pm0.7)\times10^{-17}$	&$1.42\pm0.05$	&0.455	&0.33	&46\pm7	&3.8\pm0.6	&40		\\
6-N6334I			&$2.55\pm0.22$	&$111\pm12$	&$(2.3\pm0.4)\times10^{-16}$	&$2.02\pm0.05$	&0.461	&0.31	&73\pm13	&6.0\pm1.0	&70		\\
7-N6334In			&$2.10\pm0.19$	&$98\pm12$	&$(8.8\pm1.5)\times10^{-17}$	&$1.46\pm0.05$	&0.518	&0.32	&81\pm14	&6.6\pm1.0	&52		\\
8-G34-0			&$1.96\pm0.24$	&$63\pm6$	&$(2.0\pm0.4)\times10^{-16}$	&$2.26\pm0.09$	&0.483	&0.34	&44\pm9	&3.6\pm0.7	&46		\\
9-G34-1			&$1.39\pm0.23$	&$63\pm7$	&$(8.2\pm1.5)\times10^{-17}$	&$1.76\pm0.08$	&0.488	&0.37	&42\pm8	&3.4\pm0.6	&33		\\
10-G35			&$1.86\pm0.15$	&$90\pm7$	&$(4.4\pm0.6)\times10^{-17}$	&$1.53\pm0.03$	&0.463	&0.34	&36\pm5	&3.0\pm0.4	&46		\\
11-I20126			&$1.82\pm0.24$	&$86\pm9$	&$(8.4\pm1.6)\times10^{-17}$	&$2.21\pm0.11$	&0.607   	&0.34	&20\pm4	&1.6\pm0.3	&59		\\
12-N3\supc		&$1.69\pm0.31$	&$45\pm4$	&$(4.0\pm0.5)\times10^{-17}$	&$1.58\pm0.04$	&0.648	&0.35	&29\pm4	&2.4\pm0.3	&23		\\
13-W75N			&$2.04\pm0.18$	&$112\pm12$	&$(1.4\pm0.2)\times10^{-16}$	&$1.99\pm0.05$	&0.729	&0.33	&48\pm7	&4.0\pm0.5	&67		\\
14-DR21OH		&$1.60\pm0.26$	&$73\pm7$	&$(3.1\pm0.6)\times10^{-16}$	&$1.98\pm0.08$	&0.808	&0.36	&110\pm20&8.6\pm1.7	&42		\\
15-N48			&$1.88\pm0.18$	&$58\pm5$	&$(1.0\pm0.2)\times10^{-16}$	&$1.71\pm0.05$	&0.459	&0.34	&56\pm11	&4.6\pm0.9	&31		\\
16-N53\supc		&$1.55\pm0.22$	&$45\pm4$	&$(9.7\pm1.8)\times10^{-17}$	&$1.76\pm0.07$	&0.487	&0.36	&49\pm9	&4.0\pm0.7	&24		\\
17-N63\supc		&$1.80\pm0.33$	&$45\pm3$	&$(6.5\pm1.1)\times10^{-17}$	&$2.03\pm0.07$	&0.570	&0.34	&20\pm3	&1.7\pm0.3	&27		\\
18-N7538S		&$1.74\pm0.19$	&$93\pm10$	&$(1.3\pm0.2)\times10^{-16}$	&$1.72\pm0.05$	&0.304	&0.35	&72\pm5	&5.9\pm0.5     	&49		\\
\enddata
\tablecomments{
\\
$^\mathrm{a}$ Free parameter fitted by the model: $\beta$ is the dust emissivity index; $T_0$ and $\rho_0$ are the temperature and density at the reference radius, 1000 AU; $p$ is  the density power law index; $\chi_\mathrm{r}$ is the reduced $\chi$ as defined in equation (6) of Palau et al. (2014).\\
$^\mathrm{b}$ Parameters of the massive dense cores inferred from the modeled density and temperature structures.
$q$ is the temperature power-law index.
$M_\mathrm{0.15pc}$ is the mass inside a region of 0.15~pc of diameter computed according: 
$M_\mathrm{0.15pc} = M(R=0.075\mathrm{pc})= 4\pi\,\rho_0\,r_0^p\,\frac{R^{3-p}}{3-p}$; 
$n_\mathrm{0.15pc}$ and $T_\mathrm{0.15pc}$ correspond to average H$_2$ density and temperature inside a region of 0.15~pc of diameter. $T_\mathrm{0.15~pc}$ was estimated as
$T_\mathrm{R} = \frac{\int_0^R T(r)\rho(r)r^2 dr}{\int_0^R \rho(r)r^2 dr}$, where $T(r)$ and $\rho(r)$ were calculated as power laws with temperature at the reference radius given in column (3), temperature power-law index given in column (7), density at reference radius given in column (4) and density power-law index given in column (5) of this table.
%of the form $T(r)=T_\mathrm{0}(r/r_0)^{-q}$ and $\rho(r)=\rho_0(r/r_0)^{-p}$. 
%NOTE: number density was obtained using a mean molecular weight of 2.8.
The final expression is $T_\mathrm{R}=\frac{T_0(3-p)}{3-p-q}\left(\frac{r}{r_0}\right)^{-q}$.\\
$^\mathrm{c}$ Sources for which only the radial intensity profile at 1.2~mm was fitted. 
}
\end{deluxetable*}

While the fragmentation level was assessed using only the SMA or PdBI/NOEMA {\it extended} configuration, yielding typical synthesized beams below the arcsecond ($\sim0.8''$) and filtering emission typically above $\sim3''$, the polarized emission was obtained using all available SMA configurations, including compact and/or subcompact configurations, and therefore only emission about $14''$--30$''$ was filtered out (Zhang et al. 2014). Thus, the polarized emission includes emission at much larger scales compared to the continuum emission used to study the fragmentation.

Figs.~\ref{fpol1} and \ref{fpol2} present the magnetic field segments overplotted on the images of continuum emission used to assess the fragmentation level. From these figures it is clear that I20126 and N53 do not have enough detections to calculate their magnetic field strength, and hence will not be considered further for the analysis of the polarization data.

In Figs.~\ref{fm1a} and \ref{fm1b} of the Appendix we present the first-order moment of the \htcop\,(4--3) (346.998344~GHz) transition for each region.
The observations of the \htcop\,(4--3) transition  were carried out simultaneously with the submillimeter polarization data from the SMA, with a spectral resolution of 0.7~\kms\ for all cases, except for W3H2O and N7538S, for which the spectral resolution is 1.4~\kms.
The images presented in Figs.~\ref{fm1a} and \ref{fm1b} correspond to images built using all SMA available configurations for each region, like for the polarized emission presented in Figs.~\ref{fpol1} and \ref{fpol2} (therefore again including compact and/or subcompact configurations). Figs.~\ref{fm1a} and \ref{fm1b}  also show the outflow directions reported in the literature for each massive dense core. In general, velocity gradients are present in all the regions, indicating that they could have a non-negligible contribution to the velocity line widths obtained from spectra averaged over the region, required to estimate the magnetic field strength. This is further discussed in Section~\ref{saveldisp}.

%%%%%%%%%%%%%%%%%%%%%%%%%%%%%%%%%%%%%%%%%%%%%%%%%%%%%%%%%%%%%%%%%%%%%%%%%%%%%%%%
\section{Analysis}

\subsection{Determination of the density structure}\label{sadensity}

In order to estimate the density averaged within 0.15~pc of diameter (the field of view where the fragmentation level was assessed), we inferred the radial density profile of each massive dense core. To do this, we followed the same approach described in Palau et al. (2014), where a model was developed to simultaneously fit the radial intensity profiles at 450 and 850~\mum\ SCUBA images (from the James Clerk Maxwell Telescope) from di Francesco et al. (2008)\footnote{For the cases of N3 and N63, the SCUBA data are not available and the IRAM\,30m data at 1.2~mm from Motte et al. (2007) were used (Palau et al. 2014).}, along with the Spectral Energy Distribution (SED). The constraint imposed by the SED allows to break the degeneracy between temperature and density to the intensity of the source. The model assumes spherical symmetry, takes into account opacity effects (\ie\ the emission is not assumed to be optically thin), does not assume the Rayleigh-Jeans approximation and considers that the density and temperature decrease with radius following power-laws with indices $p$ and $q$, respectively: $\rho=\rho_0(r/r_0)^{-p}$ and $T=T_0(r/r_0)^{-q}$, with $\rho_0$ and $T_0$ being the density and temperature values at a reference radius $r_0$ taken to be 1000~AU. Regarding the dust opacity law, it was assumed to follow a power-law of frequency with index $\beta$, $\kappa=\kappa_0(\nu/\nu_0)^\beta$, where $\nu_0$ is an arbitrary reference frequency. The value of $\kappa_0=0.008991$~cm$^2$\,g$^{-1}$ at $\nu_0=230$~GHz was adopted (Ossenkopf \& Henning 1994).  Given a dust cloud heated radiatively by a central luminous source, it has been shown that $\beta$ and $q$ are related according to $q = 2/(4 + \beta )$ (Scoville \& Kwan 1976; Adams 1991; Chandler et al. 1998). Thus, the final free parameters of the model are four: the dust emissivity index, $\beta$, the envelope temperature at the reference radius $r_0$, $T_0$, the envelope density at the reference radius $r_0$, $\rho_0$, and the density power-law index, $p$. 

%run pdf2ps to the pdf sent by Robert
\renewcommand{\thefigure}{3a}
\begin{figure*}
\begin{center}
\begin{tabular}[b]{c}
    \epsfig{file=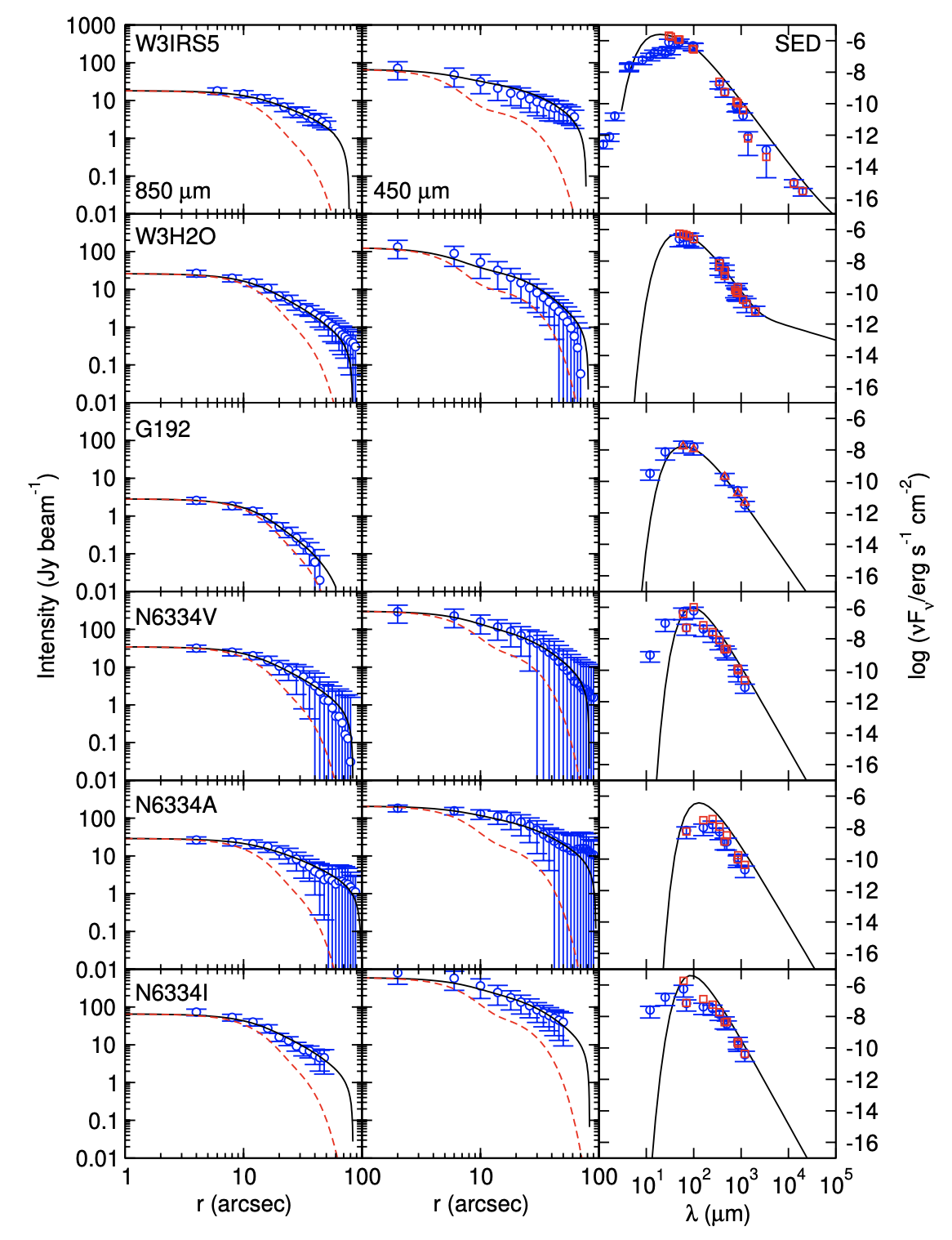, width=10.cm, angle=0}\\
\end{tabular}
\caption{Best fits for six regions of the sample (see Table~\ref{tfit} for the exact fitted parameters). Each row corresponds to one core, the left (middle) panel shows the radial intensity profile at 850 (450)~\mum, with the empty blue circles corresponding to the data, the black solid line corresponding to the model, and the dashed red line showing the beam profile; panels on the right show the SED, with blue empty circles indicating the observed fluxes, 
%the blue full circles indicating observed lower limits, 
the black solid line showing the model for a fixed aperture, and the red squares corresponding to the model for the same aperture where each flux was measured.
}
\label{fradprof1}
\end{center}
\end{figure*}

\renewcommand{\thefigure}{3b}
\begin{figure*}
\begin{center}
\begin{tabular}[b]{c}
    \epsfig{file=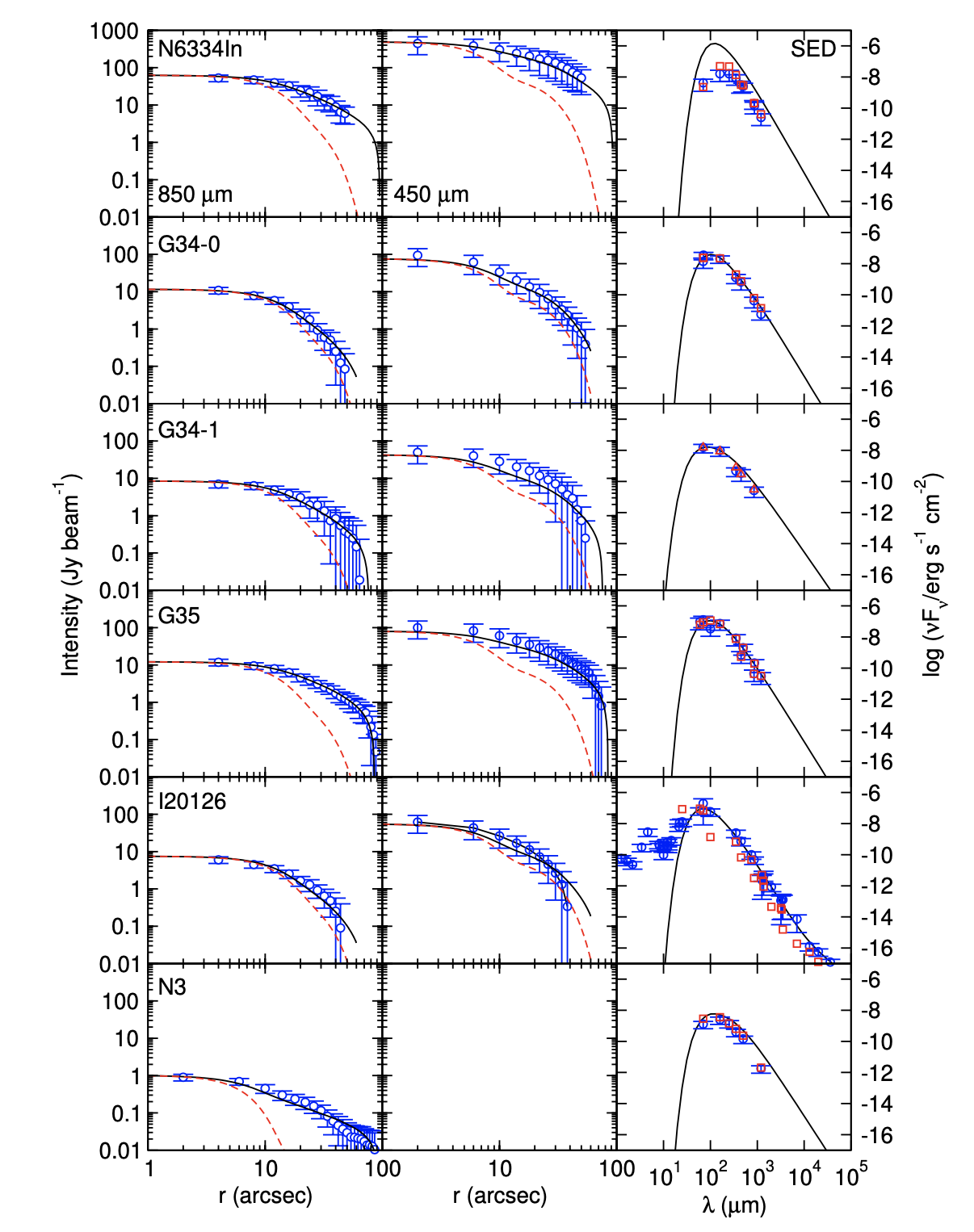, width=10cm, angle=0}\\
\end{tabular}
\caption{Same as Fig.~\ref{fradprof1}.
}
\label{fradprof2}
\end{center}
\end{figure*}

\renewcommand{\thefigure}{3c}
\begin{figure*}
\begin{center}
\begin{tabular}[b]{c}
    \epsfig{file=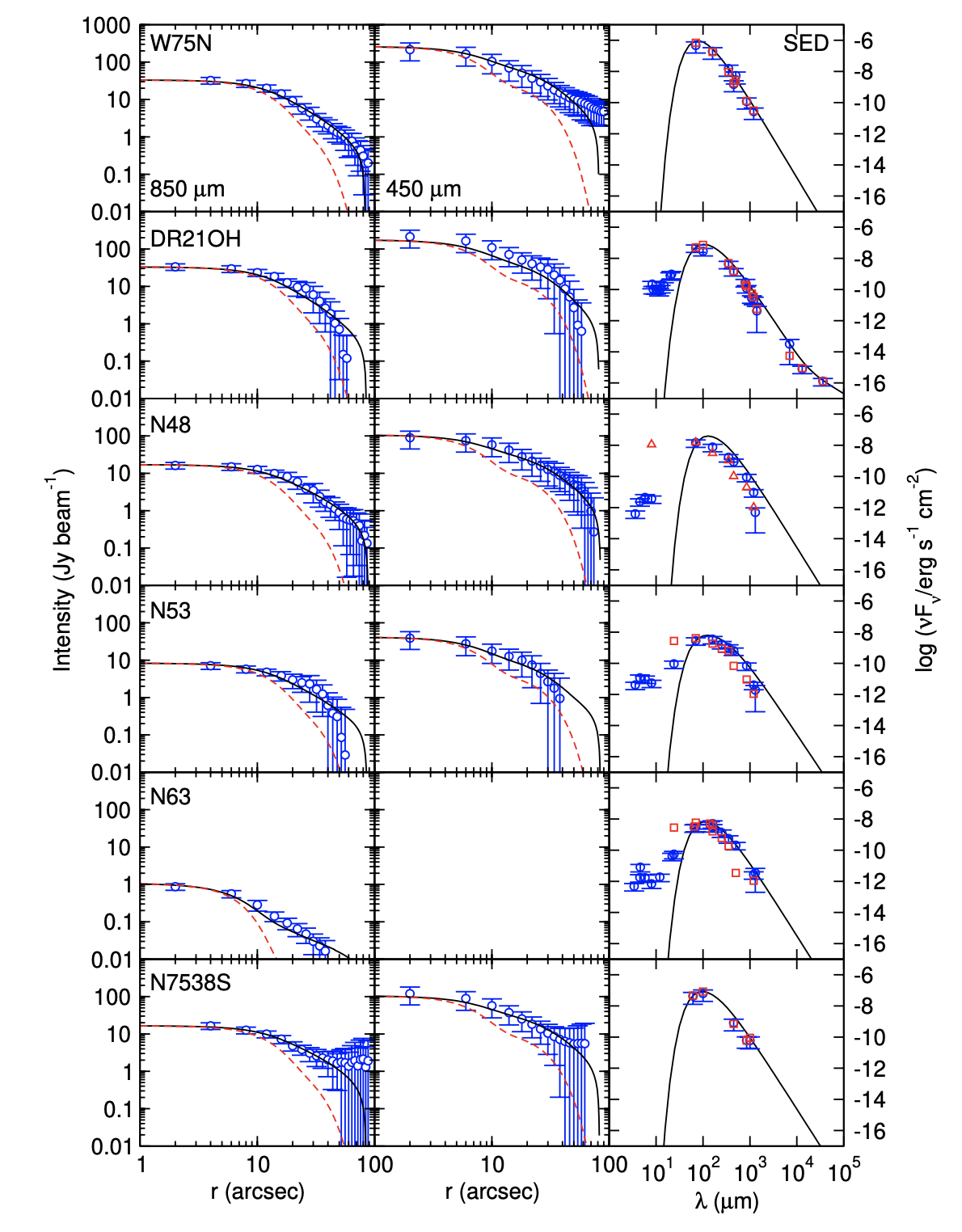, width=10cm, angle=0}\\
\end{tabular}
\caption{Same as Fig.~\ref{fradprof1}.
}
\label{fradprof3}
\end{center}
\end{figure*}

The fitting procedure was the same as the one described in Palau et al. (2014), with initial search ranges for the four parameters being $\beta=1.5\pm1.5$, $T_0=300\pm300$~K, $\rho_0 = (1.0\pm1.0)\times10^{-16}$~g\,cm$^{-3}$, and $p = 1.5\pm1.0$. The search range was reduced by a factor of 0.8 around the best-fit value found for each loop. In turn, each loop consisted of 2000 samples of the parameter space, and the final best-fit values were taken after 10 loops. Once the best-fit parameters were found, their uncertainties were estimated through the increase in $\chi^2$.  We refer to Palau et al. (2014) for further details of the model and the fitting procedure, and to Appendix~\ref{apdensity} for additional details on the SED building for two particular regions of the sample. 
In Table~\ref{tfit} we list the best-fit values for the four free parameters of the model, along with the reduced $\chi^2$, the temperature power-law index $q$, the mass within a region of 0.15~pc of diameter\footnote{The mass given in Table~\ref{tfit} is not the total mass of the core, but the mass only within a region of 0.15~pc of diameter. An estimate of the total mass of the core is given in Table~\ref{tsample}, and results from integrating the density of our model up to the observed radius for each region as seen with the JCMT. This yields values typically about a factor of 4-10 larger than the mass within 0.15 pc of diameter.} and the density averaged within the same region. The uncertainties associated with the averaged density and mass are obtained taking into account the uncertainty in the reference density $\rho_0$, which is about 10--20\% of the fitted value, while the uncertainty in the density power-law index $p$ is always $<5\%$ (Table~\ref{tfit}) and is not considered. Our results for the density power-law index are consistent, within the errors, with very recent estimates at smaller scales ($\sim2000$~AU) from the CORE sample (W3H2O, Gieser et al. 2021).
In Figs.~\ref{fradprof1}, \ref{fradprof2} and \ref{fradprof3} the observational data and the best-fit model for reach region are presented.

%, and Fig.~\ref{fNmm-nDv} shows a plot of the fragmentation level \Nmm\ vs the density averaged within 0.15~pc. In this figure, the uncertainties associated with the density are obtained taking into account the uncertainty in the reference density $\rho_0$ and in the density power-law index $p$ (Table~\ref{tfit}). The figure reveals a possible trend of \Nmm\ with density.

\subsection{Determination of the velocity dispersion}\label{saveldisp}

One of the common techniques to estimate velocity dispersions is to fit a Gaussian to the line spectrum of the region.
We extracted the \htcop\,(4--3) spectrum averaged over a region of 0.15~pc in diameter (the field of view where fragmentation was assessed), and fitted one Gaussian for each region of our sample\footnote{The \htcop\,(4--3) spectrum of W3H2O clearly presents two well-separated velocity components. The blueshifted component was found to be very well associated with the polarized emission while the redshifted component was found to be much more extended. For this reason, in this case two Gaussinas had to be fitted and the linewidth of the blueshifted component was used for our analysis. The first-order moment shown in Fig.~\ref{fm1a} for W3H2O corresponds also to the blueshifted velocity component.}. For the case of N6334I there is absorption towards the two strongest continuum sources and some emission from 5 to 10~\kms\ (velocities given with respect to the local standard of rest), but these features are very compact. The main emission is found in the velocity range from $-15$ to 0 ~\kms, and the Gaussian was fitted in this velocity range.

\renewcommand{\thefigure}{4}
\begin{figure*}
\begin{center}
\begin{tabular}[b]{c}
    \epsfig{file=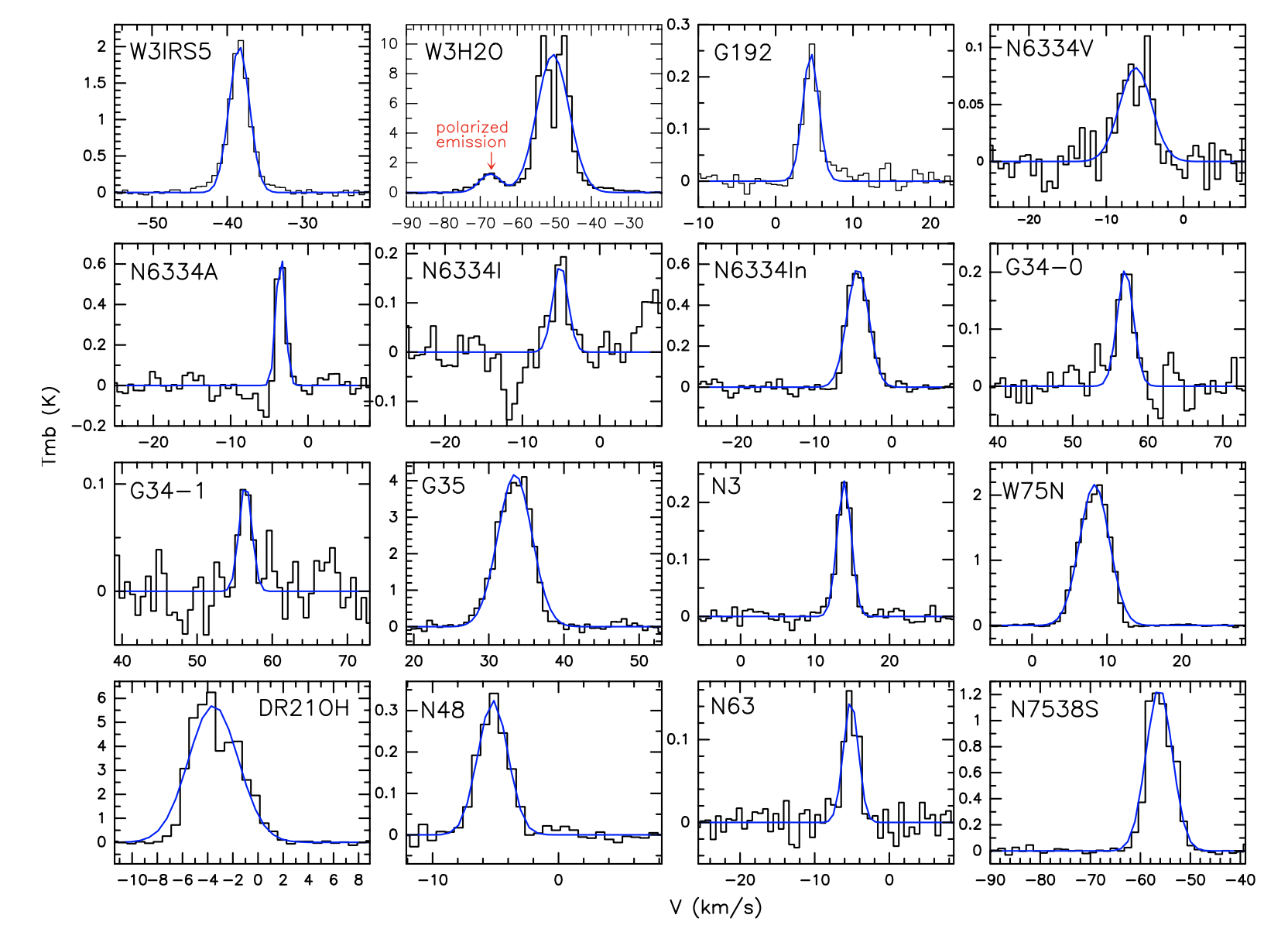, width=11cm, angle=0}\\
\end{tabular}
\caption{\htcop\,(4--3) spectra (black) with the gaussian fits (blue curves) performed to estimate the velocity dispersion in each region. For the case of W3H2O, the velocity component associated with the polarized emission is marked with a red arrow.
}
\label{fspecs}
\end{center}
\end{figure*}

The spectra and the corresponding fits are shown in Fig.~\ref{fspecs}, and the velocity line widths, $\Delta v_\mathrm{0.15pc}$, are listed in Table~\ref{tkin}. 
As can be seen from the figure, most spectra appear to be well fitted with only one Gaussian. 
%In Fig.~\ref{fNmm-sigmaPA} {\bf of Appendix~\ref{apvelsigma}}, a plot of the fragmentation level vs velocity dispersion is presented.

\begin{deluxetable*}{lCCCCCCCCCC}
%\tablenum{1}
\tablecaption{Kinematic properties of the sample from the \htcop\,(4--3) SMA data\label{tkin}}
\tablewidth{0pt}
\tablehead{
\colhead{} & 
\colhead{$\Delta v_\mathrm{0.15pc}$\supa} & 
\colhead{$\sigma_\mathrm{tot}$}\supa & 
\colhead{$\sigma_\mathrm{nonth}$\supa} & 
\colhead{$\sigma_\mathrm{turb,VDF}$\supb} &
\colhead{} &
\colhead{} & 
\colhead{$E_\mathrm{kin}$\supc} & 
\colhead{$E_\mathrm{turb}$\supc} & 
\colhead{$E_\mathrm{grav}$\supd} & 
\colhead{} \\
\colhead{Source} & 
\colhead{(\kms)} & 
\colhead{(\kms)} & 
\colhead{(\kms)} &
\colhead{(\kms)} &
\colhead{$\mathcal{M}$\supa} &
\colhead{$\frac{\sigma_\mathrm{turb}}{\sigma_\mathrm{nonth}}$}\supb & 
\colhead{($10^{45}$\,erg)}\supc & 
\colhead{($10^{45}$\,erg)}\supc & 
\colhead{($10^{45}$\,erg)}\supc & 
\colhead{$\frac{E_\mathrm{turb}}{E_\mathrm{kin}}$}
}
\colnumbers
\startdata
1-W3IRS5          		&3.20	&1.36	&{1.35}			&2.63	&3.6	&1.95	&1.24	&4.62	&0.34	&3.7  	\\
2-W3H2O 			&6.72	&2.85	&2.84			&1.30	&9.1	&0.46	&14		&2.96	&2.36	&0.2  	\\
3-G192				&2.43	&1.03	&1.03			&-		&4.6	&-		&0.35	&-		&0.08	&-	\\
4-N6334V				&4.87	&2.07	&{2.06}			&1.05	&8.0	&0.51	&6.59	&1.69	&1.81	&0.3	\\
5-N6334A				&1.34	&0.57	&0.56			&0.54	&2.6	&{0.97}	&0.45	&0.40	&1.47	&0.9	\\
6-N6334I				&2.31	&0.98	&0.97			&2.87	&3.4	&{2.96}	&2.11	&18		&3.65	&8.5 	\\
7-N6334In				&3.28	&1.39	&1.39			&0.48	&5.6	&0.35	&4.71	&0.56	&4.48	&0.1	 \\
8-G34-0				&2.54	&1.08	&1.07			&-		&4.6	&-		&1.53	&-		&1.31	&-	\\ 
9-G34-1				&1.98	&0.84	&0.84			&-		&4.2	&-		&0.89	&-		&1.20	&-	 \\ 
10-G35				&5.34	&2.27	&{2.26}			&0.64	&9.7	&0.28	&5.57	&0.44	&0.90	&0.1	\\ 
11-I20126			 	&-		&-		&-				&-		&-	&-		&-		&-		&0.26	&-	\\
12-N3 				&2.22	&0.94	&0.94			&-		&5.7	&-		&0.78	&-		&0.59	&-	\\
13-W75N 				&4.70	&2.00    	&1.99			&0.50	&7.0	&0.25	&5.78	&0.36	&1.60	&0.1	\\
14-DR21OH			&4.66	&1.98	&1.98			&0.52	&8.8	&0.26	&12.3	&0.85	&7.55	&0.1	\\
15-N48 				&2.87	&1.22	&1.22			&0.47	&6.3	&0.39	&2.50	&0.37	&2.16	&0.2	\\
16-N53				&-		&-		&-				&-		&-	&-		&-		&-		&1.67	&- 	\\
17-N63 				&2.36	&1.00	&1.00			&-		&5.5	&-		&0.61	&-		&0.28	&-	\\
18-N7538S			&6.18	&2.62	&2.62			&0.70	&11	&0.27	&15		&1.05	&3.50	&0.1	\\
\enddata
\tablecomments{
\\
$^\mathrm{a}$ $\Delta v_\mathrm{0.15pc}$ is the line width obtained from fitting a Gaussian to the \htcop\,(4--3) spectrum averaged over a region of 0.15~pc of diameter.
$\sigma_\mathrm{tot}=\Delta v_\mathrm{0.15pc}/\sqrt{8\,\mathrm{ln}(2)}$ corresponds to the 1D total (thermal+nonthermal) velocity dispersion.
$\sigma_\mathrm{nonth} =\sqrt{\sigma_\mathrm{tot}^2-\sigma_\mathrm{th}^2}$, with $\sigma_\mathrm{th}=\sqrt{k_\mathrm{B}\,T/(\mu\,m_\mathrm{H})}$ ($k_\mathrm{B}$ being the Boltzmann constant, $\mu$ the molecular weight (30 for \htcop), $m_\mathrm{H}$ the mass of the hydrogen atom and $T$ the temperature of the region, taken from column (10) of Table~\ref{tfit}).
The Mach number $\mathcal{M}$ is calculated as $\sigma_\mathrm{3D,nth}$/$c_\mathrm{s}$, with  $c_\mathrm{s}$ being the sound speed calculated as $c_\mathrm{s}=\sqrt{k_\mathrm{B}\,T/(\mu\,m_\mathrm{H})}$, using $\mu=2.3$, and $\sigma_\mathrm{3D,nth}=\sqrt{3}\,\sigma_\mathrm{nonth}$.\\
%%From Palau+15, tablenote b from Table1: $\sigma^\mathrm{NH_3}_\mathrm{1D,obs}$ and $\sigma^\mathrm{N2H^+}_\mathrm{1D,obs}$ are calculated from the measured FWHM line width, $\Delta v_\mathrm{obs}$,  as $\sigma_\mathrm{1D,obs}=\Delta v_\mathrm{obs}/(8\,\mathrm{ln}2)^{1/2}$. 
%
$^\mathrm{b}$  $\sigma_\mathrm{turb,VDF}$ is taken from the Velocity Dispersion Function at the smallest scale (Section~\ref{saveldisp}), and thus should be free of large-scale systematic motions. $\frac{\sigma_\mathrm{turb}}{\sigma_\mathrm{nonth}}$ is calculated using $\sigma_\mathrm{turb,VDF}$ and corresponds to $Q$ in Section~\ref{saveldisp} and equation~\ref{eqsigmavel}.\\
$^\mathrm{c}$ $E_\mathrm{kin}$ is the total kinetic energy calculated as $\frac{3}{2}M_\mathrm{0.15pc}\sigma_\mathrm{tot}^2$, $E_\mathrm{turb}$ is the turbulent kinetic energy calculated as $\frac{3}{2}M_\mathrm{0.15pc}\sigma_\mathrm{turb,VDF}^2$, and $E_\mathrm{grav}$ is the gravitational energy calculated as $\frac{3}{5}G\,M_\mathrm{0.15pc}^2/R$, with $R=0.15/2=0.075$~pc, and $M_\mathrm{0.15pc}$ taken from Table~\ref{tfit}.
}
\end{deluxetable*}

The velocity dispersion associated with turbulence is estimated assuming that it is a factor $Q$ of the non-thermal dispersion. The non-thermal dispersions, $\sigma_\mathrm{nonth}$, are listed in Table~\ref{tkin} and were calculated by assuming the average temperature reported in Table~\ref{tfit}, $T_\mathrm{0.15pc}$. As can be seen from Table~\ref{tkin}, in our case the non-thermal dispersions are essentially the same as the total velocity dispersions. Thus, we estimate that the 1D (along the line of sight) turbulent dispersion is:

\begin{equation}\label{eqsigmavel}
\sigma_\mathrm{turb,spec} = Q\,\Delta\,v_\mathrm{0.15pc}/\sqrt{8 \mathrm{ln}2},
\end{equation}

where the subindex `spec' is written to remind that this estimate makes use of the line width inferred from the spectrum. In this equation, we assumed that $Q\sim0.5$. The $Q$ factor, defined here as $Q\equiv\frac{\sigma_\mathrm{turb}}{\sigma_\mathrm{nonth}}$, is required to take into account the fact that systematic large-scale motions could be contributing a non-negligible part of the total non-thermal dispersion, as shown by recent simulations of gravitational contraction of turbulent cores (\eg\ Guerrero-Gamboa \& V\'azquez-Semadeni 2020). An uncertainty of $\sim10$\% is assumed for $\sigma_\mathrm{turb,spec}$ (\eg\ Guerrero-Gamboa \& V\'azquez-Semadeni 2020).

A first estimate of the $Q$ factor was performed from our \htcop\,(4--3) first-order moment maps of Figs.~\ref{fm1a} and \ref{fm1b}. These figures show that velocity gradients are clearly present in our regions. In an attempt to separate the turbulent component of the kinetic energy, $E_\mathrm{turb}$, from the total kinetic energy,  $E_\mathrm{kin}$, which could be dominated by large-scale motions (\eg\ rotation, infall), Velocity Dispersion Functions (VDFs) were built for the same regions for which the Angular Dispersion Function (ADF) were also calculated in Section~\ref{saBADF}. The VDFs were calculated as:

\begin{equation}\label{eqVDF}
\langle [\Delta V(\ell)]^2 \rangle^{1/2} \equiv \sqrt{\frac{1}{N(\ell)}  \sum_{i=1}^{N(\ell)} [  V({\bf x}) - V({\bf x}+\ell)   ]^2 },
\end{equation}

where $V$ is the velocity along the line of sight at each position of our maps. We refer the reader to Section~\ref{saBADF} for further details on the dispersion function. The results are shown in Fig.~\ref{fVDF} and $\sigma_\mathrm{turb,VDF}$, listed in Table~\ref{tkin}, was estimated from the intercept value. Table~\ref{tkin} also reports the ratio $\frac{\sigma_\mathrm{turb}}{\sigma_\mathrm{nonth}}$ or $Q$. With the exception of W3IRS5 and N6334I, $\frac{\sigma_\mathrm{turb}}{\sigma_\mathrm{nonth}}$ averages to 0.4, with a standard deviation of 0.2, very consistent with the value measured in the simulations of Guerrero-Gamboa \& V\'azquez-Semadeni (2020). The two regions with $\frac{\sigma_\mathrm{turb}}{\sigma_\mathrm{nonth}}\sim2$ could be affected by small-scale rotation or infalling motions, as found in previous works (\eg\ Zhang et al. 1998, 2002). Actually, the spectrum of N6334I presents an absorption signature due to infall (Fig.~\ref{fspecs}). Thus, our adopted value of $Q\sim0.5$ seems reasonable.

For completeness, Table~\ref{tkin} also lists the values of the Mach number, $\mathcal{M}$, ranging from 3 to 11, and the values of the total kinetic energy, $E_\mathrm{kin}$, the turbulent kinetic energy, $E_\mathrm{turb}$, their ratio, and the gravitational energy, $E_\mathrm{grav}$.

\renewcommand{\thefigure}{5}
\begin{figure*}
\begin{center}
\begin{tabular}[b]{c}
    \epsfig{file=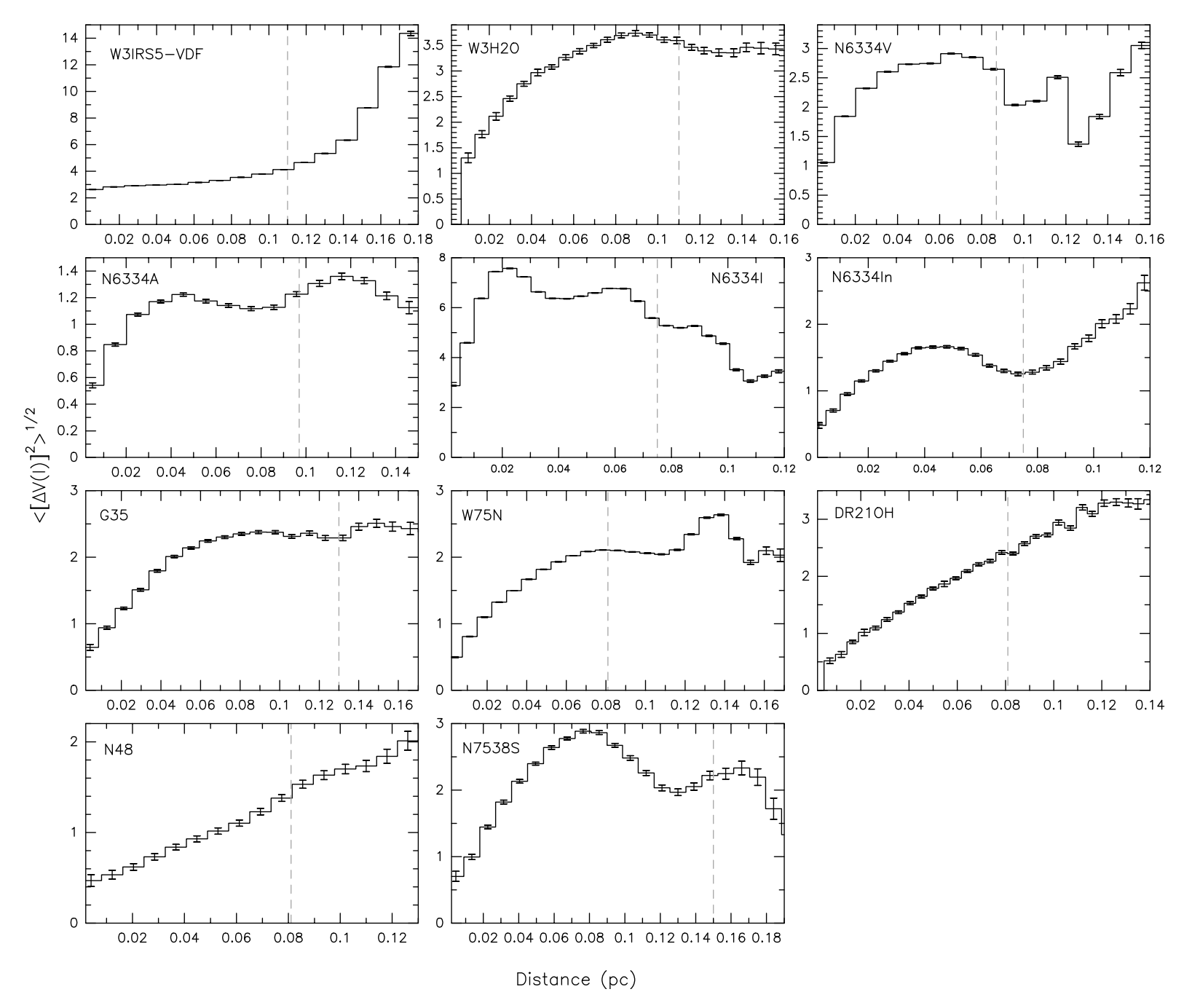, width=11cm, angle=0}\\
\end{tabular}
\caption{Velocity Dispersion Functions in \kms\ (\ie\ $\langle [\Delta V(\ell)]^2 \rangle^{1/2}$, as defined in equation~\ref{eqVDF}), for the regions for which the ADF (Section~\ref{saBADF}) was also calculated, measured using the \htcop\,(4--3) first-order moment maps of Figs.~\ref{fm1a} and \ref{fm1b}. The dashed grey vertical line indicates the largest angular scale that the SMA is able to recover for the \htcop\,(4--3) emission.
}
\label{fVDF}
\end{center}
\end{figure*}

\subsection{Determination of the magnetic field strength: the Davis-Chandrasekhar-Fermi method}\label{saB}

Polarization observations of thermal dust emission at submillimeter wavelengths constitute a powerful tool to estimate the magnetic field strength onto the plane of the sky, $B_\mathrm{pos}$. In order to estimate this, the Davis-Chandrasekhar-Fermi method (DCF, Davis 1951; Chandrasekhar \& Fermi 1953) has been widely used. In this method, it is assumed that the turbulent kinetic energy and the turbulent magnetic energy are equal, and that the turbulent gas induces the observed dispersion in the polarization position angles (PA). Therefore higher PA dispersions correspond to weak magnetic fields such that the turbulent gas can drag the field lines.
%Conversely, small dispersions in the polarization angle should be indicative of magnetic fields strong enough to force the gas to move along the field lines. 
Following Chandrasekhar \& Fermi (1953), the magnetic field strength of the ordered component of the magnetic field, $B_0$, can be estimated from the following relation, once the density, $\rho$, turbulent velocity dispersion along the line of sight, $\sigma_\mathrm{turb}$, and PA dispersion, $\sigma_\mathrm{PA}$, are known:

\begin{equation}\label{eqBpos}
B_0 \sim B_\mathrm{pos} = \sqrt{4\pi\rho}\,\frac{\sigma_\mathrm{turb}}{\delta B/B_0}
\sim
f\,\sqrt{4\pi\rho}\,\frac{\sigma_\mathrm{turb}}{\sigma_\mathrm{PA}},
\end{equation}

where it is assumed that $\delta\,B/B_0 \sim \sigma_\mathrm{PA}$, with $\delta\,B$ being the perturbed magnetic field on the top of $B_0$. In this equation, $f$ is a numerical correction factor usually adopted to be 0.5 (Ostriker et al. 2001). This numerical factor was derived for the cases where equation~\ref{eqBpos} is valid, mainly for $\sigma_\mathrm{PA} \lesssim25^\circ$. For larger dispersions, $\sigma_\mathrm{PA}$ should be replaced by $\tan{\sigma_\mathrm{PA}}$ (Falceta-Gon\c calves et al. 2008), and there is no need to apply the numerical correction factor $f$.
The tangent correction is required in our case because our measured $\sigma_\mathrm{PA}$ are in some regions large (see below), of up to 50$^\circ$, and thus $\tan{\sigma_\mathrm{PA}}$, which is what strictly corresponds to $\delta B/B_0$, cannot be approximated to $\sigma_\mathrm{PA}$.

In the following subsections we present two different approaches to estimate $\sigma_\mathrm{PA}$ following the DCF method.

\subsubsection{Polarization Position Angle Dispersion from standard deviation}\label{saBstdev}

In order to estimate a dispersion in polarization PA, we extracted the PA values, $\Phi$,  from each PA image, for the same region where fragmentation was assessed (0.15~pc of diameter), and rejected those angles with an error larger than $\sim10^\circ$ (11$^\circ$ for G34-1, and 15$^\circ$ for G192\footnote{Only for the case of G192 we included PA with errors $\sim15^\circ$ because an inspection of the PA values showed that there was two PA components, one around 20$^\circ$ and the other around 85$^\circ$. The component around 20$^\circ$ is the one with large errors in the PA, but the different values of this component are very similar, making this component more significant (see Fig.~\ref{fpol1}). 15$^\circ$ of PA error corresponds to a signal-to-noise ratio $\sim2$.}),
%to have a better defined peak of the histogram thus allowing the Gaussian fit), 
which corresponds to a S/N ratio smaller than 3 for the signal in polarization (Zhang et al. 2014).  Three PA points per beam were extracted for each region.
%QZ: the dispersion analysis of polarization angles is not sensitive to the number of data points, thus using 2x2 or 3x3 per beam sampling won't affect the results much.

A first approach to estimate $\sigma_\mathrm{PA}$ is based on the calculation of the standard deviation of the weighted mean. However, given the fact that the PA values have an error associated, it is desirable to subtract the contribution of the PA error to $\sigma_\mathrm{PA}$. Thus, given a number $N$ of PAs measured in a certain region, we have used the following expression:

\begin{equation}\label{eqsigmaPAstdev}
\sigma_\mathrm{PA,stdev} = \sqrt{\frac{N\,\sum_{i=1}^{N}\,w_i\,(\Phi_i - \overline\Phi_w)^2}{(N-1)\sum_{i=1}^{N}\,w_i}-\frac{N}{\sum_{i=1}^{N}\,w_i}},
\end{equation}

where $\overline\Phi_w=\frac{\sum_{i=1}^{N}\,w_i\,\Phi_i}{\sum_{i=1}^{N}\,w_i}$ is the weighted mean of PAs, and the second term within the square root corresponds to the contribution of the PA errors to the dispersion (equation A2 of A\~nez-L\'opez et al. 2020a). In these equations, $w_i$ are the weights of each PA $\Phi_i$, $w_i = 1/\delta\Phi_i^2$, with $\delta\Phi_i$ the PA error of each PA measurement. This final $\sigma_\mathrm{PA,stdev}$ can be considered as an intrinsic standard deviation, because the contribution from the PA errors has been removed. The magnetic field strength in the plane of the sky following this approach is calculated as:

\begin{equation}\label{eqBstdev}
B_\mathrm{stdev} = \sqrt{4\pi\rho}\,\frac{\sigma_\mathrm{turb}}{\tan{\sigma_\mathrm{PA,stdev}}},
%B_\mathrm{pos} = \sqrt{4\pi\rho}\,\frac{\sigma_\mathrm{vel,los}}{\delta B/B_0}.
\end{equation}

where $\sigma_\mathrm{turb}$ has been obtained from equation~\ref{eqsigmavel} adopting $Q\sim0.5$ (Section~\ref{saveldisp}).

The obtained values of $\sigma_\mathrm{PA,stdev}$ and $B_\mathrm{stdev}$ for each region are listed in Table~\ref{tB}. $\sigma_\mathrm{PA,stdev}$ ranges from 6 to 50$^\circ$, and their associated uncertainties were estimated from Monte Carlo simulations, to take into account the sparse sampling in our PA images (see Appendix~\ref{apMonteCarlo} for further details). The uncertainties in $B_\mathrm{stdev}$ were estimated by propagating the uncertainties in the density, velocity dispersion and $\sigma_\mathrm{PA,stdev}$.

%A plot of \Nmm\ vs $\sigma_\mathrm{PA,stdev}$ is presented in Fig.~\ref{fNmm-sigmaPA}  (Appendix~\ref{apvelsigma}). {\bf If the magnetic field strength was directly related to $\sigma_\mathrm{PA,stdev}$ (as assumed in the DCF method) and if the magnetic field was actually suppressing fragmentation, one would expect to find a positive trend in this figure}, as a stronger magnetic field should yield a small  $\sigma_\mathrm{PA,stdev}$ and this in turn should prevent fragmentation. However, such a trend is not apparent from our figure.
%%In a few cases such as W3IRS5, W3H2O, N48 and N63 the PA dispersion is rather large because there is a gradient in the PA image.

\renewcommand{\thefigure}{6}
\begin{figure*}
\begin{center}
\begin{tabular}[b]{c}
    \epsfig{file=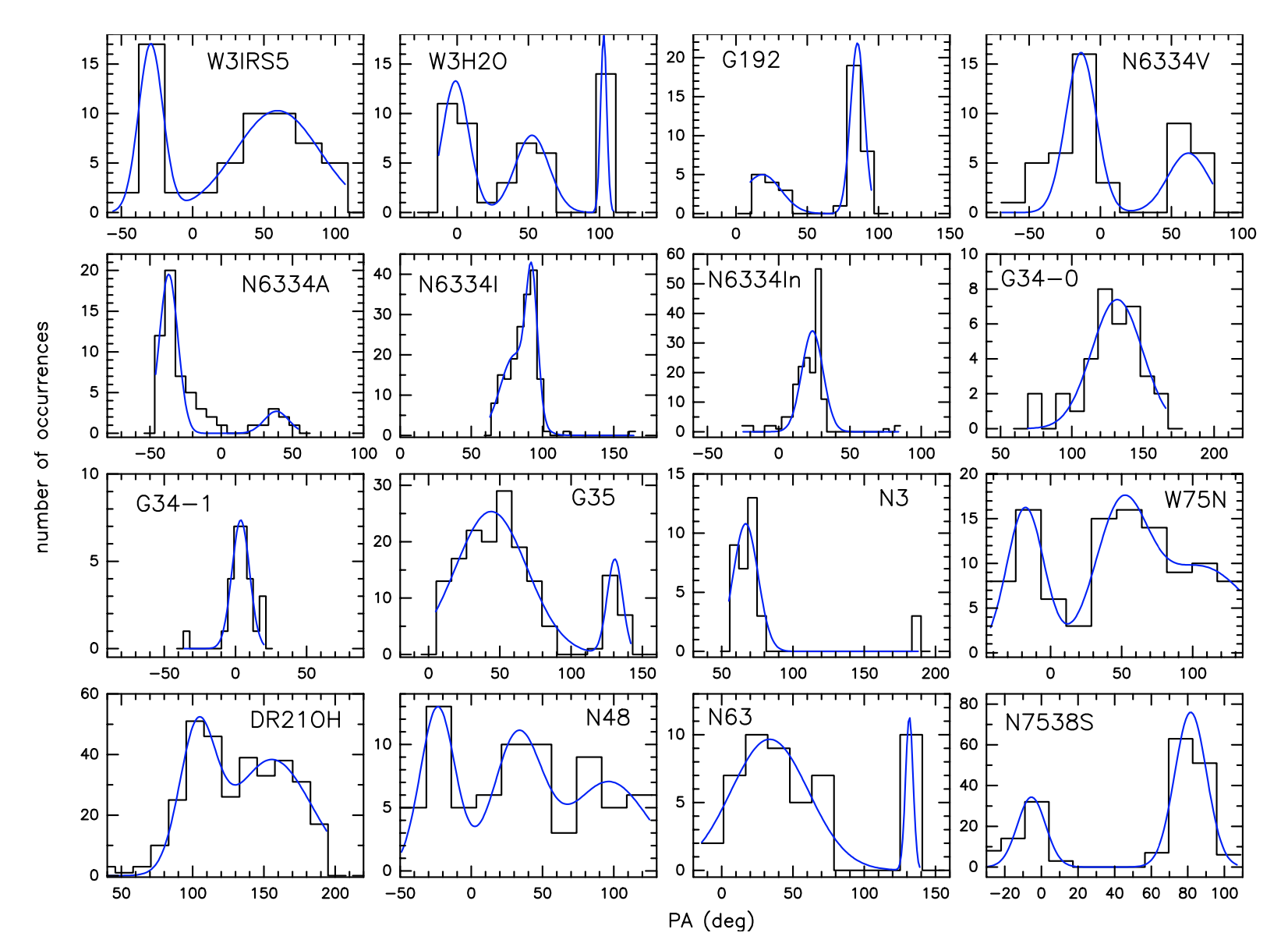, width=11cm, angle=0}\\
\end{tabular}
\caption{Polarization angle histograms (black) with the gaussian fits (blue curves) performed in Python to estimate $\sigma_\mathrm{PA,gauss}$ in each region. In all panels the full range of 180$^\circ$ is shown.
}
\label{fpahisto}
\end{center}
\end{figure*}

\subsubsection{Polarization Position Angle Dispersion from multiple gaussian fitting}\label{saBgauss}

In the previous section,  $\sigma_\mathrm{PA}$ was estimated by calculating the standard deviation of the PAs. However, the dispersion values calculated from equation~\ref{eqsigmaPAstdev} would systematically overestimate the dispersion in the cases where multiple components of the magnetic field are present. Thus, another approach to estimate the $\sigma_\mathrm{PA}$ is to fit Gaussians to the histogram of PAs. The number of bins was determined using the `auto' option of the `histogram' function of Python, and  the resulting histograms are presented in Fig.~\ref{fpahisto}. 
For the cases of G192, N3 and N63, $\sim4$ points per beam were extracted from the PA image to allow a more robust fit of the histogram. 
In some cases, the histogram could be fitted with one single Gaussian component. 
%We used number of bins = auto+1 for all regions. 
%In other cases, two or three Gaussian components were fitted, in order to avoid obtaining an artificially large PA dispersion for those cases where different groups of PA are present. However, in the cases where two or three components are present, as is the case of NGC\,6336V (Ju\'arez et al. 2017), the fit of one single Gaussian could yield too high a PA dispersion  while still being the magnetic field quite ordered, mimicking a low magnetic field strength in cases where the magnetic field can be relatively high. 
%Thus, we fitted multiple Gaussian components when required. 
In other cases, two (three) components were clearly separated in the PA histogram, and each component was fitted with a different Gaussian. However, there were cases clearly deviating from one single Gaussian but still with the two components merged so that it is ambiguous whether there are two (narrower) components or one (broader) component. In general, we considered separated PA components if the second peak of the tentative component is separated in $y$-axis by more than half of the peak of the strongest component while still being significant in number of points (around 5). This implied fitting two components for the ambiguous cases of N6334I, DR21OH, and N63\footnote{For the case of N63, we also tried to fit 3 PA components. This implied a stronger magnetic field by less than a factor of two. 
%This is true for both the DCF method and for the DCF method following the correction of Falceta-Gon\c calves et al. (2008), $B_2$ (see Figs.~\ref{fNmm} and \ref{fNmmB2}).
}, and  three components for W75N.
%(we did not fit two components in G34.4.0 and NGC6334A because the peak of the second component corresponded only to 2 and 3 (respectively) points, and the second gaussian would be made of only 3 and 5 points, respectively).

The final PA dispersions following this approach, $\sigma_\mathrm{PA,gauss}$, were computed as the average width of the different components, weighted by the area of each Gaussian, and range from 6 to 24$^\circ$. The magnetic field strength in the plane of the sky following this approach is calculated as:

\begin{equation}\label{eqBgauss}
B_\mathrm{gauss} = \sqrt{4\pi\rho}\,\frac{\sigma_\mathrm{turb}}{\tan{\sigma_\mathrm{PA,gauss}}},
%B_\mathrm{pos} = \sqrt{4\pi\rho}\,\frac{\sigma_\mathrm{vel,los}}{\delta B/B_0}.
\end{equation}

where $\sigma_\mathrm{turb}$ has been obtained from equation~\ref{eqsigmavel} adopting $Q\sim0.5$ (Section~\ref{saveldisp}).

The results are presented in Fig.~\ref{fpahisto} and listed in Table~\ref{tB}.  The uncertainty in $\sigma_\mathrm{PA,gauss}$ was  estimated from Monte Carlo simulations as in the previous section (see Appendix~\ref{apMonteCarlo} for further details).
%Fig.~\ref{fNmm-sigmaPA} (Appendix~\ref{apvelsigma}) shows the fragmentation level against $\sigma_\mathrm{PA,gauss}$, {\bf and again no trend can be appreciated in the figure, as in the previous approach}. 
%For the calculation of the magnetic field strength the correction by Falceta-Gon\c calves et al. (2008) was used, and the final values are listed in Table~\ref{tB} and shown in Fig.~\ref{fNmmBmu}. 
%Note that I also plotted Nmm vs tan(sigmaPA) and the trend is exactly the same. It seems not worthy showing this plot because it is exactly the same as Nmm vs sigmaPA :)

\begin{rotatetable*}
\begin{deluxetable*}{cCCC CCCC CCCC CCC}
\tablecaption{Magnetic field properties derived using the DCF and ADF methods\label{tB}}
\tablewidth{760pt}
\tabletypesize{\scriptsize}
\tablehead{
\colhead{} & 
\colhead{$\sigma_\mathrm{stdev}^\mathrm{PA}$\supa} & 
\colhead{$\sigma_\mathrm{gauss}^\mathrm{PA}$\supb} &
\colhead{$\sigma_\mathrm{ADFbeam}^\mathrm{PA}$\supc} &
\colhead{} &
\colhead{$B_\mathrm{stdev}$\supa} &
\colhead{$B_\mathrm{gauss}$\supb} &
\colhead{$B_\mathrm{ADFbeam}$\supc} &
\colhead{$B_\mathrm{ADFH09}$\supc} &
\colhead{$B_\mathrm{lit}$\supd} &
\colhead{} &
\colhead{} &
\colhead{} &
\colhead{} &
\colhead{} \\
%%%%%%%%%%%%%%%%%%%%%%%%%%%%%%%%
\colhead{ID} & 
\colhead{($^\circ$)} & 
\colhead{($^\circ$)} &
\colhead{($^\circ$)} &
\colhead{$\frac{\langle B_\mathrm{t}^2\rangle}{\langle B_\mathrm{0}^2\rangle}$\supc} &
\colhead{(mG)} &
\colhead{(mG)} &
\colhead{(mG)} &
\colhead{(mG)} &
\colhead{(mG)} &
\colhead{$\mu_\mathrm{stdev}$\supa} &
\colhead{$\mu_\mathrm{gauss}$\supb} &
\colhead{$\mu_\mathrm{ADFbeam}$\supc} &
\colhead{$\mu_\mathrm{ADFH09}$\supc} &
\colhead{$\mu_\mathrm{gauss}^\mathrm{S21}$\supb} 
}
\colnumbers
\startdata
1        	&50\pm7	 &23\pm4		&18\pm2		&1.37\pm0.93	&0.2\pm0.1 &0.5\pm0.1 	&0.7\pm0.1 	&0.2\pm0.1  	&-		&1.8\pm0.6  	&0.6\pm0.2 	&0.5\pm0.1	&1.8\pm0.8	&1.4\\
2		&41\pm11	 &9\pm3		&16\pm2		&1.01\pm0.69	&0.9\pm0.4 &4.7\pm1.5	&2.7\pm0.5	&0.8\pm0.3	&4.6		&1.0\pm0.5	&0.2\pm0.1	&0.3\pm0.1	&1.2\pm0.5 	&0.7\\
3		&18\pm6	 &8\pm8		&-			&-			&0.4\pm0.1 &0.8\pm0.9	&-			&-		   	&-		&0.5\pm0.2	&0.2\pm0.2	&-			&-			&0.7\\
4		&35\pm6	 &12\pm5		&14\pm1		&0.66\pm0.54	&0.7\pm0.2 &2.4\pm1.2	&2.1\pm0.5	&0.6\pm0.3	&0.7		&1.1\pm0.5	&0.3\pm0.2	&0.4\pm0.2	&1.2\pm0.7	&1.0\\
5		&16\pm6	 &7\pm3		&11\pm2		&0.16\pm0.03	&0.5\pm0.2 &1.1\pm0.5	&0.7\pm0.2	&0.3\pm0.1	&-		&1.5\pm0.7	&0.6\pm0.3	&1.0\pm0.4	&2.1\pm0.7	&2.5\\
6		&7\pm3	 &8\pm3		&6\pm1		&0.04\pm0.01	&2.6\pm1.4 &2.1\pm1.0	&2.9\pm0.7	&1.4\pm0.3	&2.8		&0.4\pm0.3	&0.5\pm0.3	&0.4\pm0.1	&0.8\pm0.3	&2.0\\
7		&6\pm3	 &7\pm4		&5\pm1		&0.08\pm0.01	&4.1\pm2.3 &3.3\pm2.1	&4.7\pm1.3	&1.6\pm0.3	&3.2		&0.3\pm0.2	&0.4\pm0.2	&0.3\pm0.1	&0.8\pm0.3	&1.4\\
8		&17\pm6	 &18\pm7		&-			&-			&0.8\pm0.4 &0.8\pm0.4	&-			&-		   	&0.5		&0.8\pm0.4	&0.9\pm0.5	&-			&-			&2.1\\
9		&6\pm2	 &6\pm5		&-			&-			&1.8\pm0.7 &1.8\pm1.5	&-			&-		   	&1.1		&0.4\pm0.2	&0.4\pm0.3	&-			&-			&1.6\\
10		&33\pm5	 &23\pm3		&9\pm2		&0.32\pm0.11	&0.7\pm0.2 &1.1\pm0.2	&2.9\pm0.6	&0.8\pm0.2	&1.4		&0.7\pm0.2	&0.5\pm0.1	&0.2\pm0.1	&0.7\pm0.2	&1.1\\
11		&-		 &-			&-			&-			&-		   &- 			&-			&-		   	&1.8		&-		   	&-			&-			&-			&-\\
12		&47\pm16	 &9\pm7		&-			&-			&0.2\pm0.1 &1.2\pm0.9	&-			&-		   	&0.3		&2.6\pm1.6	&0.4\pm0.3	&-			&-			&1.4\\
13		&35\pm16	 &24\pm13	&21\pm1		&2.08\pm0.91	&0.7\pm0.4 &1.1\pm0.7	&1.2\pm0.2	&0.3\pm0.1	&-		&1.1\pm0.7	&0.7\pm0.4	&0.6\pm0.2	&2.2\pm0.8	&1.4\\
14		&32\pm13	 &22\pm12	&19\pm2		&1.21\pm0.30	&1.1\pm0.6 &1.7\pm1.0	&2.1\pm0.4	&0.6\pm0.1	&1.8		&1.4\pm0.9	&0.9\pm0.6	&0.8\pm0.3	&2.5\pm1.0	&2.0\\
15		&43\pm12	 &19\pm9		&20\pm2		&2.73\pm0.54	&0.3\pm0.2 &0.9\pm0.5	&0.9\pm0.2	&0.2\pm0.1	&0.5		&2.5\pm1.4	&0.9\pm0.6	&1.0\pm0.4	&4.4\pm1.6	&2.2\\
16 		&-		 &-			&-			&-			&-		   &- 			&-			&-		   	&0.8		&-		   	&-			&-			&-			&-\\
17		&46\pm20	 &25\pm14	&-			&-			&0.2\pm0.1 &0.3\pm0.2	&-			&-		   	&-		&2.0\pm1.5	&0.9\pm0.7	&-			&-			&1.9\\
18		&38\pm4	 &8\pm2		&6.0\pm1		&0.51\pm0.60	&1.0\pm0.2 &5.2\pm1.5	&7.3\pm2.0	&1.1\pm0.7	&-		&1.1\pm0.4	&0.2\pm0.1	&0.2\pm0.1	&1.0\pm0.7	&0.8\\
\enddata
\tablecomments{
\\
$^\mathrm{a}$ 
$\sigma_\mathrm{stdev}^\mathrm{PA}$ corresponds to the PA dispersion calculated within a region of 0.15~pc of diameter and using the `standard deviation' approach as described in Section~\ref{saBstdev}. $B_\mathrm{stdev}$ is calculated using the DCF method and following equation~\ref{eqBstdev}, for which $n_\mathrm{0.15pc}$ is taken from Table~\ref{tfit}, and $\Delta v_\mathrm{0.15pc}$ is taken from Table~\ref{tkin}. $\mu_\mathrm{stdev}$ is the observed mass-to-flux ratio over the critical value, calculated following equation~\ref{eqmupracunits} (Section~\ref{samu}), and using $B_\mathrm{stdev}$.\\
$^\mathrm{b}$ 
$\sigma_\mathrm{gauss}^\mathrm{PA}$ corresponds to the PA dispersion calculated within a region of 0.15~pc of diameter and using the `multiple gaussian' approach as described in Section~\ref{saBgauss}. $B_\mathrm{gauss}$ is calculated using the DCF method and following equation~\ref{eqBgauss}. $\mu_\mathrm{gauss}$ is the observed mass-to-flux ratio over the critical value, calculated following equation~\ref{eqmupracunits} and using $B_\mathrm{gauss}$. $\mu_\mathrm{gauss}^\mathrm{S21}$ corresponds to the observed mass-to-flux ratio over the critical calculated following Skalidis \& Tassis (2021, see Section~\ref{ssduncintrinsic}).\\
$^\mathrm{c}$ 
$\sigma_\mathrm{ADFbeam}^\mathrm{PA}$ corresponds to the PA dispersion calculated within a region of 0.15 pc of diameter and using the `ADF beam' approach as described in Section~\ref{saBADFbeam}.
$B_\mathrm{ADFbeam}$ is calculated following equation~\ref{eqBADFbeam}. $\mu_\mathrm{ADFbeam}$ is the observed mass-to-flux ratio over the critical value, calculated following equation~\ref{eqmupracunits} and using $B_\mathrm{ADFbeam}$.
$\langle B_\mathrm{t}^2\rangle/\langle B_\mathrm{0}^2\rangle$ is the ratio of the perturbed magnetic field energy vs the ordered magnetic field energy. 
$B_\mathrm{ADFH09}$ is calculated using the `ADF H09' approach following equation~\ref{eqBADFH09}. $\mu_\mathrm{ADFH09}$ is the corresponding observed mass-to-flux ratio over the critical.\\
$^\mathrm{d}$ 
$B_\mathrm{lit}$ is the magnetic field strength for each region as reported in the literature (see Section~\ref{ssduncBlit}). The references and the method used in each case are as follows: W3H2O: Chen et al. (2012a, DCF); N6334V: Ju\'arez et al. (2017, ADF); N6334I and N6334In: Li et al. (2015, force equilibrium between gravity and magnetic tension and magnetic pressure);  G34-0 and G34-1: Tang et al. (2019, ADF for single-dish); G35: Qiu et al. (2013, ADF); I20126: Edris et al. (2005, Zeeman effect of OH masers); N3 and N53: Hezareh et al. (2013, ion-neutral); DR21OH: Girart et al. (2013, ADF) and Hezareh et al. (2010, ion-neutral); N48: Ching et al. (2017, ADF). 
}
\end{deluxetable*}
\end{rotatetable*}

\subsection{Determination of the magnetic field strength: The angular dispersion function (ADF) method}\label{saBADF}

The previous approaches to estimate the PA dispersion could be introducing biases. First, the `standard deviation' approach could be overestimating $\sigma_\mathrm{PA}$ because it is ignoring the fact that there could be different magnetic field components within the same region and takes into account the large-scale variations of the PA. Second, the `multiple gaussian' approach might be biased because of the decision of how many gaussian components should be used. Both the `standard deviation' and the `multiple gaussian' approaches might also be biased for broad PAs distributions where PAs separated about 180$^\circ$ actually correspond to the same direction. Fig.~\ref{fpahisto} shows that for 12 out the 16 regions studied here the different PA components are separated by less than 90$^\circ$ and should not be strongly affected by this problem. Only 4 regions present very broad distributions, W3H2O, W75N, N48, and N63, being W3H2O the most striking case (see also Table~\ref{tB}).
%Patrick: A fundamental issue is that for any 'broad' distribution of PAs, this effectively means averaging and weighing. e.g., a pair of PAs around -90 deg and +90 deg will add to a larger dispersion (because they are at the two wings of the Gaussian). But this is exactly not what we should get because  -90 and +90 deg are the same orientations. This issue comes from the fact that the B-field PAs are cyclic quantities but not vectors. They only carry information about orientation which is defined in a range of 180 deg. 
%In the aforementioned methods, the assumption that $\delta\,B/B_0 \sim \sigma_\mathrm{PA}$ is assumed. 

A possible way to estimate more robustly $\sigma_\mathrm{PA}$ or $\delta B/B_0$ is the statistical method proposed by Hildebrand et al. (2009) and Houde et al. (2009). This method is based on the calculation of the angular dispersion function (ADF), defined as:

\begin{equation}\label{eqADF}
\langle [\Delta \Phi(\ell)]^2 \rangle^{1/2} \equiv \sqrt{\frac{1}{N(\ell)}  \sum_{i=1}^{N(\ell)} [  \Phi({\bf x}) - \Phi({\bf x}+\ell)   ]^2 },
\end{equation}

where ${\bf x}$ is the position vector in the plane of the sky, $\ell \equiv |{\bf \ell}|$, and $N(\ell)$ is the number of pairs of vectors separated by the displacement ${\bf \ell}$. The square of equation~\ref{eqADF} is also referred to as the structure function of the second order, but the structure function does not have angle units, and for this reason we refer to $\langle [\Delta \Phi(\ell)]^2 \rangle^{1/2}$ as the angular dispersion function. 
The ADF can be calculated for those regions with a large number of detections of polarization PA so that calculating an average for each distance bin is feasible.
%the fitting of the ADF curve with three free parameters is robust. 
For our sample, we applied this approach only to 11 regions, which are those with more than 45 PA detections.
Fig.~\ref{fADF} presents the ADF for these 11 regions. The ADF can be used in two different approaches to estimate $\delta B/B_0$.

\subsubsection{The angular dispersion function at the smallest (beam) scales}\label{saBADFbeam}

From the ADF one can directly measure the PA dispersion at the smallest resolved scale in our observations, naturally separating the large-scale component of the magnetic field from the small scale perturbations (Hildebrand et al. 2009). We define $\sigma_\mathrm{PA, ADFbeam}$ as the value of the ADF at scales equal to half the spatial scale of the beam, $\ell_\mathrm{B}$:

\begin{equation}\label{eqsigmaPAADFb}
\sigma_\mathrm{PA, ADFbeam} \equiv \langle [\Delta \Phi(\ell_\mathrm{B}/2)]^2 \rangle^{1/2}.
\end{equation}

The obtained values of $\sigma_\mathrm{PA, ADFbeam}$ along with the magnetic field strength in the plane of the sky following this approach,

\begin{equation}\label{eqBADFbeam}
B_\mathrm{ADFbeam} = \sqrt{4\pi\rho}\,\frac{\sigma_\mathrm{turb}}{\tan{\sigma_\mathrm{PA,ADFbeam}}},
%B_\mathrm{pos} = \sqrt{4\pi\rho}\,\frac{\sigma_\mathrm{vel,los}}{\delta B/B_0}.
\end{equation}

are listed in Table~\ref{tB}. Here $\sigma_\mathrm{turb}$ has been obtained from equation~\ref{eqsigmavel} adopting $Q\sim0.5$ (Section~\ref{saveldisp}). 
We refer to this approach as the `ADF beam' approach.

\renewcommand{\thefigure}{7}
\begin{figure*}[ht]
\begin{center}
\begin{tabular}[b]{c}
    \epsfig{file= 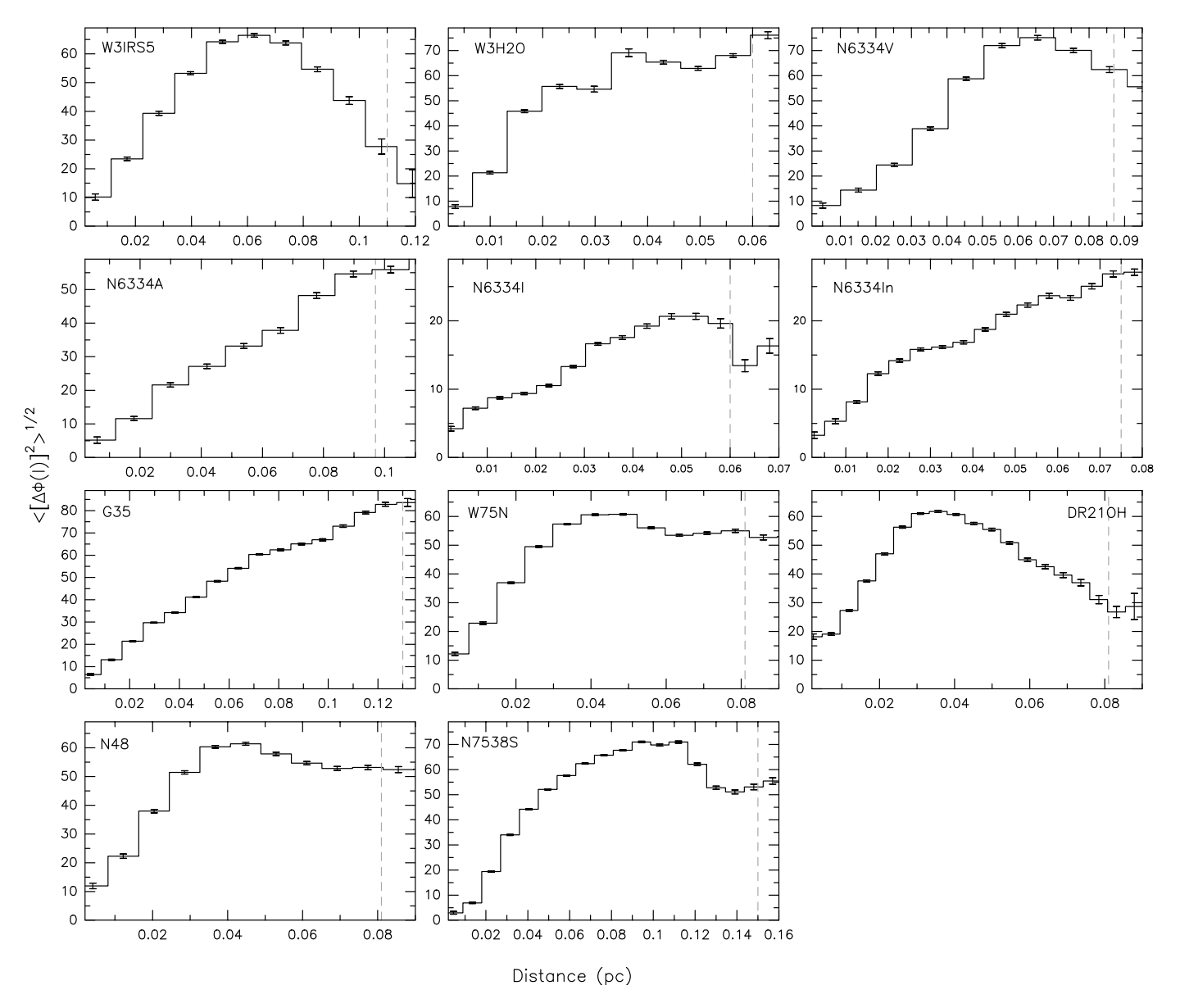, width=11cm, angle=0}\\
\end{tabular}
\caption{ADF (\ie\ $\langle [\Delta \Phi(\ell)]^2 \rangle^{1/2}$, as defined in equation~\ref{eqADF}) for each region with enough polarization detections. The dashed grey vertical line indicates the largest angular scale that the SMA is able to recover for the polarized emission.
}
\label{fADF}
\end{center}
\end{figure*}

\subsubsection{Fitting the angular dispersion function following Houde et al. (2009)}\label{saBADFH09}

Another approach to estimate $\delta B/B_0$ using the ADF was introduced by Houde et al. (2009). This approach takes into account the smoothing effect due to finite resolutions and the integration along the line of sight. In addition, it separates the contribution of large scale magnetic fields to angle dispersions because it assumes that there are two statistically independent components of the magnetic field. One component is related to the large-scale, ordered, magnetic field, $B_0$, and the other component corresponds to a perturbed or turbulent magnetic field, $B_\mathrm{t}$ (or $\delta B$). With this assumption, Houde et al. (2009) calculated the ADF to study the change of polarization PA differences with $\ell$, and related the output of this statistical analysis to the ratio of $B_\mathrm{t}$ energy to $B_0$ energy, $b\equiv \langle B_\mathrm{t}^2\rangle/\langle B_\mathrm{0}^2\rangle$, which corresponds to $(\delta B/B_0)^2$ and can thus be used in equation~\ref{eqBpos}.
%$(\delta\,B/B_0)^2$. 
We refer the reader to Houde et al. (2009) for further details on this approach. In short, Houde et al. (2009) calculate the ADF in the form:

\begin{equation}\label{eqADFH09}
1-\langle \mathrm{cos}[\Delta \Phi(\ell)] \rangle \simeq \frac{b}{N}\big(1-e^{-\ell^2/(2\delta^2+4W^2)}\big) + a'_2\,\ell^2,
\end{equation}

where $\delta$ is the magnetic field turbulent correlation length, $W$ is the `beam radius' ($W=FWHM/\sqrt{8\,\mathrm{ln}2}$, with FWHM being the full width at half-maximum of the beam), $a'_2$ is the coefficient of the parabolic approximation for the uniform part of the magnetic field in the ADF\footnote{The fact that the contribution of the uniform magnetic field in the ADF is approximated to a parabola does not necessarily mean that the morphology of the magnetic field follows a parabola, it rather corresponds to keeping the first $\ell^2$ term in the Taylor expansion of equation (42) of Houde et al. (2009), which is acceptable if $\ell$ is less than a few times the beam radius $W$ (Houde et al. 2009).}, and $N$ is the number of turbulent cells:

\begin{equation}
N \equiv \Delta'\,(\delta^2+2W^2)/(\sqrt{2\pi}\delta^3),
\end{equation}

with $\Delta'$ being the effective thickness of the cloud, expected to be slightly smaller than the cloud thickness\footnote{As explained in Section 3.2 of Houde et al. (2009), $\Delta'$ can be interpreted as the width of the large-scale autocorrelation function, and can be thought as the proportion of the cloud that contains the bulk of the polarized flux. Thus, it necessarily needs to be smaller than the physical cloud thickness.}. Here we assume that $\Delta'$ is equal to the core's thickness, taken as the diameter of the dense core in the plane of the sky as measured with the SMA (Koch et al. 2010). Thus, equation~\ref{eqADFH09} can be decomposed into a correlated component ($-\frac{b}{N}e^{-\ell^2/(2\delta^2+4W^2)})$, blue line in bottom panels of Fig.~\ref{fADFH09}) and the contribution of the large-scale uniform magnetic field ($b/N + a'_2\,\ell^2$, red dashed line in top panels of  Fig.~\ref{fADFH09}).

By fitting equation~\ref{eqADFH09} to the observational data the values for the three free parameters, $b/N$, $\delta$ and $a'_2$, are obtained. Only distances smaller than the physical distance of the largest angular scale of the SMA polarization observations ($\sim12''$\footnote{For the particular cases of N6334V and N6334A, the largest distance considered for the fit is $\sim20''$, because these two regions have slightly larger beams, of $\sim5''$, than the beams for the other regions, $\sim2''$--$3''$ and we require that at least a distance of 4 times the beam is covered. For W3H2O and N6334I, the largest distance considered for the fit was set to 0.06 pc because there were no data for larger distances.}) were taken into account to perform the fit. 
%aA was allowed to vary from 0.01 to 2 in steps of 0.002; aA=b/N/sqrt(2*pi) -> b/N is allowed to vary from 0.025 to 5 in steps of 0.005
The $b/N$ free parameter was allowed to vary from 0.025 to 5 in steps of 0.005, the $\delta$ free parameter was allowed to vary from
 5 mpc to half the effective cloud thickness\footnote{In the method described by Houde et al. (2009), $\delta$ is assumed to be much smaller than the thickness of the cloud. For this reason we adopt that the upper limit of $\delta$ must be about half the effective thickness of the cloud $\Delta'$.} (Table~\ref{tADFH09}), and the $a'_2$ free parameter was allowed to vary from $-120$ to 120 pc$^{-2}$ in steps of 1 pc$^{-2}$. The fitted parameters were determined by minimizing $\chi^2$.

\begin{deluxetable}{lCCC CC}[h]
%\tablenum{1}
\tablecaption{Magnetic field properties derived using the ADF with the Houde et al. (2009) approach `ADF H09'\label{tADFH09}}
\tablewidth{0pt}
\tablehead{
\colhead{} & 
\colhead{$\delta$\supa} & 
\colhead{} & 
\colhead{$a'_2$\supa} & 
\colhead{$\Delta'$\supa} & 
\colhead{} \\
%%%%%%%%%%%%%%%%%%%%%%%%%
\colhead{Source} & 
\colhead{(mpc)} & 
\colhead{$\frac{b}{N}$\supa} & 
\colhead{(pc$^{-2}$)} & 
\colhead{(mpc)} & 
\colhead{$N$\supa} 
}
\colnumbers
\startdata
1-W3IRS5        		&31\pm1	&1.03\pm0.69	&-77\pm1		&85\pm24		&1.3\pm0.1\\	
2-W3H2O			&14\pm1	&0.56\pm0.38	&10\pm1		&47\pm14		&1.8\pm0.2\\	
4-N6334V			&28\pm1	&0.59\pm0.47	&7\pm1		&57\pm29		&1.1\pm0.1\\	
5-N6334A			&31\pm1	&0.14\pm0.03	&25\pm1		&63\pm32		&1.1\pm0.1\\	
6-N6334I			&24\pm1	&0.05\pm0.01	&7\pm1		&50\pm10		&0.9\pm0.1 \\	
7-N6334In			&12\pm1	&0.03\pm0.01	&13\pm1		&50\pm15		&2.6\pm0.3 \\	
10-G35			&35\pm1	&0.31\pm0.10	&36\pm1		&85\pm16		&1.0\pm0.1 \\	
13-W75N			&15\pm1	&0.59\pm0.25	&-38\pm1		&81\pm20		&3.5\pm0.4 \\	
14-DR21OH		&17\pm1	&0.69\pm0.16	&-100\pm1	&68\pm10		&1.8\pm0.2 \\	
15-N48			&15\pm1	&0.59\pm0.10	&-37\pm1		&102\pm21	&4.6\pm0.5 \\	
18-N7538S		&32\pm1	&0.54\pm0.63	&1\pm1		&64\pm23		&0.9\pm0.1 \\	
\enddata
\tablecomments{
$^\mathrm{a}$ $\delta$, $b/N$, and $a'_2$ are the three free parameters of the `ADF H09' approach (Section~\ref{saBADFH09}). $\delta$ is the magnetic field turbulent correlation length, $b/N$ is the value of the correlated component at the origin, and $a'_2$ is the coefficient of the uniform parabolic approximation adopted in equation~\ref{eqADFH09}. $\Delta'$ is the effective thickness of the cloud, estimated from the size of the SMA continuum emission obtained using all configurations. $N$ is the number of turbulent cells along the line of sight. See Section~\ref{saBADFH09} for further details on each parameter.
%NOTE: b/N = f_NC(0)
The uncertainty in $\delta$, and $a'_2$ are taken equal to the step of the explored parameter space, the uncertainty in $\Delta'$ is taken equal to one beam, and the uncertainty in $N$ is assumed to be 10\%. The uncertainty in $b/N$ is estimated from Monte Carlo simulations performed to take into account the sparse sampling in our observations (Appendix~\ref{apMonteCarlo}).
}
\end{deluxetable}

\renewcommand{\thefigure}{8}
\begin{figure*}[ht]
\begin{center}
\begin{tabular}[b]{c}
    \epsfig{file= 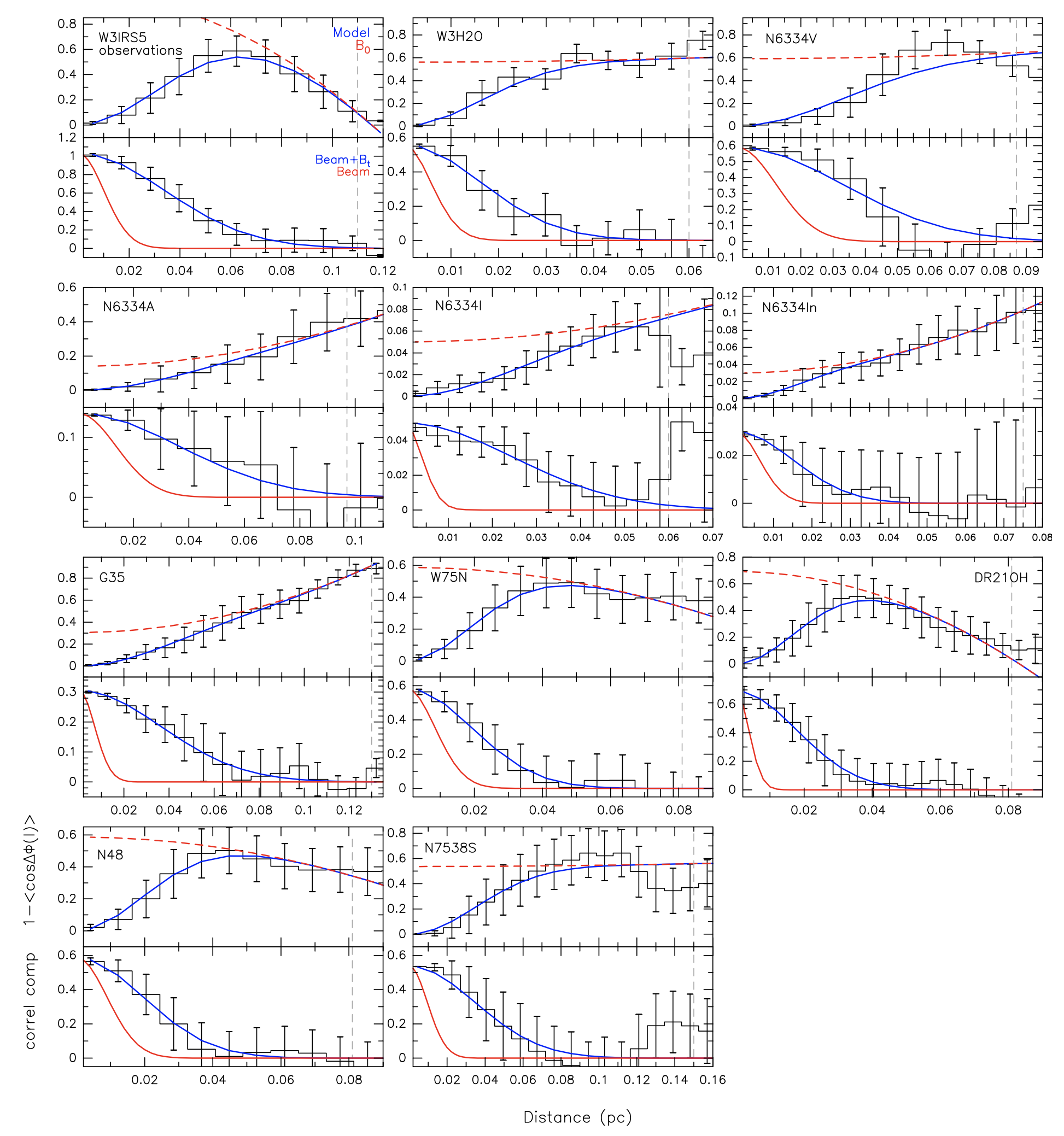, width=15cm, angle=0}\\
\end{tabular}
\caption{Results of the `ADF H09' approach (Section~\ref{saBADFH09}) for each region with enough polarization detections. For each region, the top panel corresponds to the ($1-\langle \mathrm{cos}[\Delta \Phi(l)] \rangle$) function, and the bottom panel corresponds to the correlated component (exponential term of equation~\ref{eqADFH09}). In both upper and lower panels, the black solid line and error bars correspond to the mean and the standard deviation of all the pairs in each bin. The red dashed line corresponds to the large-scale uniform magnetic field (\ie\ it does not contain the correlated component of the function and is $b/N + a'_2\,l^2$). In the upper panel, the blue line shows the fit to the data using equation~\ref{eqADFH09}, and in the bottom panel the blue line shows the correlation due to the beam and the turbulent component of the magnetic field, while the solid red line corresponds to the correlation due to the beam alone. The dashed grey vertical line indicates the largest angular scale that the SMA is able to recover for the polarized emission, below which the fit was performed.
}
\label{fADFH09}
\end{center}
\end{figure*}

Table~\ref{tADFH09} lists the three fitted parameters, $\delta$, $b/N$, and $a'_2$, along with the adopted value for $\Delta'$ (estimated from the SMA continuum images combining all available configurations), and the corresponding values for $N$. $\langle B_\mathrm{t}^2\rangle/\langle B_\mathrm{0}^2\rangle$ and $B_\mathrm{ADFH09} \equiv \langle B_\mathrm{0}^2\rangle^{1/2}$ are given in Table~\ref{tB} to ease the comparison with the other methods/approaches used in this work. $B_\mathrm{ADFH09}$ has been calculated following equation (57) of Houde et al. (2009):

\begin{equation}\label{eqBADFH09}
B_\mathrm{ADFH09} \equiv \langle B_\mathrm{0}^2\rangle^{1/2} = \sqrt{4\pi\rho}\,\frac{\sigma_\mathrm{turb}}{\sqrt{\langle B_\mathrm{t}^2\rangle/\langle B_\mathrm{0}^2\rangle}},
\end{equation}

where $\langle B_\mathrm{t}^2\rangle/\langle B_\mathrm{0}^2\rangle = b$. $\sigma_\mathrm{turb}$ has been obtained from equation~\ref{eqsigmavel} adopting $Q\sim0.5$ (Section~\ref{saveldisp}).
We refer to this approach as the `ADF H09' approach.
%
%Fig.~\ref{fNmm-sigmaPA} presents a plot of \Nmm\ vs $\langle B_\mathrm{t}^2\rangle/\langle B_\mathrm{0}^2\rangle$. 
%In this plot, a typical uncertainty of 0.02--0.05 has been adopted for the free parameter $a=b/(sqrt(2pi)N)$.
The uncertainty of the free parameter $b/N$ controling $\langle B_\mathrm{t}^2\rangle/\langle B_\mathrm{0}^2\rangle$ has been estimated from Monte Carlo simulations as described in Appendix~\ref{apMonteCarlo} (where we followed Liu et al. 2019).

%Fitting equation~\ref{eqBADFH09} naturally yields the $b/N$ parameter, directly related to $\langle B_\mathrm{t}^2\rangle/\langle B_\mathrm{0}^2\rangle$, the ratio between the turbulent and ordered magnetic fields. This method should allow to separate a turbulent component of the magnetic field, associated with the small scales, from a large-scale ordered magnetic field. However, it relies on the assumption that the large-scale magnetic field follows the parabolic approximation. 

A discussion of how these values compare to the previous methods/approaches is presented below.

\subsection{A comparison between the DCF and ADF methods}\label{sacomp}

In Sections~\ref{saBstdev}, \ref{saBgauss}, \ref{saBADFbeam}, and \ref{saBADFH09}, $\delta B/B_0$ was estimated using two different methods (DCF, ADF) with two approaches for each method.
We then calculated the magnetic field strength following equations~\ref{eqBstdev}, \ref{eqBgauss}, \ref{eqBADFbeam}, and \ref{eqBADFH09}, and using the density and velocity dispersion inferred in Sections~\ref{sadensity} and \ref{saveldisp} (equation~\ref{eqsigmavel}), with all quantities averaged within 0.15 pc of diameter. 
It is important to emphasize that, while the fragmentation level is measured in SMA/PdBI/NOEMA images obtained using only extended configurations, thus filtering out typically angular scales larger than $\sim3''$ (Table~\ref{tsample}), the polarized and \htcop\ emissions are obtained from images including also subcompact and/or compact SMA configurations, thus filtering out much larger scales, of 14$''$--$30''$. The density is obtained from the modeling of the radial intensity profiles and SEDs obtained from single-dish data with angular resolutions $\gtrsim11''$. Thus, the magnetic field strength is calculated using data sensitive to the core scale, and averaged within this same scale (of 0.15~pc of diameter), while the fragmentation level is obtained using data sensitive {\it only} to much smaller scales ($<0.03$~pc).
%, as the PA dispersion was obtained from polarization data only within the 0.15~pc region of diameter, the velocity dispersion was obtained from a spectrum averaged within this same region, and the density was averaged also within this same region.

\renewcommand{\thefigure}{9}
\begin{figure*}[]
\begin{center}
\begin{tabular}[b]{c}
    \epsfig{file=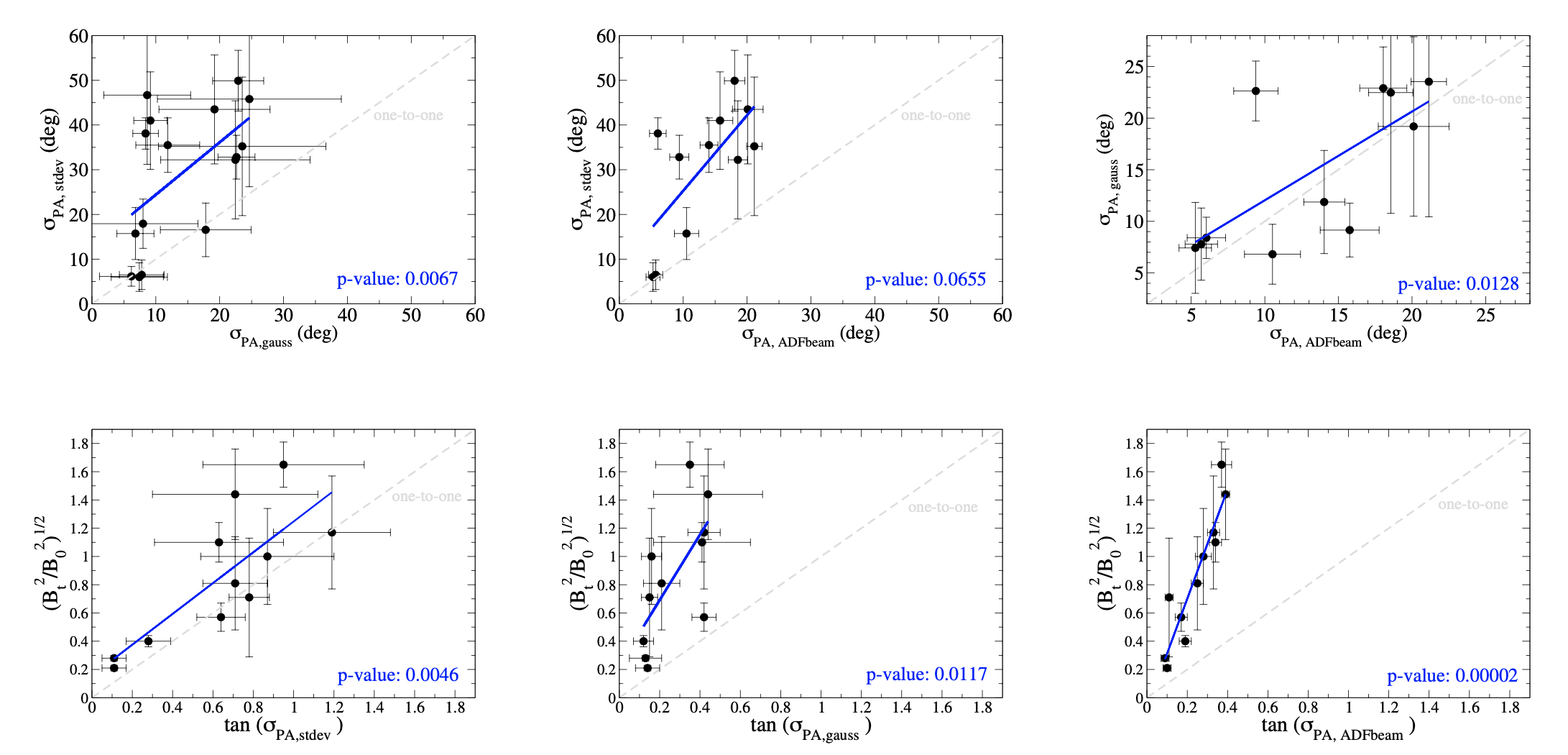, width=15cm, angle=0}\\
\end{tabular}
\caption{Comparison of the four different approaches used here to estimate $\delta B/B_0$ (Sections~\ref{saBstdev}, \ref{saBgauss}, {\bf \ref{saBADFbeam}}, and \ref{saBADFH09}), with the results of linear regression fits (blue line) and the p-values (probability that the null hypothesis is true) listed in each panel. The grey dashed line indicates the one-to-one relation.
%Top-left: $\sigma_\mathrm{PA,stdev}$ vs $\sigma_\mathrm{PA,gauss}$.
%Top-middle: $\sigma_\mathrm{PA,stdev}$ vs $\sigma_\mathrm{PA,ADFbeam}$.
%Top-right: $\sigma_\mathrm{PA,gauss}$ vs $\sigma_\mathrm{PA,ADFbeam}$.
%Bottom-left: $(\langle B_\mathrm{t}^2\rangle/\langle B_\mathrm{0}^2\rangle)^{1/2}$ vs $\mathrm{tan}(\sigma_\mathrm{PAstdev})$.
%Bottom-middle: $(\langle B_\mathrm{t}^2\rangle/\langle B_\mathrm{0}^2\rangle)^{1/2}$ vs $\mathrm{tan}(\sigma_\mathrm{PA,gauss})$.
%Bottom-right: $(\langle B_\mathrm{t}^2\rangle/\langle B_\mathrm{0}^2\rangle)^{1/2}$ vs $\mathrm{tan}(\sigma_\mathrm{PA,ADFbeam})$.
}
\label{fcomparison}
\end{center}
\end{figure*}

Now we would like to compare the results of the four different approaches followed to estimate $\delta B/B_0$ in equation~\ref{eqBpos}.
The `ADF H09' approach should be more robust than the `standard deviation' and the `multiple gaussian' approaches because it decomposes the total magnetic field into the perturbed and the uniform parts taking into account the effects of the beam smoothing and the average along the line of sight (although it also has large uncertainties associated, as shown by Liu et al. 2019). The `ADF beam' approach should also provide a very good estimate of the PA dispersion at small scales (turbulent component of the magnetic field) because it requires no assumptions with respect to the large-scale field. However, both the `ADF beam' and the `ADF H09' approaches could be applied only to 11 regions of our sample, while the other two approaches (`standard deviation' and `multiple gaussian') could be applied to 16 regions. If  $\sigma_\mathrm{PA,stdev}$ or $\sigma_\mathrm{PA,gauss}$ were shown to correlate to $\sigma_\mathrm{PA,ADFbeam}$ or $\langle B_\mathrm{t}^2\rangle/\langle B_\mathrm{0}^2\rangle$, this would suggest that such a determination of $\sigma_\mathrm{PA}$ is a reasonable approach.

In Fig.~\ref{fcomparison} we present plots comparing the four different approaches used in this work to estimate $\delta\,B/B_0$ ($\sigma_\mathrm{PA,stdev}$, $\sigma_\mathrm{PA,gauss}$, $\langle B_\mathrm{t}^2\rangle/\langle B_\mathrm{0}^2\rangle$, and $\sigma_\mathrm{PA,ADFbeam}$). In each panel of the figure the p-value is listed (probability that the null hypothesis is true, \ie\ that the correlation is due to a random process). As can be seen from the figure, the p-values are in all cases $\lesssim0.01$ (with the exception of the $\sigma_\mathrm{PA,stdev}$ vs $\sigma_\mathrm{PA,ADFbeam}$ plot).  This comparison reveals that: 
i) $\sigma_\mathrm{PA,ADFbeam}$ is very well correlated with $\langle B_\mathrm{t}^2\rangle/\langle B_\mathrm{0}^2\rangle$ of the `ADF H09' approach. This was expected because the Houde et al. (2009) method is precisely aimed at separating the PA dispersion at the small scales from the large-scale ordered field. However, 
$\langle B_\mathrm{t}^2\rangle/\langle B_\mathrm{0}^2\rangle$ is systematically deviating more for increasing $\sigma_\mathrm{PA,ADFbeam}$.
ii) $\sigma_\mathrm{PA,gauss}$ correlates quite well with $\sigma_\mathrm{PA,ADFbeam}$, and actually their relation is very close to the one-to-one relation. This indicates that the `multiple gaussian' approach is achieving a reasonable estimate of the PA dispersion at the smallest scales.
iii) while all the approaches correlate to each other, the tighter correlation is found between $\sigma_\mathrm{PA,stdev}$ and $\langle B_\mathrm{t}^2\rangle/\langle B_\mathrm{0}^2\rangle$, with a p-value of 0.0046.  $\sigma_\mathrm{PA,stdev}$  might be better related to $\langle B_\mathrm{t}^2\rangle/\langle B_\mathrm{0}^2\rangle$ than  $\sigma_\mathrm{PA,gauss}$ because of the uncertainty in the decision of how many gaussians should be fitted to the PA histograms to finally obtain $\sigma_\mathrm{PA,gauss}$.
iv) the relation between $\sigma_\mathrm{PA,stdev}$ and $\langle B_\mathrm{t}^2\rangle/\langle B_\mathrm{0}^2\rangle$ is very close to the one-to-one relation, meaning that the $\langle B_\mathrm{t}^2\rangle/\langle B_\mathrm{0}^2\rangle$ is probably including `intermediate-scale' dispersion due to maybe deviations from the parabolic approximation.

This comparison suggests that the determination of  $\sigma_\mathrm{PA,stdev}$ is a reasonable good approach to $\sigma_\mathrm{PA}$, because it presents the lowest p-value and best correlation with the results of the ADF method. In the following, we will consider the value of $\sigma_\mathrm{PA,stdev}$ as the reference value for  $\delta B/B_0$.
%There is also a relation between $B_2$ and $B_\mathrm{ADF}$, if we don't take into account the outlier DR21OH.
A consistency check was performed by plotting $B_\mathrm{stdev}$ vs $n_\mathrm{0.15pc}$ in Fig.~\ref{fBstdevn} of Appendix~\ref{apBstdevn}. The figure shows that there is a relation between these two quantities, as expected from equation~\ref{eqBpos}.
Diagrams of the fragmentation level \Nmm\ vs $B_\mathrm{pos}$ for the four approaches used here are presented in Fig.~\ref{fNmmBmu}.

\subsection{The mass-to-magnetic flux ratio $\mu$}\label{samu}

The ratio of the observed mass-to- magnetic flux over the critical mass-to- magnetic flux, $\mu$, was calculated by following equation (1) of Crutcher et al. (2004), which reads:

\begin{equation}\label{eqmu}
\mu \equiv \frac{(M/\mathrm{\Phi_\mathrm{B}})_\mathrm{observed}}{(M/\mathrm{\Phi_\mathrm{B}})_\mathrm{crit}} = \frac{m\,N(H_2)A/B_\mathrm{tot}A}{1/2\pi G^{1/2}},
\end{equation}

where $A$ refers to the area over which the mass $M$ and the magnetic flux $\Phi_\mathrm{B}$ are measured, $m=2.8m_\mathrm{H}$, with $m_\mathrm{H}$ the mass of the Hydrogen atom, and $B_\mathrm{tot}$ is the total (deprojected) magnetic field strength. In practical units, and writing the equation in terms of the magnetic field strength in the plane of the sky, for which its statistical average value is $B_\mathrm{pos}=\frac{\pi}{4}\,B_\mathrm{tot}$ (Crutcher et al. 2004), equation~\ref{eqmu} reads as:

\begin{equation}\label{eqmupracunits}
%7.6*pi/4=5.969026042
\mu = 5.969\times10^{-24} \frac{N(\mathrm{H}_2)}{B_\mathrm{pos}},
\end{equation}

with $B_\mathrm{pos}$ given in mG. $N(\mathrm{H}_2)$ was calculated as $M_\mathrm{0.15pc}/\pi\,R^2 = n_\mathrm{0.15pc}\,R\,4/3$ for $R=0.15/2=0.075$~pc.

The results are given in Table~\ref{tB} and the right-hand panels of Fig.~\ref{fNmmBmu} show \Nmm\ vs $\mu$ for the four approaches used here.
The figure reveals no apparent relation between \Nmm\ and $B_\mathrm{pos}$, or \Nmm\ and $\mu$, for any of the methods/approaches used here.

In Fig.~\ref{fNmmBmu}, the cores classified as presenting `aligned fragmentation' (Table~\ref{tfrag}) are marked with blue squares. In general, these cores with `aligned fragmentation'  present high fragmentation levels and in most cases (4 out of 6) the magnetic field follows the direction perpendicular to the filamentary structure being fragmented. These cores span a wide range of magnetic field strengths and $\mu$.

%NOTE: TURBULENCE VS B STRENGTH IN ORION KL ARE AROUND... SEE PAPER Koch, Tang, Ho 1008.0220  AND Tang, Ho, Koch, Rao arXiv:1006.2957
% Machida et al. 2005: estimate of Bcrit for I22198 and A5142
%This suggests that the magnetic field  in I22198 should dominate over rotation and turbulence. Machida et al. (2005) gave a critical value of rotational energy vs magnetic field to allow fragmentation. We made a rough estimate, following Machida \et\ (2005) of critical B and found 0.4--66~mG for I22198 and 0.8--7 mG for A5142.  

\renewcommand{\thefigure}{10}
\begin{figure*}[h]
\begin{center}
\begin{tabular}[b]{c}
    \epsfig{file=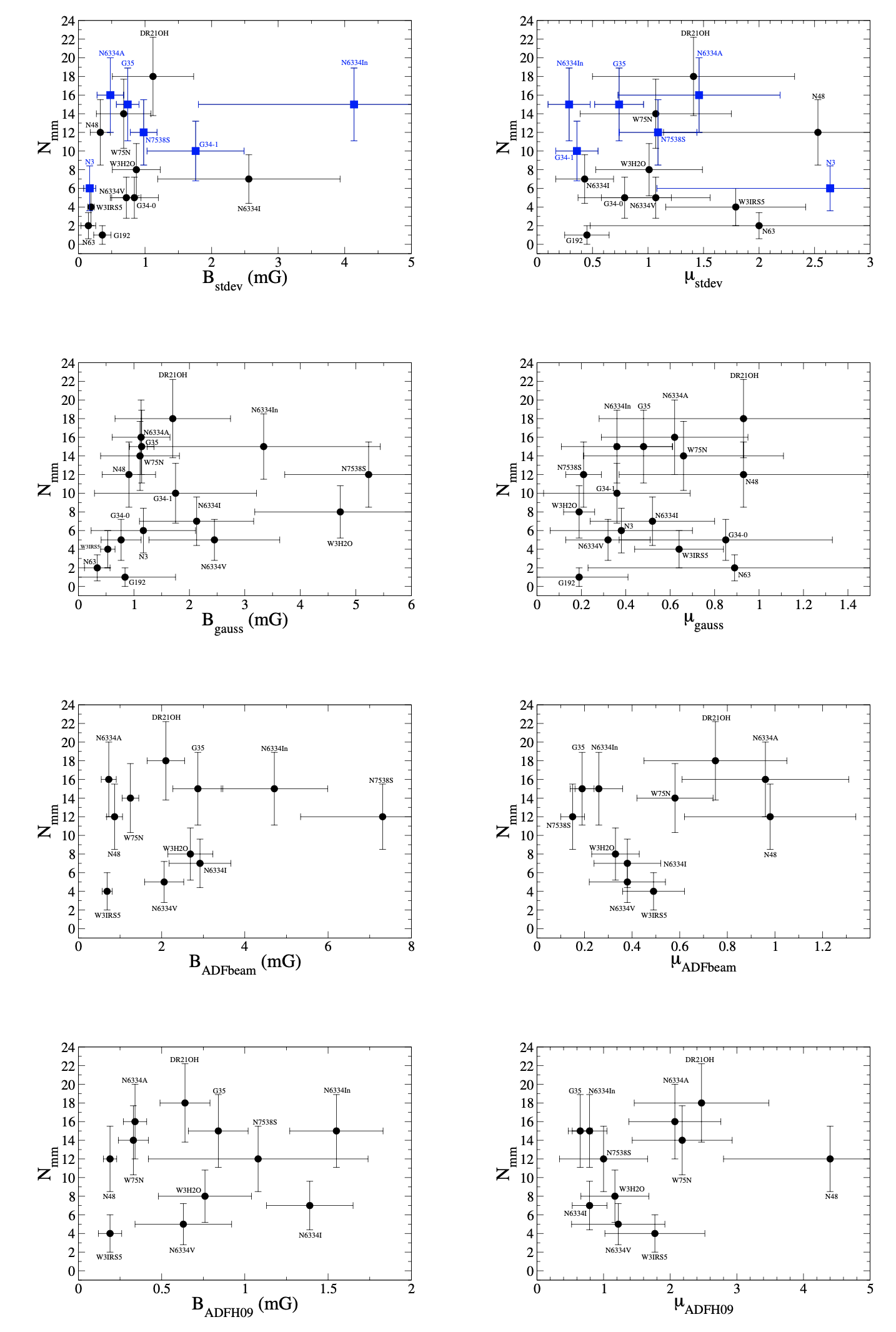, width=13cm, angle=0}\\
\end{tabular}
\caption{The four panels on the left show the plots of \Nmm\ vs the magnetic field strength for the different approaches used in this work: `standard deviation' ($B_\mathrm{stdev}$), `multiple gaussian' ($B_\mathrm{gauss}$), `ADF beam' ($B_\mathrm{ADFbeam}$), and `ADF H09' ($B_\mathrm{ADFH09}$).   
The four panels on the right show the plots of \Nmm\ vs the ratio of mass-to-flux to critical mass-to-flux, $\mu$, for the same four approaches.
In the top panels, the blue squares correspond to the cores classified as presenting `aligned fragmentation' in Table~\ref{tfrag}.
}
\label{fNmmBmu}
\end{center}
\end{figure*}

\section{Discussion}

\subsection{Uncertainties in the determination of the magnetic field strength and mass-to-flux ratio $\mu$}\label{sdunc}

In previous sections we estimated the magnetic field strength in the plane of sky following the DCF and ADF methods, and searched for a possible trend of this quantity with the fragmentation level. We found no clear trend between these two quantities. Before discussing the physical implications of this result, we should consider how robust our determination of the magnetic field strength is. 
In Fig.~\ref{fNmmBmu} we plotted the magnetic field strength and $\mu$ with the corresponding uncertainties after taking into account the uncertainties in the density (Section~\ref{sadensity}), velocity dispersion (Section~\ref{saveldisp}) and polarization PA dispersion (Appendix~\ref{apMonteCarlo}). 
%Only three regions with large relative errors in the PA dispersions (N6334I, N6334In and G34-1) present large uncertainties, of $\sim2$~mG, while the other regions present smaller uncertainties in the range 0.1--0.4~mG.

\subsubsection{Uncertainty in the density}

The estimate of the density uncertainty was described in Section~\ref{sadensity} and was obtained by increasing the $\chi^2$ in our model. However, an additional uncertainty could come from the different spatial filtering of single-dishes (used to assess density) vs interferometers (used to assess velocity and PA dispersions). To assess how much the different spatial filtering of each telescope could affect our determination of the magnetic field strength, the average density was estimated using the SMA continuum flux densities, including all the configurations available as for the case of the polarization and \htcop\ data. For each region, we estimated the total mass recovered by the SMA (Appendix~\ref{apsma}) and found that the amount of mass filtered out by the SMA is on average only $\sim25$\% of the mass inferred from the modeling of the single-dish data presented in Section~\ref{sadensity}. 
In Fig.~\ref{fnsma} of Appendix~\ref{apsma}, we present a plot of the SMA average density vs single-dish average density, showing a relation close to 1-1, with the SMA densities only slightly below the single-dish densities. 
%THIS PLOT SHOULD BE UPDATED IF YOU WANT TO SHOW IT: In addition, a plot presenting the fragmentation level vs the magnetic field strength inferred using the SMA average density is also shown in the Appendix. 
Therefore, the slightly different spatial filtering between the single-dish and the SMA is not heavily affecting our results. 
The figure also indicates that the deviation from the spherical assumption of our model should not strongly affect our results either.
The advantage of using single-dish telescopes to infer the density structure is that the temperature structure can be better determined thanks to the simultaneous fitting of the SED and the radial intensity profiles.
%, a requirement to properly determine the density structure. On the other hand, determining the temperature structure using interferometric data would be by far beyond the scope of this paper.

\subsubsection{Uncertainty in the velocity dispersion}

Regarding the estimate of the velocity dispersion, the \htcop\,(4--3) transition was used. This is a good tracer of dense regions, which should correlate well with the polarized emission, as shown in Figs.~\ref{fm1a} and \ref{fm1b} of Appendix~\ref{apm1}. However, both the velocity dispersion and the PA dispersion could still be affected by the presence of outflows. As shown in Figs.~\ref{fm1a} and \ref{fm1b}, the magnetic field is, for most of the cases, perpendicular to the outflows directions, with the only clear exception of N6334In and N7538S. Other cases where the polarized emission seems to follow the outflow directions are DR21OH and N48. In none of the regions there is evidence of the magnetic field segments being especially perturbed along the outflow directions. Thus, it is unlikely that the velocity and PA dispersions are strongly affected by outflows. 
It should also be noticed that both the velocity dispersion and the PA dispersion could be affected by large-scale systematic motions such as gas inflows. We discuss this possible effect in Section~\ref{sdVDF} and in Appendix~\ref{apsimulations}.

\subsubsection{Uncertainty in the PA dispersion}

Here we list the main contributions to the PA dispersion uncertainty.

\paragraph{Sparse sampling of the data/poor sensitivity} In many cases the polarized emission is detected only in certain portions of the entire continuum emission, preventing us from fully sampling it. This could be due to, for example, a lack of sensitivity. A poor sensitivity would hinder the detection of the polarized emission from low-density gas, which would probably add to the PA dispersion because the turbulent power should be larger in larger scales (Heitsch et al. 2001). This typically tends to overestimate the magnetic field strength. As already mentioned above, our adopted uncertainties take into account the sparse sampling effect (Appendix~\ref{apMonteCarlo}). 

\paragraph{Beam smoothing, average along the line of sight} Both effects imply an overestimation of the magnetic field strength because they tend to blur out the PA dispersion (Heitsch et al. 2001).
%this should most strongly affect the two regions with larger beam in our observations (N6334V and N6334A) and these are not precisely the regions with largest magnetic field strengths.
%QZ: This does not prove a lack of the effect of beam smoothing. Magnetic field strengths could vary from region to region. And we don't know the intrinsic value of the magnetic field strength in NGC 6334V and 6334A. They could still be over estimated.
In our case, both effects are taken into account in the `ADF H09' approach and it was shown in Section~\ref{sacomp} that $\sigma_\mathrm{PA,stdev}$ correlates very well with $\langle B_\mathrm{t}^2\rangle/\langle B_\mathrm{0}^2\rangle$. Thus, it does not seem likely that our inferred values of the magnetic field strength are strongly affected by this.
 
\paragraph{Small-angle approximation} Given that some of our PA dispersions are large, the correction by Falceta-Gonçalves et al. (2008) was applied (Section~\ref{saB}), but there other alternatives in the literature such as the one proposed by Heitsch et al. (2001). We applied equation (12) of Heitsch et al. (2001) for the `standard deviation' case and in general the values of the B-field strength are smaller by ~10\%, except in a few cases, implying an average magnetic field with the Heitsch+01 equation which is about 40\% smaller than the average value with Falceta-Gonçalves equation. Other corrections have been proposed (Hildebrand et al. 2009; Houde et al. 2009, 2011; Franco et al. 2010; Koch et al. 2010), but they typically  imply a factor well below 4 (see Cort\'es et al. 2016, 2019 for a comparison among the values obtained using the different corrections).

\paragraph{Very strong ordered magnetic fields} The superposition of $\delta\,B$ with a strong and uniform large-scale field, could produce an underestimation of the small-scale turbulent dispersion, because the weight of almost no large-scale dispersion will effectively reduce the small-scale turbulent dispersion. Such extremely ordered configurations are not typical in our sample.

\paragraph{Very weak magnetic fields} In the `ADF H09' approach, $\langle B_\mathrm{t}^2\rangle/\langle B_\mathrm{0}^2\rangle$ could be underestimated for the cases of very weak magnetic fields. In these cases the magnetic field could be so strongly perturbed that it could resemble a random field with no important changes with distance, implying an overestimation of the large-scale ordered field. However, such an extreme disordered and random-like magnetic field is not seen in our observations (Figs.~\ref{fpol1} and \ref{fpol2}).

\subsubsection{Intrinsic uncertainty of the methods applied}\label{ssduncintrinsic}

In addition to all the specific uncertainties mentioned above, a number of caveats have been raised in the literature regarding the use of equation~\ref{eqBpos}, summarized below. 

\paragraph{Additional MHD modes to Alfv\'en modes} A recent work by Skalidis \& Tassis (2021) suggests a new equation to estimate the magnetic field strength to take into account not only the Alfv\'en modes but also other additional magnetosonic compressive modes which must be present in molecular clouds. This requires a modification of equation~\ref{eqBpos} to $B_\mathrm{pos} = f\,\sqrt{2\pi\rho}\,\frac{\sigma_\mathrm{turb}}{\sqrt{\sigma_\mathrm{PA}}}$. We applied this new equation using $\sigma_\mathrm{PA,gauss}$ (the dispersion from the approach yielding largest magnetic field strengths) and the average magnetic field in our sample decreased by a factor of 3 (from 1.8 to 0.6 mG) while $\mu$ increased by the same factor. The new values of $\mu_\mathrm{gauss}$ after applying this method are listed in the last column of Table~\ref{tB} for comparison. 

\paragraph{Deviation from equipartition} One of the basic assumptions of equation~\ref{eqBpos} could be violated if the perturbed magnetic field with energy $E_\mathrm{\delta\,B}$ is not in equipartition with the turbulent kinetic energy, $E_\mathrm{turb}$. In this case, the magnetic field should be multiplied by a factor (equal to the root square of $E_\mathrm{\delta\,B}/E_\mathrm{turb}$) ranging from 0.4 to 1 in the simulations of Heitsch et al. (2001) and $\mu$ could increase up to a factor of 2.5. 
There could also be deviations from equipartition if the total kinetic energy of the gas, $E_\mathrm{kin}$, typically assumed to be equal to $E_\mathrm{turb}$, has a non-negligible contribution from systematic motions such as inflow/infall motions, \ie\ the gravitational energy is not negligible. This could easily be the case in our cores because bulk motions of gas flowing towards the center of massive dense cores have been observed (\eg\ Csengeri et al. 2011, Lee et al. 2013, Battisti \& Heyer 2014, Liu et al. 2015, Motte et al. 2018, Schw{\"o}rer et al. 2019). Actually, the morphology of the magnetic field segments in several regions studied here suggests such kind of motions (see, for example, the cases of W3IRS5, N6334V, W75N, DR21OH, N48 and N7538S in Figs.~\ref{fpol1} and \ref{fpol2}). In these cases it would be required to separate not only the turbulent component of the magnetic field out of the large-scale uniform field, but also the systematic motions should be separated from the line width to finally have the true turbulent kinetic energy $E_\mathrm{turb}$. Since in many cases it is assumed that $E_\mathrm{turb}\sim\,E_\mathrm{kin}$,  there is an overestimation of the magnetic field strength and an underestimation of $\mu$. In our case we used the $Q\equiv\frac{\sigma_\mathrm{turb}}{\sigma_\mathrm{nonth}}$ factor to take this into account. However, assuming the same $Q$ factor for all regions might not be correct, and we further discuss this in Section~\ref{sdVDF}. A comparison with simulations for one of our regions which presents converging flows (N6334V, Ju\'arez et al. 2017) is presented in Appendix~\ref{apsimulations}, showing that the `standard deviation' approach yields results comparable to those in the simulations.
%Separating the turbulent component of the PA dispersion from an ordered component is important because the DCF method, as well as equation (8) used to obtain the magnetic field strength from the ADF method, both rely on the assumption that the kinetic energy $E_\mathrm{kin}$ is of the order of the perturbed magnetic field energy $E_{\delta\mathrm{B}}$. This in turn requires that the gravitational energy $E_\mathrm{grav} \ll E_\mathrm{kin} \sim E_{\delta\mathrm{B}}$. In our sample, an estimate of the ratio $E_\mathrm{kin}/E_\mathrm{grav}$ resulted in average values of $\sim 2.5$, but there are regions with the ratio very close to 1 or even smaller, as expected because these are active star-forming regions. Thus, because in our sample the gravitational energy plays a non-negligible role, it is mandatory to properly separate the turbulent component of the PA dispersion from the ordered component so that we can still assume that  $E_\mathrm{kin} \sim E_{\delta\mathrm{B}}$.

\paragraph{Averaged quantities} our measured magnetic field strengths and $\mu$ are {\it averaged} values: within the studied area, densities can change by orders of magnitudes and the magnetic field also scales with density to some power. However, while the density structure is much better resolved and the mass can be more accurately estimated, the DCF and ADF methods give statistical average values for the magnetic field strength for which uncertainties are not easy to quantify. Actually, in a recent paper by A\~nez-L\'opez et al. (2020a) it was found $\mu < 1$ in a star-forming massive core. But for the same core the technique of Koch et al. (2012a) was applied to locally assess the force ratio between the magnetic field and gravity, revealing specific portions within the initially studied area which clearly were supercritical (while the average $\mu$ was below 1).

\subsubsection{Comparison to other determinations of the magnetic field strength in the literature}\label{ssduncBlit}

A final way to test how robust is our determination of the magnetic field strength is to compare it to other values reported in the literature, especially when completely independent methods are used, such as the `ion-neutral drift' technique.
In Table~\ref{tB}, the magnetic field strengths obtained in other works in the literature are reported (see table notes of Table~\ref{tB} for a reference to the different methods), and in Appendix~\ref{apBlit} more details are given about the comparison between the values determined here and the values obtained for the same regions in previous works.

Fig.~\ref{fBlit} of Appendix~\ref{apBlit} shows a plot of the magnetic field strength reported in the literature and the strengths derived in this work. The figure reveals a relation quite close to the one-to-one relation. The cases of N6334I, N6334In, DR21OH, and N3 are particularly significant, as the methods used in the literature for these regions are independent to the method used here. This figure indicates that the method used here to infer the magnetic field strength seems to be reasonable.
%OLD: One of the uncertainties could be that we used \htcop\,(4--3) to measure the line widths, and this could be tracing only a portion of the gas (denser and relatively hotter), preventing us from including gas at larger scales which could present higher line widths. For example, Li et al. (2015) measured a CO\,(2--1) line width of 12--14~\kms\ for NGC\,6334I/In at scales comparable to ours ($\sim0.2$~pc), while we measured line widths of 3--6~\kms, a factor $\sim3$ smaller.   

%INCLUDING ALSO STDEV-HEITCH+01 AND GAUSS APPROACHES
%\renewcommand{\thefigure}{5}
%\begin{figure*}[]
%\begin{center}
%\begin{tabular}[b]{cc}
%    \epsfig{file= Nmm_density015pc.pdf, width=5.5cm, angle=0}&
%    \epsfig{file= Nmm_Bstdev_simdens.pdf, width=5.5cm, angle=0}\\
%        \epsfig{file= Nmm_mustdev_simdens.pdf, width=5.5cm, angle=0}&
%        \epsfig{file= Nmm_mustdevSkalidis21_simdens.pdf, width=5.5cm, angle=0}\\
%        \epsfig{file= Nmm_mustdevH01_simdens.pdf, width=5.5cm, angle=0}&
%            \epsfig{file= Nmm_mustdevH01Skalidis21_simdens.pdf, width=5.5cm, angle=0}\\
%        \epsfig{file= Nmm_mugauss_simdens.pdf, width=5.5cm, angle=0}&
%            \epsfig{file= Nmm_mugaussSkalidis21_simdens.pdf, width=5.5cm, angle=0}\\
%        \epsfig{file= Nmm_muadf_simdens.pdf, width=5.5cm, angle=0}&
%            \epsfig{file= Nmm_muadfSkalidis21_simdens.pdf, width=5.5cm, angle=0}\\
%\end{tabular}
%\caption{}
%\label{fNmmBsimdens}
%\end{center}
%\end{figure*} 

\renewcommand{\thefigure}{11}
\begin{figure*}[]
\begin{center}
\begin{tabular}[b]{c}
        \epsfig{file=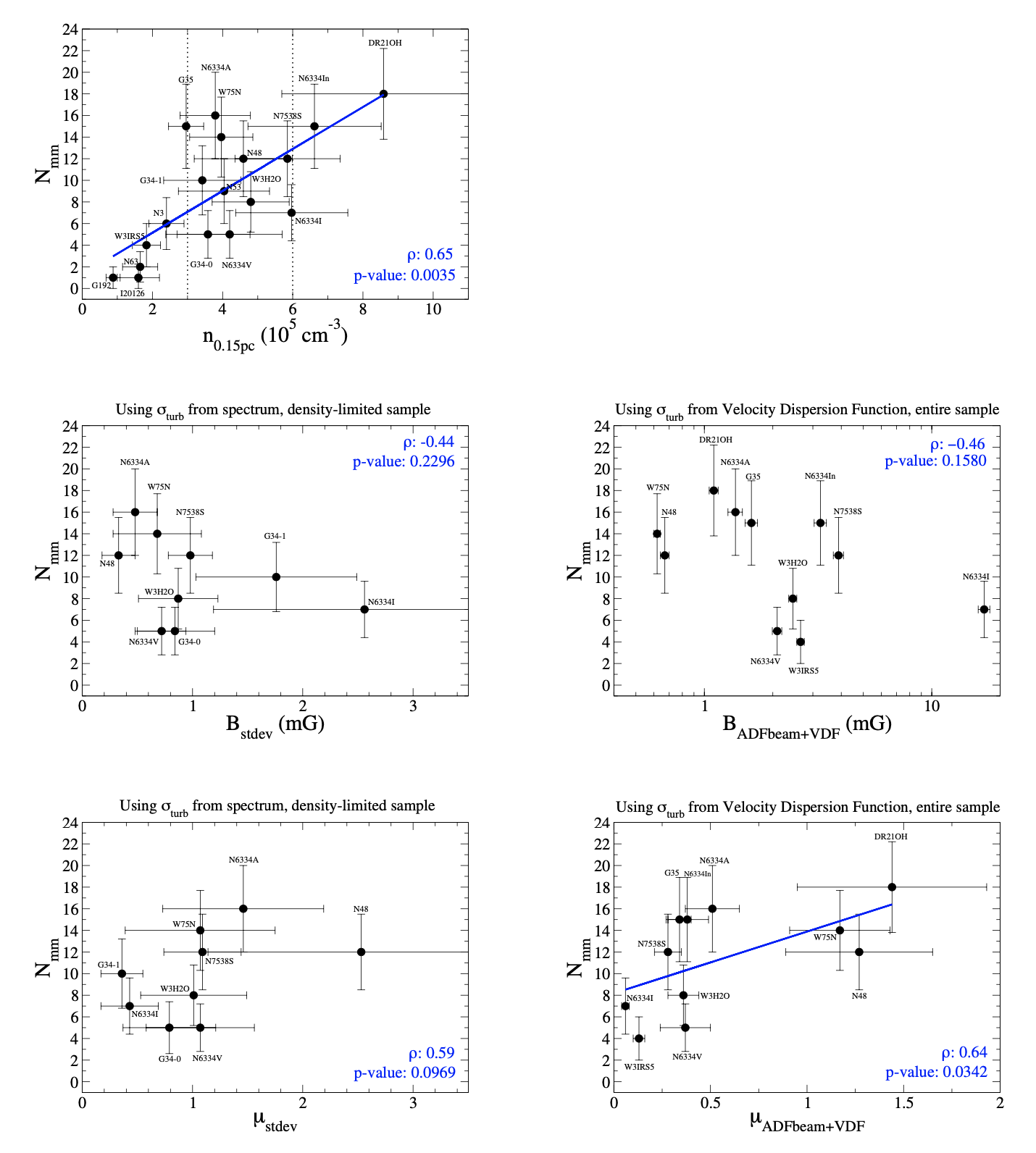, width=11cm, angle=0}\\
\end{tabular}
\caption{
Top: Fragmentation level vs density averaged within the same field of view where fragmentation, line width and PA dispersion were assessed (0.15 pc of diameter). The blue line indicates the result of a linear regression of the form $N_\mathrm{mm} = (1\pm2) + (1.9\pm0.5)\,[\frac{n_\mathrm{0.15pc}}{10^5\,\mathrm{cm}^{-3}}]$, with a correlation coefficient of 0.71.
Middle-left and bottom-left: Fragmentation level vs $B_\mathrm{stdev}$ (middle) and vs $\mu_\mathrm{stdev}$ (bottom) for regions with similar density (in the range (3--6)$\times10^5$~\cmt, marked with vertical dotted lines in the top panel), and using $\sigma_\mathrm{turb}=\sigma_\mathrm{turb,spec}$ calculated from a Gaussian fit to the \htcop\,(4--3) spectrum following equation~\ref{eqsigmavel}.
Middle-right and bottom-rigth: Fragmentation level vs $B_\mathrm{ADFbeam+VDF}$ (middle) and vs $\mu_\mathrm{ADFbeam+VDF}$ (bottom) for all the regions in our sample where ADF was applied, and using $\sigma_\mathrm{turb}=\sigma_\mathrm{turb,VDF}$ calculated from the Velocity Dispersion Function (column (5) of Table~\ref{tkin}), thus separating in velocity the small-scale turbulent motions from the systematic large-scale motions.
The blue line indicates the result of a linear regression of the form $N_\mathrm{mm} = (8\pm2) + (6\pm3)\,\mu_\mathrm{ADFbeam+VDF}$, with a correlation coefficient of 0.58.
In all panels, the Spearman's rank correlation coefficient $\rho$ and the p-value are annotated in the bottom/top right corner.
}
\label{fNmmBsimdens}
\end{center}
\end{figure*} 

%\begin{figure*}
%\gridline{\fig{Nmm_density015pc.pdf}{0.3\textwidth}{(a)}}
%\gridline{\fig{Nmm_Badf_simdens.pdf}{0.3\textwidth}{(b)}          \fig{Nmm_BadfSkalidis21_simdens.pdf}{0.3\textwidth}{(c)}}
%\gridline{\fig{Nmm_muadf_simdens.pdf}{0.3\textwidth}{(d)}          \fig{Nmm_muadfSkalidis21_simdens.pdf}{0.3\textwidth}{(e)}}
%\caption{Inverted pyramid figure of six individual files. \label{fig:pyramid}}
%\end{figure*}

\subsubsection{Uncertainty in $\mu$}

In Table~\ref{tB}, $\mu$ has values very close to 1 or even $<1$, especially for the `multiple gaussians' and `ADF beam' approaches. This is kind of unexpected because all the cores in our sample are known to undergo active star formation and should thus be supercritical.
However, our calculation of $\mu$ is obviously affected by all the aforementioned uncertainties associated with the calculation of the magnetic field strength, which in some cases would imply a factor of 2.5 or even 3 larger $\mu$. In addition, the absolute value of the column density might also be affected by the fact of not taking into account the mass already blocked in stars. The core masses within 0.15 pc of diameter, used to estimate the column density in equation~\ref{eqmupracunits}, range from 20 to 100~\mo\ (Table~\ref{tfit}). If half of this mass is considered to be in stars, this would imply an additional factor of 1.5 larger $\mu$ (see also Section 5.7 in Girart et al. 2013). Therefore, the absolute values of $\mu$ reported in Table~\ref{tB} would be probably shifted to higher values if the aforementioned caveats could be quantified and taken into account. While providing accurate absolute values of $\mu$ is well beyond the scope of this paper, the {\it relative} values of $\mu$ between the regions of our sample should not be that strongly affected and are probably a good measure of this quantity.

Finally, it is worth noting that measuring $\mu$ smaller than 1 should not be necessarily implying that the magnetic field is dominating over gravity because $\mu$ can actually depend on the spatial scale where it is measured (\eg\ Koch et al. 2012b). For example, in G\'omez et al. (2021, see Section 4.3) it is shown that if the density is a power law of the radius `r', and the magnetic field strength follows a power-law of the density with index $2/3$, then $\mu (r) \sim r^{1-p/3}$, where $p$ is the density power-law index. Thus, for the values of $p$ reported in this work, which average to $p\sim1.8$, $\mu (r) \sim r^{0.4}$ and thus $\mu$ should decrease for smaller radii. Such a decrease of $\mu$ for smaller radii has actually been measured by Crutcher et al. (2009), Tang et al. (2019) and Arzoumanian et al. (2021). In addition to this, there is still the fact that the superficial terms in the virial theorem have been neglected in the standard definition of $\mu$, while these could yield up to a factor of 2 smaller critical masses and therefore a factor of 2 larger $\mu$ (Strittmatter 1966).

\subsubsection{Summary of uncertainties}

In spite of all the  caveats mentioned in this section, the crucial aspect of the analysis presented here is that it is performed uniformly for the entire sample, measuring each parameter using exactly the same method and within the same field of view for all the regions. 
For the validity of equation~\ref{eqBpos}, one of the most crucial aspects probably is the separation of the perturbed/turbulent component of the velocity dispersion, $\sigma_\mathrm{turb}$, and of the magnetic field, $\delta B$, from the ordered or large-scale component, $B_0$, and this can be specifically done using the ADF method (any of the two approaches presented in Section~\ref{saBADF}) as it allows to calculate the PA dispersion as function of distance.
The fact that $\sigma_\mathrm{PA,stdev}$ was found to correlate well with $\langle B_\mathrm{t}^2\rangle/\langle B_\mathrm{0}^2\rangle$ (Fig.\ref{fcomparison}) is indicative that this separation was successfully done in terms of {\it relative} variations of $\delta B/B_0$ in our sample.
%Such a correlation is a significant result because the ADF method does not suffer from the biases described in Section~\ref{saBADF} and its robustness has been tested by Liu et al. (2019). These authors find that, although this approach has large uncertainties, it yields no more than a factor of 2 of difference with respect to other approaches (such as the structure function and unsharp masking). 

Given the goal of our work, it is important to use $B_\mathrm{stdev}$ instead of $B_\mathrm{ADF}$ because $B_\mathrm{stdev}$ could be calculated for 16 regions while  $B_\mathrm{ADF}$ was only calculated to 11 regions, and this allowed us to improve the statistics to test the \Nmm\ vs $B_\mathrm{pos}$ or \Nmm\ vs $\mu$ relations.
%%It is worth mentioning that the ADF approach does not suffer from biases related to the presence of different components of the magnetic field. In addition, Liu et al. (2019) compare the magnetic field strength obtained using different approaches (structure function, ADF, and unsharp masking), and find that, although ADF has large uncertainties associated, there is no more than a factor of 2 of difference between the different approaches, indicating that the ADF approach is quite robust. The advantage of using the `standard deviation' approach in this work is that it can be applied to more regions than the ADF approach.
Therefore, far from intending to provide accurate absolute values, our reported values of the magnetic field strength and $\mu$ should be useful to assess the {\it relative} variation of these quantities in this sample.

\subsection{A strong correlation of fragmentation level with density within 0.15~pc}\label{sddensity}

%QZ: We should also discuss limitation in our sample: For example, the targets have a range of masses, and more massive ones should produce more fragments.
In the top panel of Fig.~\ref{fNmmBsimdens} we present a plot of the fragmentation level vs the density averaged within 0.15 pc. As can be seen from this plot, a correlation between the fragmentation level and the averaged density is apparent. Because we determined the density for a fixed size, such a relation with density is equivalent to a relation with mass. A linear regression fit gives a correlation coefficient of 0.71, while the Spearman's rank correlation coefficient $\rho$ is 0.65 and the p-value is 0.0035. The relation is consistent with previous observational studies (\eg\ Gutermuth et al. 2011; Palau et al. 2014; Lee et al. 2015; Liu et al. 2016; Nguyen-Luong et al. 2016; Pokhrel et al. 2016, 2018, 2020; Mercimek et al. 2017; Alfaro \& Rom\'an-Z\'u\~niga 2018; Mendigut\'ia et al. 2018; Murillo et al. 2018; Li et al. 2019; Lin et al. 2019; Orkisz et al. 2019; Sanhueza et al. 2019; Sokol et al. 2019; Svoboda et al. 2019; Zhang et al. 2019) and theoretical/numerical studies (Guszejnov et al. 2018; Burkhart et al. 2018; Dobbs et al. 2019) reporting an important role of density in determining fragmentation of massive dense cores, and is expected for the case of thermal Jeans fragmentation. 

%It is worth noting that such a trend was not expected a priori, as there are many other environmental factors, in addition to density or mass, that could also determine the fragmentation level, such as turbulence level, rotational energy, stellar feedback and/or magnetic field strength (Section~\ref{si}), and it is precisely the aim of this work to try to disentangle which of these factors is dominant. For example, there could exist a very massive core which is not fragmenting because the magnetic field is strong enough. In this sense, one could think that a very massive core should strongly fragment and that it would require a strong magnetic field to prevent this fragmentation. Thus, a relation might be expected of fragmentation level vs mass-to-flux over critical mass-to-flux, $\mu$, to 'normalize' the magnetic field to the mass of the core. However, no hints of a relation between fragmentation level and $\mu$ were found in the right panels of Fig.~\ref{fNmmBmu} for any of the approaches used in this work.

\subsection{Interplay between density and the magnetic field strength}\label{sdinterplay}

The lack of correlation between the fragmentation level and the magnetic field strength or $\mu$ could be due to the fact that our sample is including massive dense cores with a too broad range of densities. In other words, since density and magnetic field could both affect the fragmentation level simultaneously, we consider here whether a relation is found when only regions with very similar densities are considered. Looking at the top panel of Fig.~\ref{fNmmBsimdens}, the range (3--6)$\times10^5$~\cmt\ includes a large number of regions and could be good to perform the aforementioned test. 

In the middle-left panel of Fig.~\ref{fNmmBsimdens} we present a plot of \Nmm\ vs $B_\mathrm{stdev}$,  only for the regions within the narrow density range given above. In the bottom-left panel of the same figure the plot of \Nmm\ vs $\mu_\mathrm{stdev}$ is also shown.
%However, if N6334In is considered as an outlier (\eg\ Sadaghiani et al. 2020), 
While statistically these samples are too small (implying p-values larger than 0.09) and the uncertainties are high, a possible connection between \Nmm\ and the magnetic field strength or $\mu$ cannot be ruled out.
%, and such a connection is consistent with the magnetic field being partially suppressing fragmentation in these regions. Obviously, this requires of much larger and more sensitive samples to definitely confirm or reject the relation. If the relation turns out to be confirmed in future studies, the different magnetic field strengths or $\mu$ in the cores with similar density would explain the considerable scatter in the  \Nmm\ vs $n_\mathrm{0.15pc}$ relation. 
What we find here is in full agreement with a recent work towards the infrared dark cloud G14.225$-0.506$, where the two main hubs of the cloud have very similar densities, and their different fragmentation levels can be explained with the different measured magnetic field strengths (A\~nez-L\'opez et al. 2020b).

\subsection{A tentative trend of fragmentation level with mass-to-flux ratio}\label{sdVDF}

In the measurements of $B_\mathrm{pos}$ and $\mu$ calculated so far, a constant factor $Q\sim0.5$, was adopted. $Q$ was defined in Section~\ref{saveldisp}
as $Q\equiv\frac{\sigma_\mathrm{turb}}{\sigma_\mathrm{nonth}}$. However, in the same Section~\ref{saveldisp} the $Q$ factor was also estimated from the VDFs for the same regions where the ADFs were calculated in Section~\ref{saBADFbeam}. We consider here the fact that $Q$ might actually vary from one region to the other (as shown in Table~\ref{tkin}), and recalculated $B_\mathrm{pos}$ and $\mu$ for the `ADF beam' approach. We chose the `ADF beam' approach because this is the approach for which $\sigma_\mathrm{PA}$ is calculated with exactly the same technique as $\sigma_\mathrm{turb,VDF}$, taking the value of the dispersion function (structure function) at the smallest scales, both in velocity and polarization PA. By doing this, the effects of large-scale motions (\eg\ due to gravity) should be avoided and equation~\ref{eqBpos} (DCF) should be fully valid.

In the right panels of Fig.~\ref{fNmmBsimdens} we present the result for \Nmm\ vs $B_\mathrm{pos}$ and $\mu$ following this technique, and the figure reveals a tentative trend for the case of \Nmm\ vs $\mu$, with a p-value of $\sim0.03$, where \Nmm\ increases for regions with larger $\mu$, as expected from numerical simulations and theoretical work. According to this figure, the magnetic field in our sample seems to play a non-negligible role in the determination of the fragmentation level of massive dense cores. 

If confirmed with new more sensitive observations carried out in larger samples, our work would strongly indicate that, when the DCF method is applied (equation~\ref{eqBpos}), it is crucial to properly separate the large-scale motions from the small-scale turbulent motions in the velocity dispersion, and that the velocity dispersions inferred from averaging spectra could have a non-negligible part due to systematic motions. Thus, the common assumption made in the literature that the non-thermal velocity dispersions are due entirely to turbulence is not probably correct at least in our sample.

\subsection{Comparison of average fragment masses with Jeans and magnetic critical masses}\label{sdjeans}

The fact that the \Nmm\ vs $n_\mathrm{0.15pc}$ relation is stronger than the \Nmm\ vs $B_\mathrm{stdev}$ relation suggests that thermal Jeans fragmentation has a non-negligible role in the fragmentation of our sample. If this is the case, the mass of the fragments should be comparable to the Jeans mass. Using the density averaged within 0.15~pc reported in Table~\ref{tfit}, we calculated the Jeans mass following equation (6) of Palau et al. (2015):

{
\begin{equation}\label{eqMjeans}
\bigg[\frac{M_\mathrm{Jeans}}{M_{\sun}}\bigg] = 
0.6285\,\bigg[\frac{T}{10\,\mathrm{K}}\bigg]^{3/2}
\bigg[\frac{n_\mathrm{H_2}}{10^5\,\mathrm{cm}^{-3}}\bigg]^{-1/2}.
\end{equation}
}

The values of the Jeans mass for each massive dense core, $M_\mathrm{Jeans}$, are reported in Table~\ref{tfrag}. The table gives a range of $M_\mathrm{Jeans}$ corresponding to the range of temperatures assumed, from 20~K (lower limit) to $T_\mathrm{0.15pc}$ (Table~\ref{tfit}, upper limit). As can be seen from this table, $M_\mathrm{Jeans}$ is of the order of 1--5~\mo, very similar to the average mass of the fragments in each core.

In order to asses whether the measured magnetic field in each region is able to prevent the collapse of the detected fragments, the critical masses for magnetic support were calculated following equation (16) of McKee \& Ostriker (2007):
 
\begin{equation}\label{eqMcrit}
M_\mathrm{crit} \equiv c_\mathrm{\Phi}\frac{\Phi_\mathrm{B}}{G^{1/2}},
\end{equation}

where $G$ is the gravitational constant, $\Phi_\mathrm{B}$ is the magnetic field flux threading the core, and $c_\mathrm{\Phi}$ is a numerical coefficient 
%that depends on the internal distribution of density and magnetic fields. 
adopted as $c_\mathrm{\Phi}=1/2\pi$, which corresponds to the value for an infinite cold sheet and is nearly identical to the value for a core with a poloidal field and a constant mass-to-flux ratio (McKee \& Ostriker 2007). The magnetic flux was calculated as $\Phi_\mathrm{B}=\pi R^2\,B_\mathrm{pos}$, yielding to the equation:

\begin{equation}
M_\mathrm{crit} = \frac{R^2\,B_\mathrm{pos}}{2\,G^{1/2}}.
\end{equation}

In this equation, $R$ was taken equal to $\langle R_\mathrm{fragm}\rangle$ (Table~\ref{tfrag}), and $B_\mathrm{pos}$ was estimated by scaling $B_\mathrm{stdev}$ (Table~\ref{tB})  in density to the average density of all the fragments, of $3\times10^7$~\cmt\ (Table~\ref{tfrag}), assuming that $B_\mathrm{frag}=B_\mathrm{stdev}\big[\frac{3\times10^7~\mathrm{cm}^{-3}}{n_\mathrm{0.15pc}}\big]^{0.4}$ (Li et al. 2015).

The resulting values of $M_\mathrm{crit}$ are listed in Table~\ref{tfrag} and are within the range calculated for $M_\mathrm{Jeans}$. Therefore, in general the fragments studied in our sample have enough masses to overcome both the thermal and the magnetic support.

\renewcommand{\thefigure}{12}
\begin{figure}[]
\begin{center}
\begin{tabular}[b]{c}
    \epsfig{file=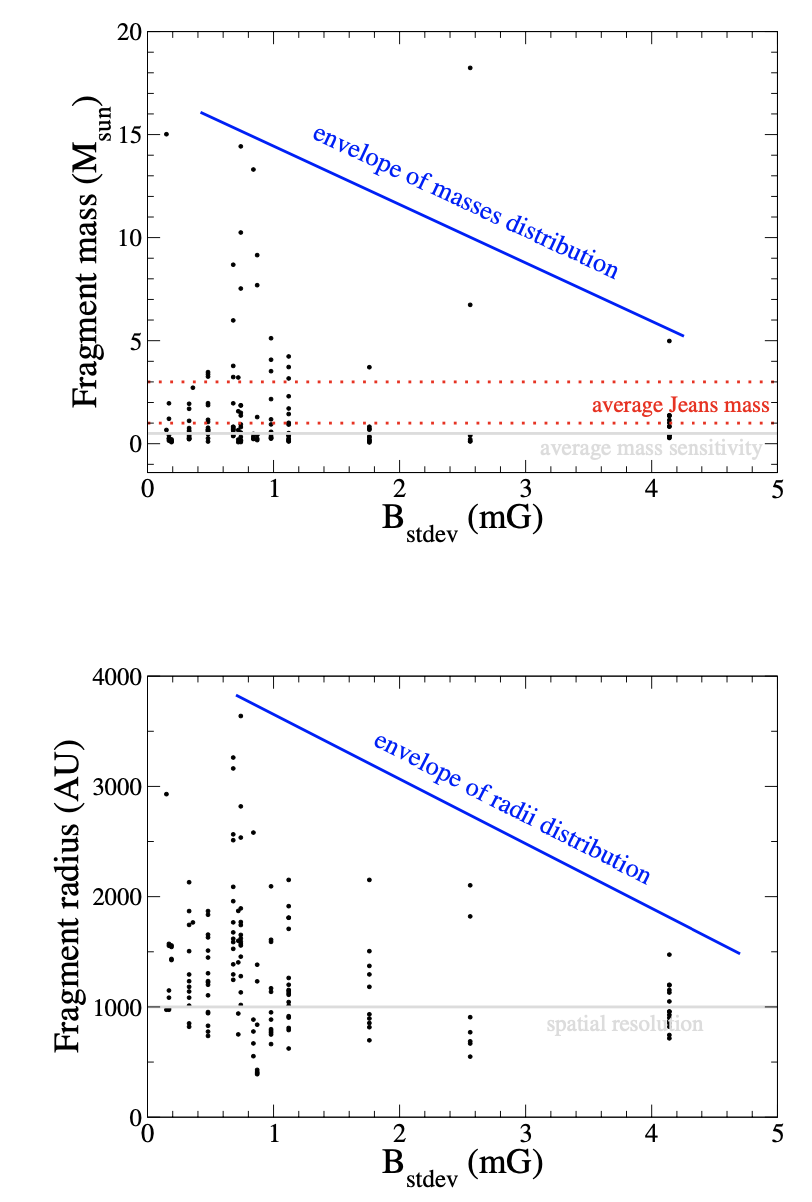, width=7cm, angle=0}\\
\end{tabular}
\caption{
Top: Masses of the fragments identified within each massive dense core vs $B_\mathrm{stdev}$. The horizontal red dotted lines indicate the range of average $M_\mathrm{Jeans}$ found for our sample assuming $T=20$~K and $T=T_\mathrm{0.15pc}$ (Table~\ref{tfrag}). The horizontal grey line indicates the typical mass sensitivity of our observations (Table~\ref{tsample}).
Bottom: Radii (at the 3$\sigma$ contour) of the fragments identified within each massive dense core vs $B_\mathrm{stdev}$. The horizontal grey line indicates the typical spatial resolution of our observations (Table~\ref{tsample}).
Oblique lines are drawn to guide the eye.
}
\label{fradiiB}
\end{center}
\end{figure}

\subsection{Fragment sizes vs magnetic field strength}\label{sdfragsizes}

We explore here whether there is any relation between the sizes or masses of the fragments and the magnetic field strength of their parental core.
Fig.~\ref{fradiiB} presents a plot with the masses (top) and 3$\sigma$ radii (bottom) for all the 160 detected fragments vs the magnetic field strength. Both panels indicate that 
there could be an upper envelope with the more massive/largest fragments tending to occur where the magnetic field is weaker. In addition, the scatter in the masses and sizes of the fragments appears to decrease with growing field strength. For the largest field strength (corresponding to N6334In), the scatter appears to be very small (although statistics are not very large, see the standard deviations of the fragment sizes for each region in Table~\ref{tfrag}). 
Assuming that a fragment under a strong magnetic field is not accreting material from its surroundings as efficiently as a fragment under a weak magnetic field (because the magnetic field should slow the collapse down), the smaller sizes observed for regions with larger magnetic fields could indicate that the magnetic field is preventing these fragments from a fast growing in both mass and size, consistent with theoretical work (\eg\ Hennebelle \& Inutsuka 2019).
%This could be suggesting that a more dominating field strength holds material together more tightly, making more difficult for a fragmentation process to start and accrete more material from its surrounding to grow into a larger fragment.

In addition, {two of the} regions with strongest magnetic field, N6334In, and G34-1, have their fragments aligned along a prevailing direction, and as such they were classified as undergoing `aligned fragmentation' in Table~\ref{tfrag}. This finding is fully consistent with the work of Fontani et al. (2018),  who also report that a filamentary distribution of the fragments is favored for strong magnetic fields. For these cases, the magnetic field morphology is rather uniform and perpendicular to the axis of the aligned fragments. With such a geometry, and with a relatively strong magnetic field strength, the fragmentation process should happen along the field lines with material moving more easily along the field lines from the outer regions towards a mid-plane. This could explain the relatively small scatter in size for such a configuration. Note that G35 also presents `aligned fragmentation', but has the magnetic field along the main axis of the filamentary structure, its magnetic field strength is relatively low, and presents a large scatter in the fragment masses and sizes (Table~\ref{tfrag}), thus also  fitting within this picture. In summary, the magnetic field strength, the small scatter in the sizes of the fragments, and the field morphology all suggest that the magnetic field in these cases is regulating at least partially the fragmentation process.

\subsection{Implications of our results}

As mentioned in the introduction, a number of theoretical and numerical studies suggest that magnetic fields could be crucial to determine the fragmentation level of molecular clumps and cores, because strong magnetic fields should suppress fragmentation. 
In a recent review about the role of magnetic fields in the formation of molecular clouds, Hennebelle \& Inutsuka (2019), present the assumptions leading to ideal MHD equations, taking into account ion-neutral drift. They consider the influence that the magnetic field may have on the interstellar filaments and the molecular clouds, and its role on the formation of stellar clusters.  They argue that the magnetic field could be responsible for reducing the star formation rate and the numbers of clumps, cores and stars. 

In this paper we aimed at testing this from direct observations. 
%For completeness, we plotted the fragmentation level \Nmm\ vs magnetic field strength derived using the ADF method, $B_\mathrm{ADF}$, and \Nmm\ vs $\langle B_\mathrm{t}^2\rangle/\langle B_\mathrm{0}^2\rangle$ (see bottom panels of Fig.~\ref{fcomparison}). Although the number of regions with measured $B_\mathrm{ADF}$ is half of our original sample, and better statistics would be desirable, the panels reveal a possible trend of higher fragmentation level with stronger magnetic field. This is opposite to the trend expected from the simulations, as a stronger magnetic field is predicted to reduce fragmentation. Similarly, there seems to be a possible trend of higher fragmentation level for smaller  $\langle B_\mathrm{t}^2\rangle/\langle B_\mathrm{0}^2\rangle$, but again this implies more fragmentation for higher ordered large-scale magnetic fields or for smaller turbulent magnetic fields, which was not expected either.
%\eg\ Boss 2004; V\'azquez-Semadeni et 2005, 2011; Ziegler et 2005; Banerjee \& Pudritz 2006; Price \& Bate 2007; Hennebelle et al. 2008; Commer\c con et al. 2011; Peters et 2011; Bailey \& Basu 2012; Myers et al. 2013; Boss \& Keiser 2013, 2014; Girichidis et al. 2018; Hennebelle \& Inutsuka 2019
From our uniform analysis of the entire sample of 18 massive dense cores at $\sim0.15$~pc scales, a correlation between fragmentation level and density (within 0.15~pc) is clear from our analysis (Section~\ref{sddensity} and Fig.~\ref{fNmmBsimdens}), 
%and the average mass of the fragments in each region is comparable to the Jeans mass (Table~\ref{tfrag}). These two observational findings 
indicating that the fragmentation process in our sample is mainly dominated by gravity.
This is  consistent with very recent numerical simulations from Krumholz \& Federrath (2019), who find that the magnetic field strength should not strongly affect the star formation rate or initial mass function in star-forming clouds at their earliest stages of formation. 
%(while the magnetic field could affect significantly through its interaction with star formation feedback). 

The lack of a strong correlation of the fragmentation level with the magnetic field strength or $\mu$ could arise from the decoupling (diffusion) of the magnetic field at the scales studied in this work. However, this is difficult to assess because there are little works in the literature reporting signs of diffusion on the scales and densities studied in this work. On one hand, Yen et al. (2018) looked at the ion-neutral drift velocity in the B335 Class 0 protostar which should result from ambipolar diffusion. Although these authors concentrated on much smaller scales (~100 AU) and higher densities than the ones studied in this work, where ambipolar diffusion is supposed to be efficient (Tassis \& Mouschovias 2007), no clear drift velocity was detected, suggesting that the magnetic field is still well coupled to the gas even on these small and dense scales. On the other hand, for the case of DR21OH, Girart et al.  (2013, Section 5.5) suggest that magnetic flux diffusion or dissipation is taking place at the scales studied here via fast magnetic reconnection in the presence of turbulence (Lazarian \& Vishniac 1999; Santos-Lima et al. 2010). Clearly this needs to be further explored.

It is worth mentioning that the magnetic field strengths inferred in this work cover a range of about one order of magnitude (from 0.2 to 4 mG, or $\mu$ from 0.3 to 2.5), and are also subject to a number of uncertainties (see Section~\ref{sdunc}), while simulations showing very different fragmentation levels correspond to setups differing by two orders of magnitude in $\mu$ (\eg\ from 2 to 130, Hennebelle et al. 2011; Commer\c con et al. 2011).
This suggests that the typical magnetic fields in the massive dense cores of our sample do not probably cover a sufficiently large range to leave a clear trace on the fragmentation level and that this is rather determined by other environmental factors such as density.
%Since the massive dense cores of our sample are in very early evolutionary stages, where no strong and significant stellar feedback is at play yet, our main result of no hints of a relation between fragmentation level and magnetic field strength is consistent with the results of Krumholz \& Federrath (2019), at least at the studied scales of $\sim0.15$~pc.

However, finer details of our observational dataset seem to be consistent with the magnetic field affecting the fragmentation process at least partially. First, when large-scale systematic motions are separated from the velocity dispersion and only the small-scale (turbulent) contribution is taken into account, a tentative correlation is found between \Nmm\ and the mass-to-flux ratio, as expected theoretically and numerically. This could explain the significant scatter found in the \Nmm\ vs density relation. Second, regions with strongest magnetic field (N6334In, and G34-1) undergo fragmentation along a preferential direction, which is perpendicular to the magnetic field lines. Third, regions with strong magnetic fields are also the regions with small fragments and with almost all fragments with similar sizes. These three findings suggest that the magnetic field, at least in these cases, is somehow affecting the fragmentation process. It is therefore necessary to test this in a larger sample to strengthen the hints found here.

\section{Conclusions}

We have compiled a sample of 18 massive dense cores for which submillimeter polarization observations from the Legacy Program of the SMA (Zhang et al. 2014), as well as submillimeter continuum images at high angular resolution were available. The sample was built to strictly fulfill constraints of spatial resolution of $\sim1000$~AU and mass sensitivities (from the submillimeter continuum) around $\sim0.5$~\mo, so that a fragmentation level can be measured in a uniform and reliable way (and within the same field of view of $0.15$~pc) for all the cores. The polarization images were analyzed to infer polarization position angle (PA) dispersions using four different approaches. In addition, \htcop\,(4--3) data from the SMA observations were used to infer velocity dispersions for each core. Finally, the temperature and density structure were modeled for each massive dense core using submillimeter continuum emission from single-dish telescopes and the Spectral Energy Distribution, following Palau et al. (2014). All the quantities were measured in a uniform way and within the same field of view of 0.15~pc. The aforementioned inferred properties of the massive dense cores allowed us to calculate magnetic field strengths using the DCF and ADF methods, and search for possible trends between the fragmentation level and any of the derived properties of the parental cores. Our main conclusions can be summarized as follows:

\begin{itemize}

\item[-] A total number of 160 fragments have been identified within the 18 massive dense cores. We have assigned a fragmentation level within a field of view of 0.15 pc, \Nmm, to each massive dense core.
%as the number of fragments above a 6$\sigma$ threshold in the submillimeter continuum SMA images obtained using extended configurations, and 
We found a variety of fragmentation levels, with 17\% of the cores presenting almost no fragmentation, and 39\% of the cores presenting a high fragmentation level. Additionally, cores were classified according to their fragmentation type, mainly `aligned fragmentation' (7 cores), `clustered fragmentation' (8 cores) and `no fragmentation' (3 cores).

\item[-] The inferred power-law indices for the density of the massive dense cores range from 1.46 to 2.26. The densities, masses, and temperatures, all (averaged) within 0.15~pc, range from 1.1 to 10.5 $\times10^5$~\cmt, from 11 to 105~\mo, and from 23 to 120~K, respectively. 

\item[-] The line widths of the \htcop\,(4--3) transition measured in each core range from 1.3 to 6.7~\kms, and no clear trend was found between \Nmm\ and these line widths.

\item[-] Four approaches were used to estimate polarization PA dispersions. First, the PA dispersion was estimated from the standard deviation of the PA corrected for the PA uncertainties. Second, different Gaussians were fitted to the PA histograms. Additionally, the Angular Dispersion Function analysis was performed following Houde et al. (2009), and the PA dispersion was estimated from the smallest value (beam scale) of the ADF. It was found that the PA dispersion, inferred from the standard deviation, correlates with the corresponding quantity from the ADF analysis, and the first one was used as reference to calculate the magnetic field strengths. In combination with the line widths and average densities, this yielded magnetic field strengths ranging from 0.2 to 4.1~mG.

\item[-] When considering the entire sample, a strong correlation of \Nmm\ with density averaged within 0.15pc is found although with significant scatter. In addition, \Nmm\ seems to tentatively correlate with the mass-to-flux ratio, once the large-scale systematic motions are properly separated from the velocity dispersion in the magnetic field strength calculation. These findings clearly need to be studied in larger and more sensitive samples.

\item[-] The separation of the large-scale systematic motions from the small-scale (turbulent) motions was performed through the analysis of the Velocity Dispersion Function, allowing us to calculate that the turbulent velocity dispersion is typically 40\% of the non-thermal velocity dispersion.

\item[-] The sizes and masses of each fragment were measured. The average masses of the fragments ranged from 2--4~\mo\ in most cases, comparable to the thermal Jeans mass. Regarding the sizes of the fragments, hints of more compact and less massive fragments for stronger magnetic fields were found, suggesting that in the cases of strong magnetic field this might slow down the accretion process compared to the non-magnetic case. In addition, for the strong magnetic field cases, fragmentation seems to take place along a preferred direction perpendicularly to the magnetic field.

\end{itemize}

In summary, our entire sample of massive dense cores presents a  strong correlation of the fragmentation level with the density of the parental core, and a tentative trend of the fragmentation level with the mass-to-flux ratio. In addition, hints were found of the magnetic field influencing the fragmentation process (size and mass of the fragments) for the cores with strongest magnetic fields. 
%Overall, an interplay between magnetic field and gravity must be taken into account to explain the details of the fragmentation process in the massive dense cores of our sample.
Overall, the observed properties of our sample are consistent with thermal Jeans fragmentation, and the magnetic field seems to act as a modulating process required to explain the finer details of the fragmenting cores.

%From CONACyT proposal: Gravity is the main agent defining the global collapse and fragmentation of molecular clouds, but magnetic fields, shear, and rotation act as modulating processes."

\acknowledgments

The authors are deeply grateful to the anonymous referee, whose insightful comments have undoubtedly improved the quality of this paper.
AP is grateful to Javier Ballesteros-Paredes, Raphael Skalidis, Konstantinos Tassis and Enrique V\'azquez-Semadeni for estimulating and thoughtful discussions, to Manuel Zamora-Avil\'es for calculating the magnetic field strength within a region of 0.15~pc of diameter in the simulations of Ju\'arez et al. (2017) of N6334V, and to Daniel J. D\'iaz Gonz\'alez for his support with Python. AP acknowledges financial support from CONACyT and UNAM-PAPIIT IN113119 grant, M\'exico.
JMG and RE are supported by the Spanish grant AYA2017-84390- C2-R (AEI/FEDER, UE).
JL and KQ are supported by National Key R\&D Program of China No. 2017YFA0402600. JL and KQ acknowledge the support from National Natural Science Foundation of China (NSFC) through grants U1731237, 11590781, and 11629302.
PMK acknowledges support from Ministery of Science and Technology (MoST) grants 
MOST 108-2112-M-001-012 and MOST 109-2112-M-001-022 in Taiwan, and from an Academia Sinica Career Development Award.
HBL is supported by the Ministry of Science and Technology (MoST) of Taiwan (Grant Nos. 108-2112-M-001-002-MY3).
ZYL is supported in part by NSF AST-1716259 and 1815784.
LAZ is grateful to CONACyT, M\'exico, and DGAPA, UNAM for their financial support.
H.B. acknowledges support from the European Research Council under the European Community's Horizon 2020 framework program (2014-2020) via the ERC Consolidator Grant ‘From Cloud to Star Formation (CSF)' (project number 648505). HB also acknowledges funding from the Deutsche Forschungsgemeinschaft (DFG) via the Collaborative Research Center (SFB 881) ‘The Milky Way System' (subproject B1).

Part of the data used in this paper were obtained in the SMA legacy project: Filaments, Magnetic Fields, and Massive Star Formation (PI: Qizhou Zhang).
This  research  made  use  of  Astropy\footnote{http://www.astropy.org}, a  community--developed  core  Python  package  for  Astronomy  (AstropyCollaboration et al. 2013, 2018), and in particular the numpy (van der Walt et al. 2011).
The Karma software was used for part of the analysis of this paper (Gooch 1996).

%We thank all the people that have made this AASTeX what it is today.  This includes but not limited to Bob Hanisch, Chris Biemesderfer, Lee Brotzman, Pierre Landau, Arthur Ogawa, Maxim Markevitch, Alexey Vikhlinin and Amy Hendrickson. Also special thanks to David Hogg and Daniel Foreman-Mackey for the new "modern" style design. Considerable help was provided via bug reports and hacks from numerous people including Patricio Cubillos, Alex Drlica-Wagner, Sean Lake, Michele Bannister, Peter Williams, and Jonathan Gagne.

\vspace{5mm}
\facilities{SMA, PdBI, NOEMA, JCMT(SCUBA)}


\software{astropy (AstropyCollaboration et al. 2013),  
GILDAS,
Karma
%         Cloudy \citep{2013RMxAA..49..137F}, 
%          SExtractor \citep{1996A&AS..117..393B}
}



\appendix

\section{Tests for biases of fragmentation level \Nmm}\label{apbias}

Fig.~\ref{fbias} presents the plots of \Nmm\ (Section~\ref{sr}) vs the mass sensitivity, column density sensitivity, the evolutionary indicator $L_\mathrm{bol}/M_\mathrm{core}$ (Molinari et al. 2016) and the spatial resolution for each region (Table~\ref{tsample}). If \Nmm\ were biased with the mass sensitivity or the spatial resolution one would expect a large \Nmm\ for smaller (better) mass sensitivities, and a large \Nmm\ for smaller (better) spatial resolutions. As can be seen from the figure, none relation can be appreciated of \Nmm\ with any of the tested quantities.

\renewcommand{\thefigure}{13}
\begin{figure}[h]
\begin{center}
\begin{tabular}[b]{c}
        \epsfig{file=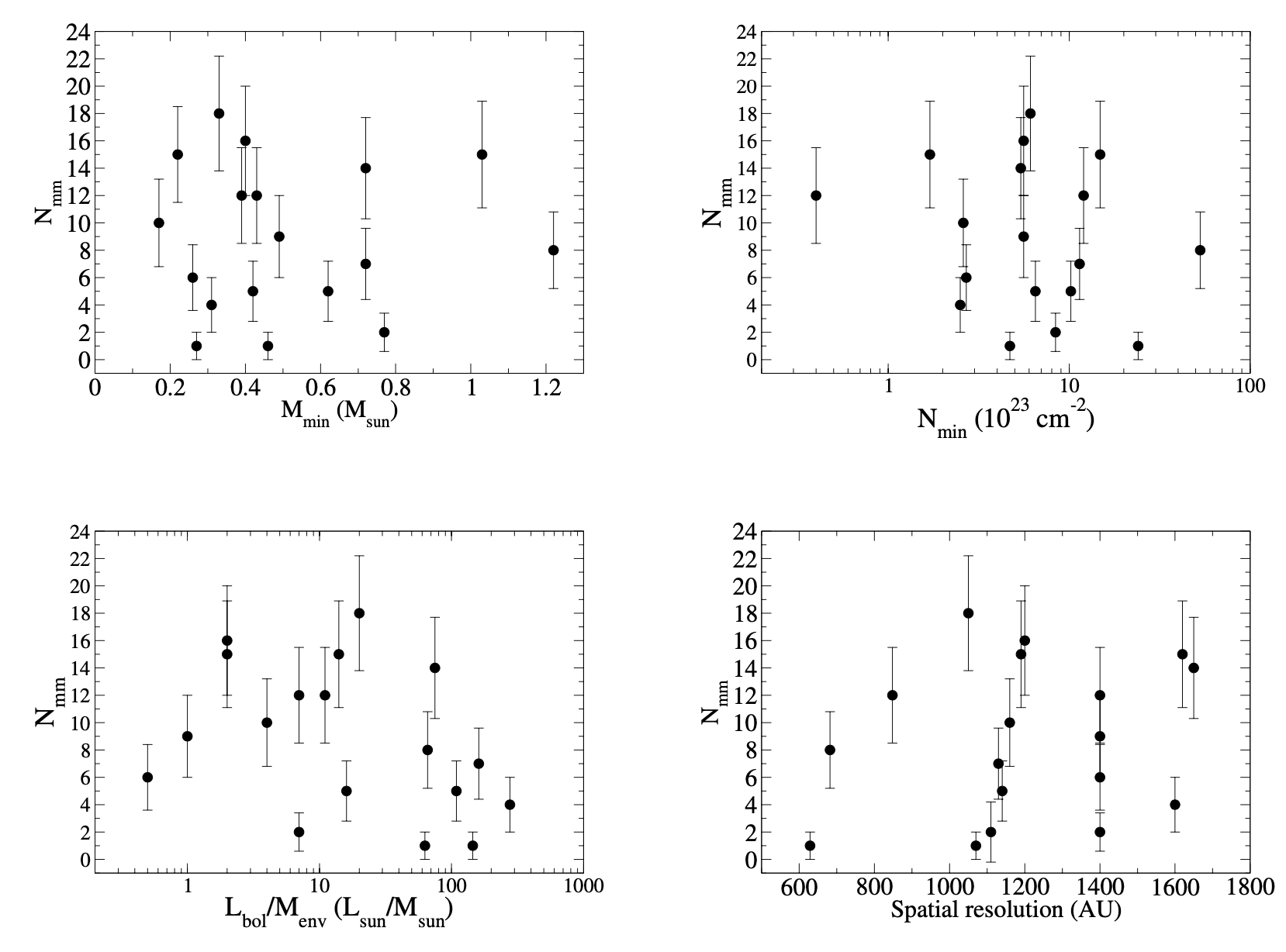, width=11cm, angle=0}\\
\end{tabular}
\caption{
Top-left: Fragmentation level \Nmm\ vs the mass sensitivity for each massive dense core of our sample.
Top-right:  Fragmentation level \Nmm\ vs the column density sensitivity (calculated from the mass sensitivity and the beam of our observations).
Bottom-left: Fragmentation level \Nmm\ vs $L_\mathrm{bol}/M_\mathrm{core}$.
Bottom-right: Fragmentation level \Nmm\ vs the spatial resolution. These figures show that there are no biases of  \Nmm\ with respect to any of these observational parameters.
}
\label{fbias}
\end{center}
\end{figure}

\section{First-order moments for \htcop\,(4--3) data}\label{apm1}

Figs.~\ref{fm1a} and \ref{fm1b} show the first-order moments of the \htcop\,(4--3) transition in colorscale, with the magnetic field segments overplotted as well as arrows for the known outflows in each core (taken from the literature).

\renewcommand{\thefigure}{14a}
\begin{figure*}[ht]
\begin{center}
\begin{tabular}[b]{c}
    \epsfig{file=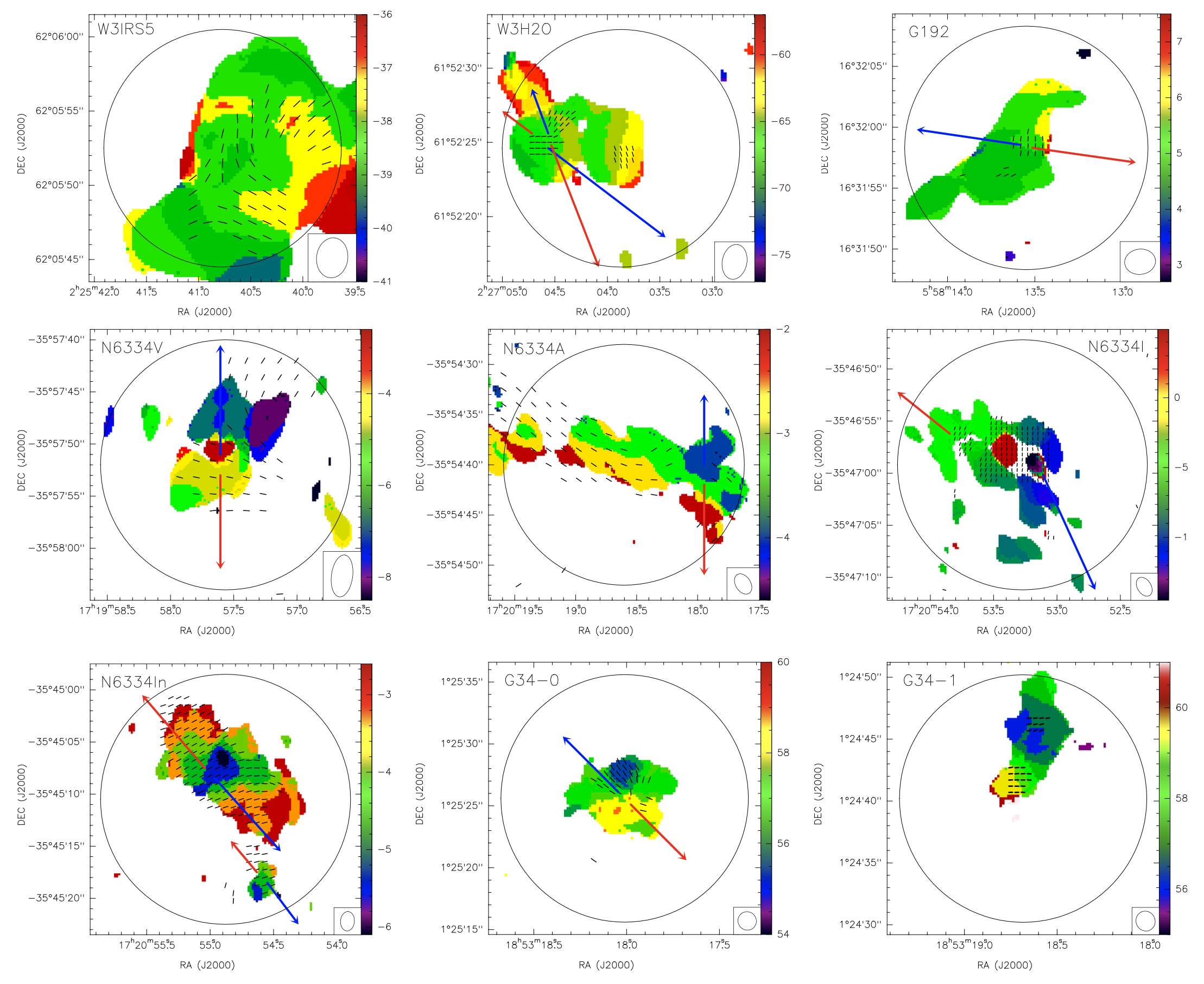, width=15cm, angle=0}\\
\end{tabular}
\caption{First-order moments for \htcop\,(4--3) data with the magnetic field segments (black) overplotted. Red and blue arrows indicate the approximate orientations of the redshifted and blueshifted outflow emission according to Zapata et al. (2011), and Zhang et al. (2014). The synthesized beam is shown in the bottom-right corner. The circle corresponds to the field of view of 0.15 pc of diameter used in this work to assess the magnetic field strength. Wedge units are \kms.
}
\label{fm1a}
\end{center}
\end{figure*}

\renewcommand{\thefigure}{14b}
\begin{figure*}[ht]
\begin{center}
\begin{tabular}[b]{c}
    \epsfig{file=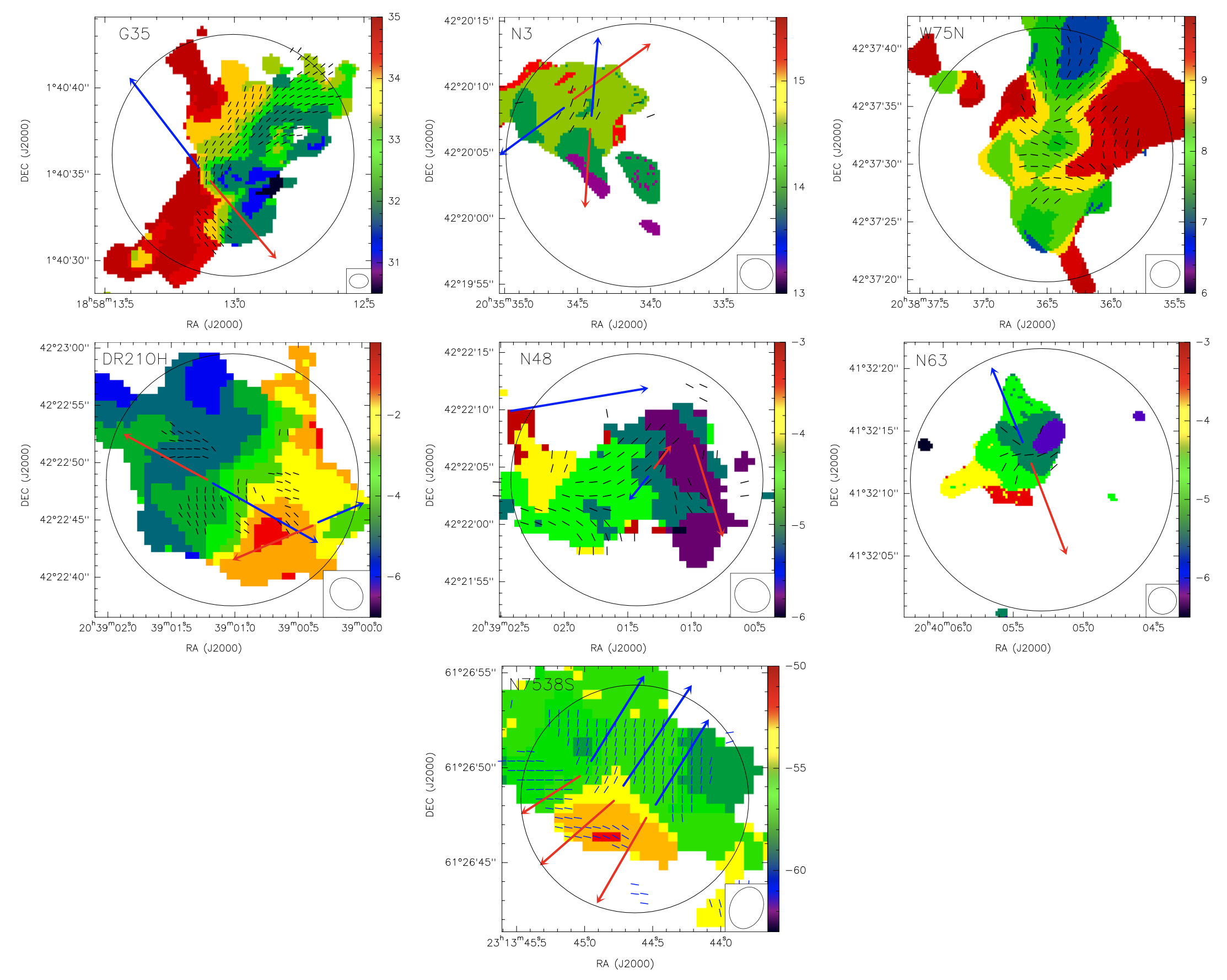, width=15cm, angle=0}\\
\end{tabular}
\caption{First-order moments for \htcop\,(4--3) data with the magnetic field segments (black) overplotted. Red and blue arrows indicate the approximate orientations of the redshifted and blueshifted outflow emission according to Naranjo-Romero et al. (2012), Duarte-Cabral et al. (2013, 2014), Girart et al. (2013), and Zhang et al. (2014). The synthesized beam is shown in the bottom-right corner. The circle corresponds to the field of view of 0.15 pc of diameter used in this work to assess the magnetic field strength. Wedge units are \kms.
}
\label{fm1b}
\end{center}
\end{figure*}

\section{Particular cases in the density and temperature structure modeling of Section~4.1}\label{apdensity}

In Section~\ref{sadensity} a density and temperature structure model was fit to observational data for each region of our sample. The observational data consisted of radial intensity profiles and the spectral energy distribution, and the model allowed us to infer an average density within 0.15 pc for each region. We comment here on particularities of some of the sources to perform the fit.

\paragraph{W3H2O} The model includes an optically thin radio source of 2.5 Jy at 30~GHz (Dreher \& Welch 1981).

\paragraph{N6334A} The assumption of the core being centrally heated might not be fulfilled because there is a Herschel core (core number 38 from Tig\'e et al. 2017) that only lies about $\sim13''$ from the SCUBA peak at 450~\mum\ and for which the flux density at 70~\mum\ is a factor of $\sim20$ larger than the flux at 70~\mum\ for the core directly associated with the 450~\mum\ peak. In this case we did several tests to fit the model. First, to build the SED we considered only the Herschel intensities associated with core number 41 (core associated with the 450~\mum\ peak). This yielded an average density within 0.15~pc of $4.8\times10^5$~\cmt. This should be a reasonable approach as long as we are considering only peak intensities of the core directly associated with the SCUBA peak, and the excess due to heating by core 38 (and 10) should not be strong because its effect should be only in specific directions compared to the entire radially averaged profile. Second, we considered only the contribution of core 41 and used peak intensities to build the SED, except for Herschel wavelengths where the beam cannot separate the different sources (\ie\ 250, 350 and 500~\mum, where the beam is $>12''$). For these wavelengths we included in the SED the flux density of all three cores and used the core size as aperture radius for the model to compute the flux. This method yielded an average density within 0.15~pc of $3.8\times10^5$~\cmt, and in Table~\ref{tfit} and Fig.~\ref{fNmmBsimdens} we use the fitted values corresponding to this second method, to be conservative.

\paragraph{G35} The IRAS flux at 100~\mum\ is a factor of 3 smaller than the Herschel-PACS measurements at 70 and 160~\mum. We calculated the model including both Herschel and IRAS fluxes (best-fit values reported in Table~\ref{tfit}, yielding an average density within 0.15~pc of $2.96\times10^5$~\cmt).

%\section{Fragmentation level vs velocity dispersion and PA dispersion using different methods and approaches}\label{apvelsigma}

%The fragmentation level vs the velocity dispersion and the PA dispersion are presented in Fig.~\ref{fNmm-sigmaPA}.

%\renewcommand{\thefigure}{15}
%\begin{figure*}[ht]
%\begin{center}
%\begin{tabular}[b]{cc}
%    \epsfig{file= Nmm_Dv_errors.pdf, width=6.5cm, angle=0}&
%    \epsfig{file= Nmm_sigmaPAstdev_errors.pdf, width=6.5cm, angle=0}\\
%    \epsfig{file= Nmm_sigmaPAgauss_errors.pdf, width=6.5cm, angle=0}&
%    \epsfig{file= Nmm_BtBo_errors.pdf, width=6.5cm, angle=0}\\
%\end{tabular}
%\caption{
%Top-left: Fragmentation level vs line width of the \htcop\,(4--3).
%Top-right: Fragmentation level vs polarization PA intrinsic standard deviation of the weighted mean (Section~\ref{saBstdev}).
%Bottom-left:  Fragmentation level vs polarization PA dispersion after fitting multiple PA components in the PA histograms (Section~\ref{saBgauss}).
%Bottom-right: Fragmentation level vs energy ratio between the perturbed magnetic field and the large-scale ordered magnetic field (Section~\ref{saBADF}).
%}
%\label{fNmm-sigmaPA}
%\end{center}
%\end{figure*} 

\section{Uncertainty estimates running Monte Carlo simulations}\label{apMonteCarlo}

One of the main uncertainties associated with the derivation of the magnetic field strength or $\mu$ from polarized submillimeter emission is the sparse sampling of the data. In order to have a first estimate of these uncertainties, Monte Carlo simulations were run as described in the Appendix of Liu et al. (2019). For each region, we modeled the large-scale field as a parabola of the form $y = g + g\,C\,x^2$, with $C$ being the curvature parameter set as $C=\sigma_\mathrm{PA,stdev}/3000$. A random dispersion equal to twice the average error in PA of each region was introduced. This was run 10 times for each region. In each run, two PA maps were produced: the unbiased (fully sampled) PA map, and the sparsely sampled map specific to each region. Then, for each run and for each unbiased/sparsely sampled map, we applied our four approaches used in the analysis to estimate the PA dispersions (Sections~\ref{saBstdev}, \ref{saBgauss}, \ref{saBADFbeam}, and \ref{saBADFH09}), and measured the difference in the dispersion between the unbiased and the sparsely sampled images. This gave us an idea of how far the measured dispersion might be from the unbiased 'real' dispersion. We then averaged these differences for the 10 runs (for each approach) and took this average difference as the uncertainty in each case. The uncertainties in dispersion were propagated to the magnetic field strength and mass-to-flux ratio. In general, we obtained larger uncertainties for the poorly sampled regions, as expected.

\section{Magnetic field strength vs density}\label{apBstdevn}

In Fig.~\ref{fBstdevn} we present a plot of the magnetic field strength as calculated in Section~\ref{saBstdev}, $B_\mathrm{stdev}$ (Table~\ref{tB}),  vs density averaged within 0.15~pc of diameter (Table~\ref{tfit}). The log-log plot shows a trend, with a slope of $1.1\pm0.3$.

\renewcommand{\thefigure}{15}
\begin{figure}[ht]
\begin{center}
\begin{tabular}[b]{c}
    \epsfig{file=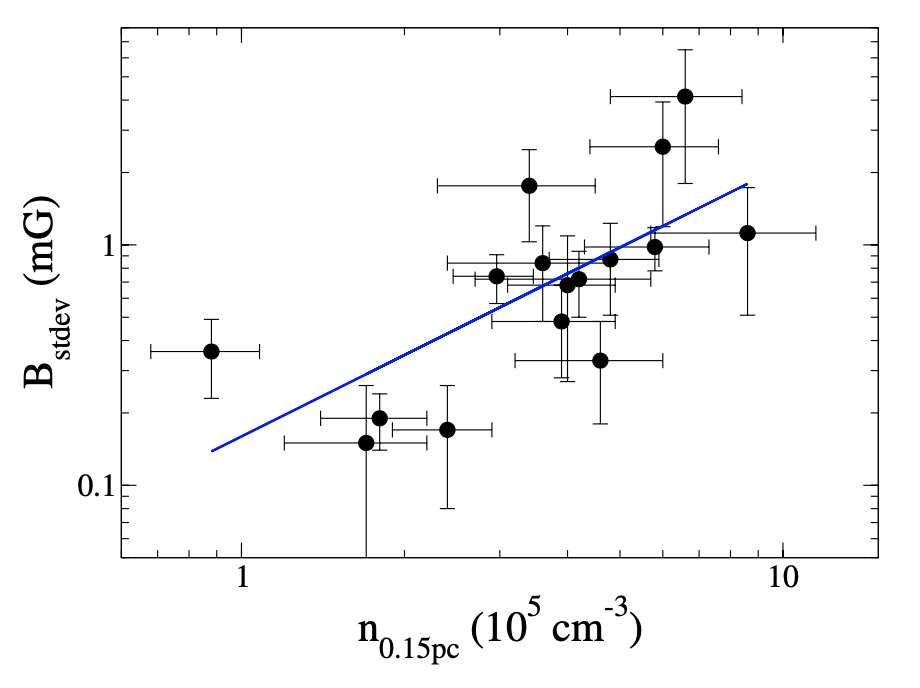, width=7cm, angle=0}\\
\end{tabular}
\caption{Magnetic field strength calculated with the `standard deviation' approach, $B_\mathrm{stdev}$, vs density averaged within 0.15~pc. 
The blue line corresponds to a linear fit in logarithmic scales with a slope of $1.1\pm0.3$ and a correlation coefficient of 0.69.
}
\label{fBstdevn}
\end{center}
\end{figure}

\section{Comparison of SMA and single-dish densities}\label{apsma}

In Section~\ref{sadensity}, the density structure for each massive dense core was inferred using data from single-dish telescopes, mainly the James Clerk Maxwell Telescope (with main beams of $19.5''$ at 850~\mum\ and $11.0''$ at 450~\mum) and the IRAM\,30m telescope (with a main beam of $11.0''$ at 1.2~mm). From this model, the average density within 0.15 pc was derived. In order to asses if the different spatial filtering of the single-dish telescopes and the SMA is affecting our determination of the density and the magnetic field strength, we compare here how the inferred densities in Section~\ref{sadensity} compare to the densities inferred using the SMA data.

The average density (within 0.15~pc of diameter) was estimated using the SMA continuum flux densities, including all the configurations available as for the case of the \htcop\ and PA data. This includes in many cases the subcompact configuration and in all cases the compact configuration of the SMA. These configurations allow to recover angular scales as large as $30''$ and $14''$, respectively (following the appendix of Palau et al. 2010). Therefore, using these SMA data we are sensitive to scales comparable to the scales of the single-dish telescopes.  

To infer the masses and densities from the SMA data, we measured the flux density within the region of 0.15 pc of diameter, and assumed the average temperature within the same diameter inferred from our modeling (given in Table~\ref{tfit}), as well as the opacity law of Ossenkopf \& Henning (1994, grains covered by thin ice mantles at 10$^6$~\cmt, 0.0175 cm$^2$ per gram of gas and dust at 870~$\mu$m). Fig.~\ref{fnsma} presents the relation between the SMA average density (within 0.15~pc of diameter) vs the average density inferred using the modeling of the single-dish data presented in Section~\ref{sadensity}. As can be seen from the figure, by doing this we recover with the SMA $\sim75$\% of the mass inferred from the modeling of the single-dish data presented in Section~\ref{sadensity}. 

\renewcommand{\thefigure}{16}
\begin{figure}[ht]
\begin{center}
\begin{tabular}[b]{c}
    \epsfig{file=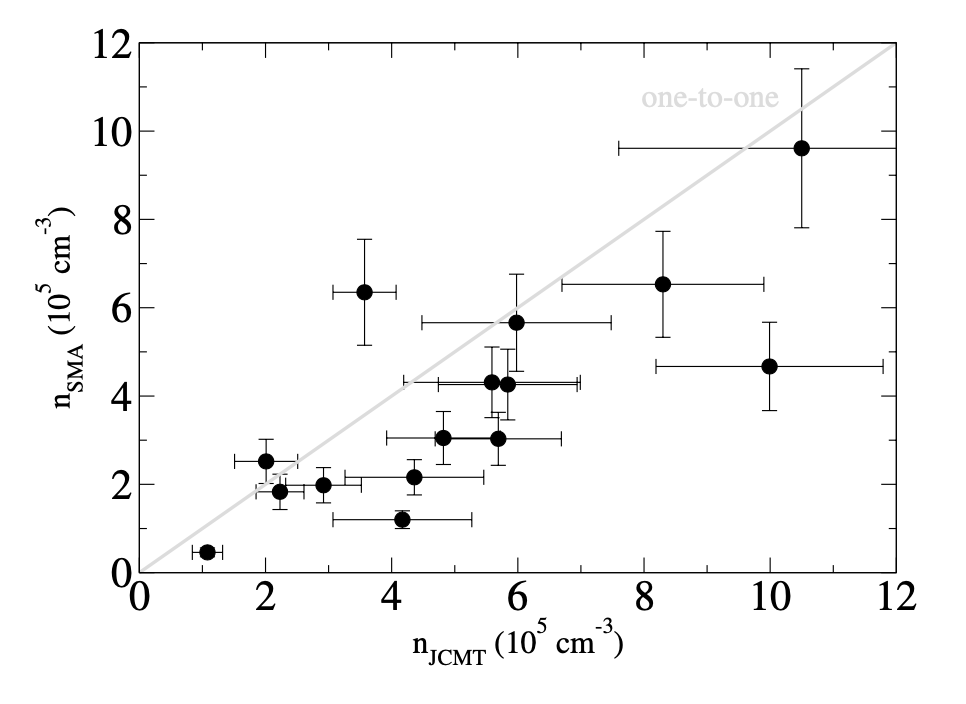, width=7cm, angle=0}\\
\end{tabular}
\caption{Plot of the SMA average density (within 0.15~pc of diameter) vs the average density inferred using the modeling of the single-dish data presented in Section~\ref{sadensity}.
The light grey line indicates the one-to-one relation to guide the eye. To make the comparison, we converted the densities reported in Table~\ref{tfit}, $n_\mathrm{0.15pc}$, from density of H$_2$ particles to total density of particles. The uncertainties for the SMA average density have been adopted to be 20\%, the typical uncertainty in the flux absolute scale.
%Fragmentation level vs magnetic field strength calculated using the SMA average density.
}
\label{fnsma}
\end{center}
\end{figure}

\section{Comparison of magnetic field strength inferred from the simulations of a globally collapsing and hierarchical cloud and one of the cores studied here}\label{apsimulations}

We compared the magnetic field strength inferred in simulations dominated by gravity with the measured strength from observations for the particular case of N6334V, where simulations of a collapsing magnetized molecular cloud were specifically tuned to explain the magnetic field and dynamics of the region (Ju\'arez et al. 2017). In these simulations, the gas flows from the large scales towards the center of the massive dense cores, and the magnetic field is dragged by the gas. The magnetic field in the simulations was measured within 0.15 pc of the massive dense core and a value of  $0.94$~mG was found (M. Zamora-Avil\'es, priv. communication), very similar to the values obtained here for N6334V using the `standard deviation' or `ADF H09' approaches, of 0.6--0.7~mG, and only a factor of 2 smaller than the values obtained with the `multiple gaussians' or `ADF beam' approaches. Thus, for the particular case of N6334V, the magnetic field measured in the simulations is very comparable to the magnetic field inferred in our work. 
%This good match could be due to the fact that the velocity dispersion in our data is also arising from the large-scale flows.
%(Magnetic Field Average =  0.000937632539225765 gauss).

\section{Comparison of magnetic field strength inferred in this work and the magnetic field strength in other works of the literature}\label{apBlit}

Here we provide details, for each region, of previous works reporting magnetic field strengths in regions of our sample.

\paragraph{DR21OH, N3, N53} Hezareh et al. (2010, 2013) estimated the magnetic field in the DR21 region, including DR21OH, N3 and N53, by comparing the velocity dispersions of ion and neutral pairs at different length scales, and find values in the range 0.33--1.8~mG (for densities very similar to the ones we obtained here). 
%$10^6$~\cmt\ for DR21OH, $3.6\times10^5$~\cmt\ for CygX-N3, and $2.2\times10^5$~\cmt\ for N53.

\paragraph{NGC\,6334} Li et al. (2015) present evidence that magnetic fields regulate the dynamics in NGC\,6334, based on their findings of hourglass-shaped field lines at gas column density peaks and from the fact that the field strength is found to be proportional to the 0.4-power of the density. They infer the magnetic field strength by assuming force equilibrium between gravity, magnetic tension and magnetic pressure at different scales, and derive a relation between magnetic field strength and density. The value given in Table~\ref{tB} for the magnetic field strength in N6334I and N6334In is the one corresponding to the density given in Table~\ref{tfit} and applying the aforementioned relation between field strength and density.
%Comparison of Li+15 to Palau+20:
%for source I i obtained: 8.3x10^5 cm-3 ->
%8.3e5*2.8*1.6733e-24*pow(3.0857e18,3)/1.9899e33
%=57416.90575293729 Msun/pc3 -> 574 (100Msun/pc-3)
%log10(574)=2.7589118923979736
%corresponds to ~ log(B)=0.45 mG -> 2.8 mG
%for source In i obtained: 1x10^6 cm-3 ->
%1e6*2.8*1.6733e-24*pow(3.0857e18,3)/1.9899e33
%=69176.99488305698 Msun/pc3 -> 692 (100Msun/pc-3)
%log10(692)=2.8401060944567575
%corresponds to ~ log(B)=0.5 mG -> 3.16 mG

\paragraph{G34} Tang et al. (2019) report submillimeter polarization observations using the Caltech Submillimeter Observatory and infer the magnetic field strength using the `ADF beam' approach for both G34-0 and G34-1 (Table~\ref{tB}; Tang et al. (2019) report densities of $1.6\times10^5$~\cmt\ for these two cores).

\paragraph{W3H2O} Chen et al. (2012a) estimate the magnetic field strength using the DCF method with a density of $1.5\times10^7$~\cmt, and obtained 17 mG. To compare this value to our measurement, we scaled the magnetic field strength assuming a dependence with density as a power law with index 0.4 (Li et al. 2015), and used our estimate for the density reported in Table~\ref{tfit}, obtaining a value of 4.6~mG.
%B = B0*(n/n0)^0.4 = 17*pow(5.8e5/1.5e7,0.4) = 4.6 mG

\paragraph{G35} Qiu et al. (2013) report a value of the magnetic field strength of 0.9--1.4 using the same SMA dataset used in this work and the `ADF H09' method. 
%(the range of values corresponds to the different adopted effective depths).

\paragraph{I20126} This region has been observed with the SMA in submillimeter polarization but no sufficient detections were found to estimate a reliable PA dispersion (H. Shinnaga et al., in preparation). However, Edris et al. (2005) perform estimates of the magnetic field at scales of $\sim1000$~AU through Zeeman splitting of OH masers, and obtain 11 mG within $0.5''$. Assuming that the density at $\sim1000$~AU is around $8.4\times10^{-17}$~g\,\cmt\ (Table~\ref{tfit}), we estimate a magnetic field strength of 1.8~mG for the density we have calculated within 0.15 pc.
%n(1000AU)=8.4e-17/(2.8*1.6733e-24)=17928644 ~ 1.8e7 cm-3
%B = B0*(n/n0)^0.4 = 11*pow(2e5/1.8e7,0.4) = 1.8 mG

Fig.~\ref{fBlit} presents a plot comparing the magnetic field strength reported in the literature with the magnetic field strength derived in this work for the `standard deviation' approach.

\renewcommand{\thefigure}{17}
\begin{figure}[ht]
\begin{center}
\begin{tabular}[b]{c}
    \epsfig{file=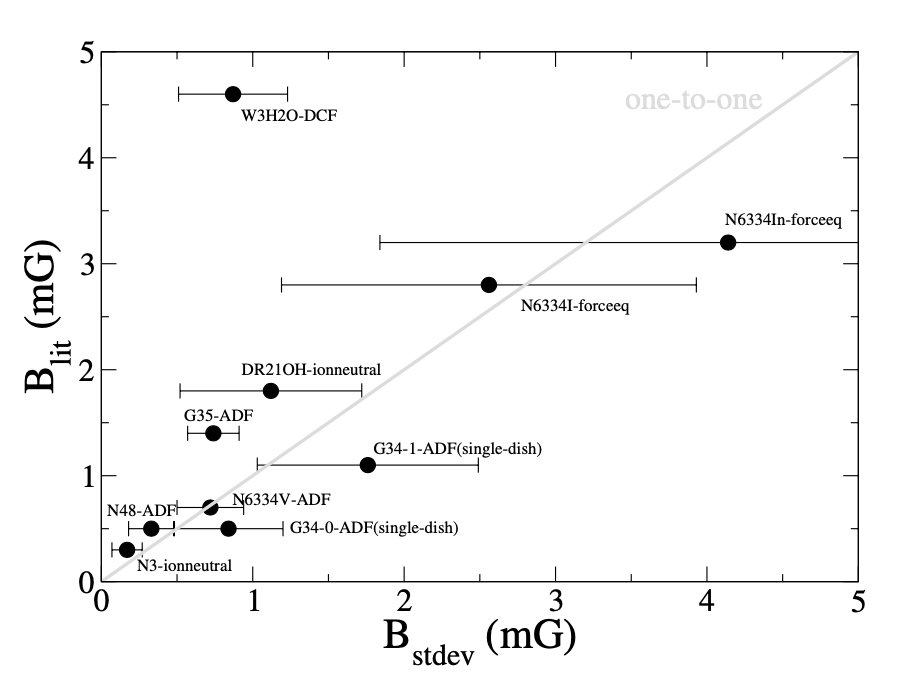, width=7cm, angle=0}\\
\end{tabular}
\caption{Comparison of the magnetic field determined in the literature and the magnetic field determined in this work using the `standard deviation' approach, $B_\mathrm{stdev}$. The light grey line corresponds to the one-to-one relation to guide the eye. Next to the name of each source, there is a short-name for the method used to determine the magnetic field strength. See Table~\ref{tB} for further details.
}
\label{fBlit}
\end{center}
\end{figure}

%\section{Fragmentation level vs normalized mass-to-flux ratio}\label{apNmm-munorm}

%\renewcommand{\thefigure}{18}
%\begin{figure}[ht]
%\begin{center}
%\begin{tabular}[b]{c}
%    \epsfig{file= Nmm_mustdev-normalized.pdf, width=6.5cm, angle=0}\\
%\end{tabular}
%\caption{{\bf The inverse trend is reflecting the dependence of the fragmentation level with density.}
%}
%\label{fBlit}
%\end{center}
%\end{figure} 

%% bibtex sample63

%\bibliography{sample63}{}
%\bibliographystyle{aasjournal}

%%%%%%%%%%%%%%%%%%%%%%%%%%%%%%%%%%%%%%%%%%%%%%%%%%%%%%%%%%%%%%%%%%%%%%%%%%%%%%%%



\end{document}